\documentclass[12pt,english]{article}
\usepackage[T1]{fontenc}
\usepackage[utf8]{inputenc}
\usepackage[pdftex]{geometry}
\usepackage{amsmath}
\usepackage{amssymb}
\usepackage[pdftex]{graphicx}
\usepackage{setspace}
\makeatletter

\providecommand{\tabularnewline}{\\}

\numberwithin{equation}{section}
\newcommand{\lyxaddress}[1]{
	\par {\raggedright #1
	\vspace{1.4em}
	\noindent\par}
}

\allowdisplaybreaks[1]

\usepackage{braketmod}

\usepackage{color}
\usepackage[pdftex]{hyperref}
\hypersetup{
colorlinks,
linktoc=page,
citecolor=blue,
linkcolor=blue,
urlcolor=blue
}

\date{}

\makeatother

\usepackage{babel}
\usepackage[style=phys,biblabel=brackets, chaptertitle=false, pageranges=false, eprint=true]{biblatex}
\begin{document}
\noindent\begin{minipage}[t]{1\columnwidth}%
\title{\textbf{Five-brane webs, 3d $\mathcal{N}=2$ theories}\\
\textbf{and quantum curves}}
\author{Naotaka Kubo\,\footnotemark}
\maketitle

\lyxaddress{\begin{center}
\vspace{-18bp}
$^{*}$\,\textit{Center for Joint Quantum Studies and Department of Physics, School of Science,}\\
\textit{Tianjin University, 135 Yaguan Road, Tianjin 300350, China}\vspace{-10bp}
\par\end{center}}
\begin{abstract}
We propose a relation between the brane configurations consisting of D3-branes and 5-brane webs which realize 3d $\mathcal{N}=2$ supersymmetric Chern-Simons theories and quantum curves by focusing on the $S^{3}$ partition functions. In particular, we conjecture that the Newton polygons of the quantum curves are equal to the toric diagrams which are dual to the 5-brane webs. For brane configurations whose worldvolume theories have Lagrangian descriptions, we show an explicit derivation of the relation by using the supersymmetric localization and the Fermi gas formalism. We also provide some evidence of the conjecture for non-Lagrangian theories. We see that our conjecture gives us new matrix models for 5-brane webs including a $\left(p,q\right)$5-brane with arbitrary $p$. This leads to explicit relations between the brane configurations, the matrix models and genus one quantum curves.
\end{abstract}
\end{minipage}

\renewcommand{\thefootnote}{\fnsymbol{footnote}}
\footnotetext[1]{\textsf{naotaka.kubo@yukawa.kyoto-u.ac.jp}}
\renewcommand{\thefootnote}{\arabic{footnote}}\medskip{}
\thispagestyle{empty}

\newpage{}

\setcounter{page}{1}

\global\long\def\bra#1{\Bra{#1}}%

\global\long\def\bbra#1{\Bbra{#1}}%

\global\long\def\ket#1{\Ket{#1}}%

\global\long\def\kket#1{\Kket{#1}}%

\global\long\def\braket#1{\Braket{#1}}%

\global\long\def\bbraket#1{\Bbraket{#1}}%

\global\long\def\brakket#1{\Brakket{#1}}%

\global\long\def\bbrakket#1{\Bbrakket{#1}}%

\global\long\def\bw{\mathsf{W}}%

\global\long\def\lp{\mathsf{TD}}%

\global\long\def\wdl{-}%

\global\long\def\wdr{+}%

\global\long\def\df{\mathrm{D5}}%

\global\long\def\dfl{\mathrm{D5}^{-}}%

\global\long\def\dfr{\mathrm{D5}^{+}}%

\tableofcontents{}

\section{Introduction\label{sec:Introduction}}

Curves have the potential to provide a bridge between two or more different theories and give new perspectives. For example, the Seiberg-Witten curves of 4d $\mathcal{N}=2$ or 5d $\mathcal{N}=1$ supersymmetric theories \cite{Seiberg:1994rs,Witten:1997sc} play an important role in the connection with integrable systems. The discovery of the instanton partition function on the $\Omega$-background \cite{Nekrasov:2002qd} by the equivariant localization gives a realization of the connection. The instanton partition function was first linked to quantum integrable systems in \cite{Nekrasov:2009rc}. More recently, it has been found that the $\mathrm{SU}\left(2\right)$ Nekrasov-Okounkov partition functions \cite{Nekrasov:2003rj} in the self-dual $\Omega$-background can be regarded as the $\tau$-functions of the Painlev\'e equations \cite{Gamayun:2012ma,Gamayun:2013auu,Iorgov:2013uoa,Iorgov:2014vla,Its:2014lga,Bershtein:2014yia,Gavrylenko:2016zlf,Nagoya:2016mlj,Bonelli:2016qwg,Gavrylenko:2017lqz}. The 5d uplift of this correspondence has been considered and investigated in \cite{Bonelli:2016qwg,Bonelli:2016idi,Bershtein:2016aef,jimbo2017cft,Mironov:2017sqp}. On the Painlev\'e equation side, the 5d uplift corresponds to a lift from differential to difference equations, which is called $q$-Painlev\'e equations (see \cite{Kajiwara_2017} for a review). On the other hand, the uplift of the 4d theories, which are connected with topological strings via the geometric engineering \cite{Katz:1996fh}, corresponds to topological strings on toric Calabi-Yau threefolds. In this correspondence, the free energies of the topological strings are identified with the $\tau$ functions of the $q$-Painlev\'e equations. The symmetries of the 5d Seiberg-Witten curves correspond to the symmetries of the $q$-Painlev\'e equations classified in \cite{2001CMaPh.220..165S}, and the 5d Seiberg-Witten curves correspond to the mirror curves of the toric diagrams of the toric Calabi-Yau threefolds.

The connections between the integrable systems, the 4d or 5d supersymmetric theories and the topological strings are further expanded. The free energy of the topological strings can be computed by the spectral determinant of quantized mirror curves via the TS/ST correspondence \cite{Grassi:2014zfa,Codesido:2015dia,Moriyama:2020lyk} (see also the review \cite{Marino:2015nla} and references therein). Through this correspondence, the $\tau$ functions of the $q$-Painlev\'e equations are described by the spectral determinants. Interestingly, the TS/ST correspondence was originally motivated by the ABJM theory. The ABJM theory is a 3d $\mathcal{N}=6$ $\mathrm{U}\left(N_{1}\right)\times\mathrm{U}\left(N_{2}\right)$ superconformal Chern-Simons (CS) theory and describes the worldvolume theory of M2-branes on $\mathbb{C}^{8}/\mathbb{Z}_{k}$ \cite{Aharony:2008ug,Hosomichi:2008jb,Aharony:2008gk}. The $S^{3}$ partition function reduces to a matrix model via the supersymmetric localization \cite{Pestun:2007rz,Kapustin:2009kz}. The ABJM matrix model is identified as a CS matrix model on a lens space via an analytic continuation, and it is further related to the free energy of the topological string via the relation between the CS theories and the topological strings \cite{Marino:2009jd}. On the other hand, it was found that the ABJM matrix model can be written in terms of the spectral determinant via the Fermi gas formalism \cite{Marino:2011eh}, and thus the topological string is related to the spectral theory. The topological string free energy associated with the ABJM theory satisfies the $q$-Painlev\'e $\mathrm{III}_{3}$ equation, and thus the ABJM matrix model (in the grand canonical ensemble) can be interpreted as a $\tau$-function of the $q$-Painlev\'e $\mathrm{III}_{3}$ \cite{Bonelli:2017gdk}.

A conceptual importance of the TS/ST correspondence is that although it started from the specific case, the ABJM theory, once we regard it as a correspondence between the mirror curves and the quantum curves, we can immediately generalize the correspondence. Namely, the correspondence claims that the free energies of the topological strings on arbitrary toric Calabi-Yau threefolds can be computed from the spectral determinant of associated quantum curves which are quantized mirror curves. Combining the connection between the 5d Seiberg-Witten curves and the mirror curves, we find that the $q$-deformed integrable systems and the spectral problems are related by the (quantum) curves. For example, when we focus on genus one curves, this relation is expected to give explicit Fredholm determinant representations of the $\tau$-functions of the $q$-Painlev\'e equations \cite{Bonelli:2017gdk,Bonelli:2022dse}. Furthermore, when we consider higher genus curves, the corresponding integrable systems become $\mathrm{SU}\left(N\right)$ $q$-Toda equations \cite{Bonelli:2017ptp,Nosaka:2020tyv}, or by taking a 4d limit the correspondence gives explicit Fredholm determinant representations for the 4d $\mathrm{SU}\left(N\right)$ theories \cite{Bonelli:2017ptp}. In this way, the (quantum) curves provide the unified perspective of the correspondence and the generalization of it.

In spite of this story, the generalization in the 3d supersymmetric theory side has not been known yet except for several cases. Namely, our question is whether there are 3d supersymmetric theories whose $S^{3}$ partition functions are written by the spectral determinant of arbitrary quantized mirror curves. The Newton polygons of quantum curves appearing in the 3d theory side have so far been limited to rectangle cases \cite{Marino:2011eh,Nosaka:2015iiw,Kubo:2019ejc,Kubo:2020qed}. The curves having the rectangle Newton polygons arise from a generalization of the ABJM theory in the following way. In type IIB string theory, the ABJM theory is the worldvolume theory of a brane configuration consisting of D3-branes, an NS5-brane and a $\left(1,k\right)$5-brane, which can be regarded as a generalization of the Hanany-Witten setup \cite{Hanany:1996ie}. In this brane picture, we obtain the rectangle cases by increasing the number of 5-branes. For example, a genus one quantum curve associated with a brane configuration consisting of four 5-branes played a crucial role for studying a relation between the $S^{3}$ partition function of the worldvolume theory of this brane configuration and the $q$-Painlev\'e $\mathrm{VI}$ equation \cite{Bonelli:2022dse}.

For answering the question, in this paper we study more general brane configurations where 5-branes form 5-brane webs, which we call $\left(p,q\right)$ webs. The brane configurations then lead to 3d $\mathcal{N}=2$ supersymmetric CS theories. We propose that the $S^{3}$ partition function of the 3d theories can be written by a quantum curve whose Newton polygon is given by a dual toric diagram of the $\left(p,q\right)$ webs. By using this proposal, one can obtain a 3d theory computed by a quantum curve having an arbitrary Newton polygon as the worldvolume theory of a brane configuration which is the dual $\left(p,q\right)$ web of the Newton polygon.

After explaining our conjecture in detail in section \ref{subsec:Conjecture}, we confirm it for a wide class of brane configurations. In particular, we show an explicit derivation of the conjecture for Lagrangian theories. We first reduce the $S^{3}$ partition function to a matrix model by using the supersymmetric localization, and then we rewrite it as a partition function of an ideal Fermi gas system by using the Fermi gas formalism. Finally, we show that the inverse of the one-particle density matrix is a quantum curve whose Newton polygon is dual to the $\left(p,q\right)$ web. We also give some checks of our proposal from the viewpoint of the web deformations and the $\mathrm{SL}\left(2,\mathbb{Z}\right)$ dualities in type IIB string theory.

We emphasize that the conjecture connects the quantum curves with not only the 3d theories but also the brane constructions of them. Conceptually, this adds these theories to the known relations between the integrable systems, the 4d or 5d theories, the topological strings and the spectral theories. As another concrete application, we show that this connection gives us matrix model representations even for a class of non-Lagrangian theories. More explicitly, we suggest new matrix models for $\left(p,q\right)$ webs including a $\left(p,q\right)$5-brane with arbitrary $p$.

Finally, we study genus one curves related to the $q$-Painlev\'e equations in detail. We give brane contractions and matrix model representations for the genus one curves. Note that in the conjecture the commutation relation between the position and the momentum operators is fixed to be $i\hbar=2\pi i$. Nevertheless, we show that we can obtain the genus one quantum curves with $\hbar=2\pi\ell$ with $\ell\in\mathbb{N}$. The new matrix models discussed above give the matrix model representation for these curves. We also discuss the relation between a quantum curve from our conjecture and the ABJM quantum curve where they share the same Newton diagram.

This paper is organized as follows. In section \ref{sec:Reviews}, we review the brane configurations in type IIB brane string theory and 3d $\mathcal{N}=2$ gauge theories which are worldvolume theories of them. We also review the supersymmetric localization and resulting matrix models. In section \ref{sec:BW-QC}, we explain our conjecture and give some evidence of it. In section \ref{sec:pqMF-QC}, we suggest new matrix models for non-Lagrangian theories. In section \ref{sec:QC-genh}, we study genus one quantum curves in detail. We also discuss a relation between a genus one curve in our setup and a quantum curve associated with the ABJM theory. Finally, in section \ref{sec:Conclution}, we summarize our results and show future directions. In appendix \ref{sec:DoubleSine}, we enumerate properties of the double sine function. In appendix \ref{sec:QMnotation}, we provide notation and formulas of quantum mechanics.

\section{Branes, 3d theories and matrix models\label{sec:Reviews}}

In this section we first review brane configurations in type IIB string theory. The low-energy effective theories on D3-branes which are finite in a direction are described by 3d $\mathcal{N}=2$ gauge theories. We then review the supersymmetric localization technique which reduces the path integral computing the partition function on $S^{3}$ to a matrix integral. We carefully identify the relation between the position of 5-barnes relative to the D3-branes and the mass and Fayet-Iliopoulos (FI) parameters in the gauge theory. We also summarize notations.

\subsection{Brane configurations\label{subsec:BraneConfig}}

In this paper we consider brane configurations consisting of D3-branes, NS5-branes and D5-branes. Since the dimension of the NS5-brane and the D5-brane is the same, they can form the bound state. A bound state of $p$ NS5-branes and $q$ D5-branes is written as $\left(p,q\right)$5-brane, where $p$ and $q$ must be coprime. The directions 012 are common to all the branes and identified with the coordinates of the worldvolume theories. The D3-branes are along 0126, while 5-branes are along 01234$\left[5,9\right]_{\theta}$. The angle $\theta$ depends on the charge as $\tan\theta=q/p$. The direction 6 is compact, and 5-branes are separated along this direction while the D3-branes are compactified in this direction.
\begin{table}[t]
\begin{centering}
\begin{tabular}{|c|c|c|c|c|c|c|c|c|}
\hline 
 & 012 & 3 & 4 & 5 & 6 & 7 & 8 & 9\tabularnewline
\hline 
\hline 
D3 & $\bigcirc$ &  &  &  & $\bigcirc$ &  &  & \tabularnewline
\hline 
NS5 & $\bigcirc$ & $\bigcirc$ & $\bigcirc$ & $\bigcirc$ &  &  &  & \tabularnewline
\hline 
D5 & $\bigcirc$ & $\bigcirc$ & $\bigcirc$ &  &  &  &  & $\bigcirc$\tabularnewline
\hline 
$\left(p,q\right)$5 & $\bigcirc$ & $\bigcirc$ & $\bigcirc$ & $/_{\theta}^{59}$ &  &  &  & $/_{\theta}^{59}$\tabularnewline
\hline 
\end{tabular}
\par\end{centering}
\caption{The brane setup for 3d ${\cal N}=2$ Chern-Simons theories. The direction 6 is periodic. $/_{\theta}^{59}$ means that a 5-brane has an angle $\theta$ in the $59$ plane, where $\tan\theta=q/p$.\label{tab:Brane}}
\end{table}
The directions where 5-branes extend are summarized in table \ref{tab:Brane}.\footnote{The 5-branes appearing in this paper is different from ones appearing in the $\mathcal{N}\geq3$ brane setup, where $\left(p,q\right)$5-branes extend along 012$\left[3,7\right]_{\theta}$$\left[4,8\right]_{\theta}$$\left[5,9\right]_{\theta}$.} Although the number of D3-branes of each segment between 5-branes can be chosen independently, in this paper we restrict the number to be uniform $N$. Brane configurations of this kind were first studied by Hanany and Witten for studying the dynamics of the 3d supersymmetric gauge theories \cite{Hanany:1996ie} (see also \cite{Giveon:1998sr} for a review). A curious brane configuration which is related to M2-branes is the one of the ABJM theory \cite{Aharony:2008ug}.

As one can see in table \ref{tab:Brane}, an NS5-brane, a D5-brane and a $\left(p,q\right)$5-brane share the directions 01234. Hence the 5-branes can form a brane web in the 59 plane, which we call a $\left(p,q\right)$ web \cite{Aharony:1997ju,Aharony:1997bh}. We also use the word $\left(p,q\right)$ web for the case when single $\left(p,q\right)$5-brane sits on a fixed $x^{6}$. One of the simplest non-trivial $\left(p,q\right)$ web would be a $\left(p,q\right)$ web of an NS5-brane and a D5-brane. Although the 5-branes are separated along the direction 6 in general, we can put a D5-brane on the top of an NS5-brane. 
\begin{figure}[t]
\begin{centering}
\includegraphics[scale=0.55]{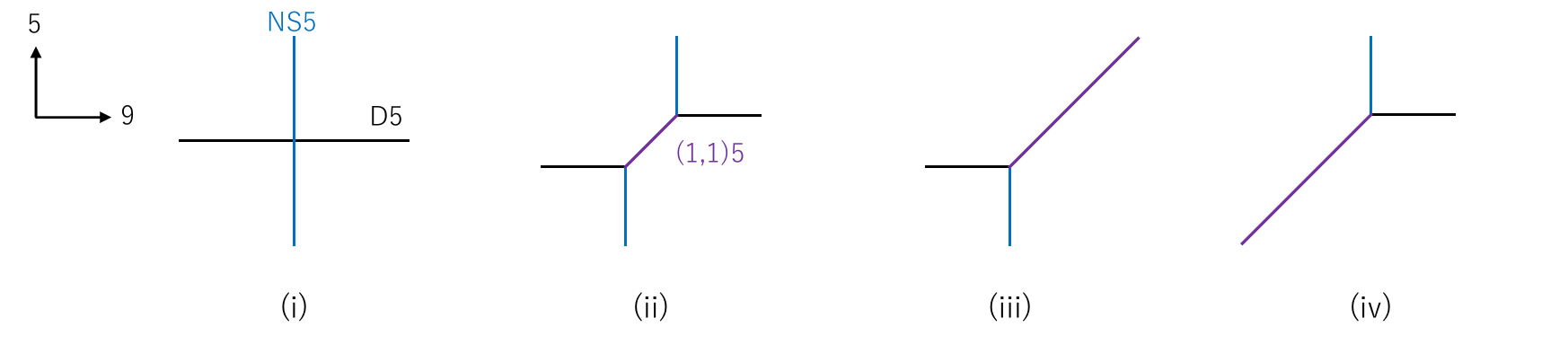}
\par\end{centering}
\caption{Four $\left(p,q\right)$ webs. (i): A $\left(p,q\right)$ web consisting of an NS5-brane and a D5-brane. (ii): A $\left(1,1\right)$5-brane is generated at the center. (iii), (iv): Junctions of an NS5-brane and a D5-brane merging into a $\left(1,1\right)$5-brane. The $\left(p,q\right)$ web in (iii) can be obtained by deforming the right D5-brane in (ii) to $+x^{5}$, while the $\left(p,q\right)$ web in (iv) can be obtained by deforming the left D5-brane in (ii) to $-x^{5}$.\label{fig:ND_NDd_k}}
\end{figure}
 Figure \ref{fig:ND_NDd_k} (i) shows this situation. At the crossing point, they can be combined to be a $\left(1,1\right)$5-brane. Figure \ref{fig:ND_NDd_k} (ii) shows this situation. Here the charge must be conserved, and to preserve the $\mathcal{N}=2$ supersymmetry the angle must be adjusted. Figure \ref{fig:ND_NDd_k} (iii), (iv) show junctions of an NS5-brane and a D5-brane merging into a $\left(1,1\right)$5-brane, but the directions of the external legs are opposite.

Various $\left(p,q\right)$ webs are related by web deformations \cite{Brodie:1997sz,Aharony:1997ju,Bergman:1999na}. Let us consider figure \ref{fig:ND_NDd_k} as an example. By breaking the D5-brane into two parts by ending it on the NS5-brane in figure \ref{fig:ND_NDd_k} (i), we can move the position of the D5-branes in the direction 5 separately. Then, by moving the right (left) D5-brane (which extends in the $\pm x^{9}$ direction) in the $\pm x^{5}$ direction, we obtain the $\left(p,q\right)$ web in figure \ref{fig:ND_NDd_k} (ii). Furthermore, by moving the right (left) D5-brane in positive (negative) infinite, we obtain the $\left(p,q\right)$5-brane of figure \ref{fig:ND_NDd_k} (iii) (or (iv)).

In general, as we will discuss in section \ref{subsec:3dTh}, the Lagrangian of the worldvolume theory is known if the whole brane configuration consists only of $\left(p,q\right)$ webs of the following two types. The first type is a junction of a $\left(1,q\right)$5-brane and $F_{\mp}$ left (right) D5-branes merging into a $\left(1,q+F_{\mp}\right)$5-brane with additional $F$ D5-branes, and the second type is an isolated D5-brane. Here $F$ and $F_{\pm}$ are non-negative integers. When $F_{\wdl}\neq0$ and $F_{\wdr}\neq0$, $\min\left\{ F_{\wdl},F_{\wdr}\right\} $ D5-branes extending in the $-x^{9}$ direction and $\min\left\{ F_{\wdl},F_{\wdr}\right\} $ D5-branes extending in the $+x^{9}$ direction can be interpret as $\min\left\{ F_{\wdl},F_{\wdr}\right\} $ D5-branes extending in the both directions. Hence we restrict at least one of $F_{\pm}$ to be zero without loss of generality. We further slightly generalize $\left(1,q\right)$5-brane to $\left(p,q\right)$5-brane with $p\geq1$.
\begin{figure}[t]
\begin{centering}
\includegraphics[scale=0.6]{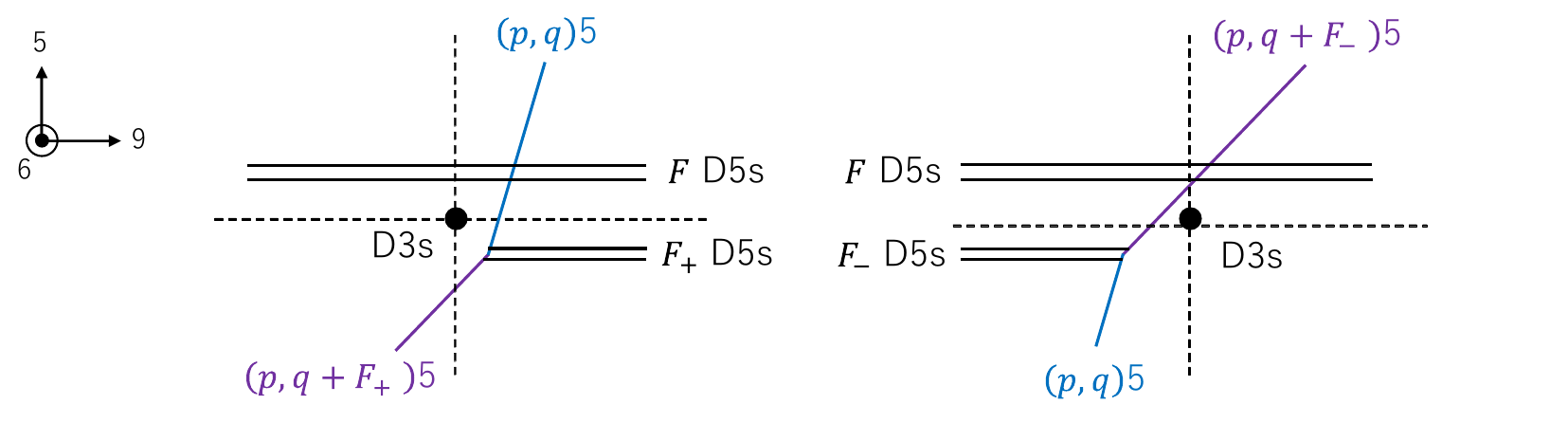}
\par\end{centering}
\caption{Two $\left(p,q\right)$ webs at a fixed $x^{6}$. The left and right figures show $F_{\protect\wdl}=0$, $F_{\protect\wdr}\protect\geq0$ and $F_{\protect\wdl}\protect\geq0$, $F_{\protect\wdr}=0$ cases, respectively. $F$ is also non-negative integer.\label{fig:BW-gen}}
\end{figure}
Then, a general $\left(p,q\right)$ web considered here can be described in two ways as in figure \ref{fig:BW-gen}. In this figure, $F$ D5-branes can get finite web deformations. Namely, we identify $\left(p,q\right)$ webs in figure \ref{fig:ND_NDd_k} (i) and (ii). This is because in the field theory they describe the same theory except for mass parameters.

\subsubsection*{Notation of the brane configuration}

We express the $\left(p,q\right)$ webs in the left and right of figure \ref{fig:BW-gen} as
\begin{equation}
\left[\left(p,q\right)+F\df+F_{\wdr}\dfr\right],\quad\left[\left(p,q\right)+F\df+F_{\wdl}\dfl\right].
\end{equation}
Namely, $\df$ denotes a D5-brane which forms a $\left(p,q\right)$ web with a $\left(p,q\right)$5-brane, and $\df^{\pm}$ denotes a D5-brane extending in the $\pm x^{9}$ direction and ending on a $\left(p,q\right)$5-brane. Similarly, the $\left(p,q\right)$ web of an isolated D5-brane is written as
\begin{equation}
\left[\left(0,1\right)\right].
\end{equation}
The whole brane system is a sequence of the $\left(p,q\right)$ webs, and we express it, for example, as
\begin{equation}
\left[\left(p,q\right)+\dfl\right]-\left[\left(0,1\right)\right]-_{\mathrm{p}}.\label{eq:BCex1}
\end{equation}
The horizontal line expresses $N$ D3-branes and thus 5-brane webs separated by the horizontal line are separated along the direction 6. The last horizontal line with $\mathrm{p}$, $-_{\mathrm{p}}$, is used when we want to emphasize that the direction 6 is periodic. (Thus \eqref{eq:BCex1} is equal to $\left[\left(0,1\right)\right]-\left[\left(p,q\right)+\dfl\right]-_{\mathrm{p}}$.) We denote the total number of $\left(p,q\right)$ webs by $R$, and we label each $\left(p,q\right)$ web by $r$. We also label an interval between $\left(r-1\right)$-th and $r$-th $\left(p,q\right)$ webs by $r$ as
\begin{equation}
\underset{r-1}{-}\left[\left(p,q\right)\right]^{\left(r-1\right)}\underset{r}{-}\left[\left(p',q'\right)\right]^{\left(r\right)}\underset{r+1}{-}.
\end{equation}
If we focus on a $\left(p,q\right)$ web, we omit the label for the $\left(p,q\right)$ web and label the left and right intervals by $1,2$ as
\begin{equation}
\underset{1}{-}\left[\left(p,q\right)\right]\underset{2}{-}.\label{eq:BCnot1}
\end{equation}

\subsubsection*{Toric diagrams, curves and combined $\left(p,q\right)$ webs}

\begin{figure}[t]
\begin{centering}
\includegraphics[scale=0.3]{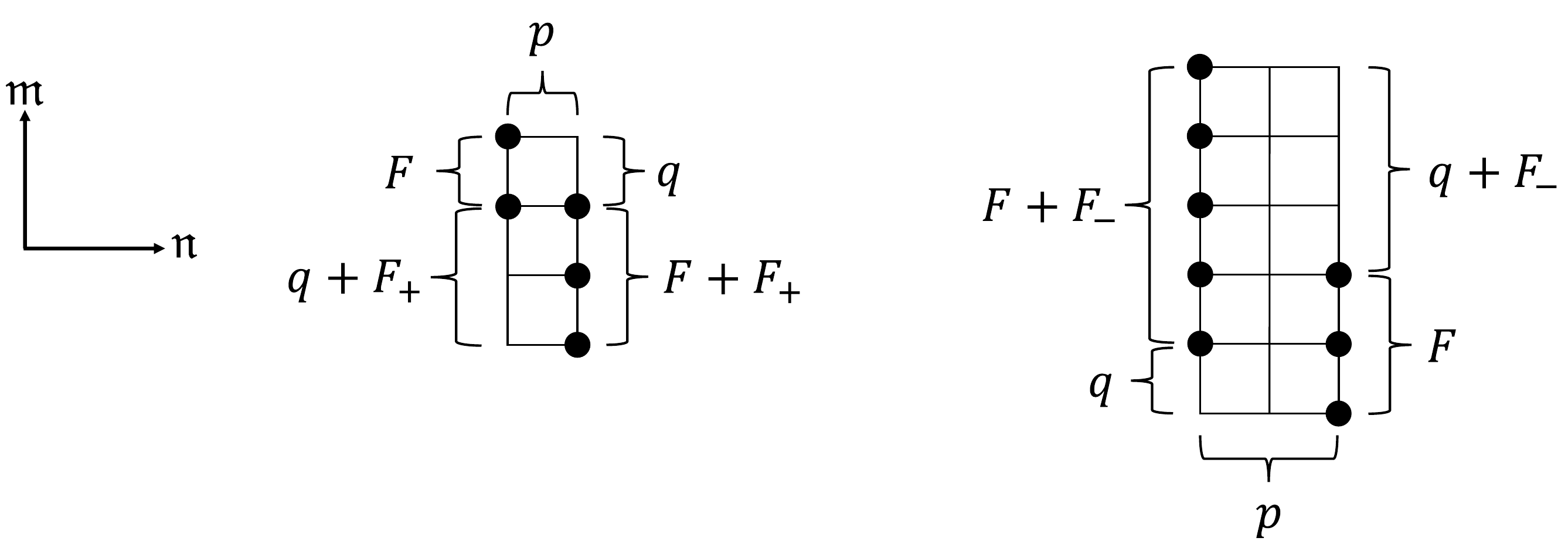}
\par\end{centering}
\caption{Two toric diagrams which are dual to the $\left(p,q\right)$ webs in figure \ref{fig:BW-gen}. The points are labeled by $\left(\mathfrak{m},\mathfrak{n}\right)$, where $\mathfrak{m}$ denotes the vertical direction and $\mathfrak{n}$ denotes the horizontal direction.\label{fig:DualPoly-gen}}
\end{figure}

Let us write a set of $\left(p,q\right)$ webs as
\begin{equation}
\bw=\left\{ \mathsf{w}^{\left(1\right)},\mathsf{w}^{\left(2\right)},\ldots,\mathsf{w}^{\left(R\right)}\right\} ,
\end{equation}
where $\mathsf{w}^{\left(r\right)}$ denotes the $r$-th $\left(p,q\right)$ web. We can read off a dual toric diagram for each $\left(p,q\right)$ web $\mathsf{w}$ with the ordinary rule \cite{Aharony:1997bh,Leung:1997tw}. We write a toric diagram as $\lp$ and a dual toric diagram of $\mathsf{w}$ as $\lp\left(\mathsf{w}\right)$. The graph of a $\left(p,q\right)$ web and a toric diagram are dual to each other in the sense of exchanging faces and vertices. For example, the dual diagrams for the two $\left(p,q\right)$ webs in figure \ref{fig:BW-gen} are described in figure \ref{fig:DualPoly-gen}.
In this figure, the toric diagrams are dual to the $\left(p,q\right)$ webs of
\begin{align}
\left(p,q\right) & =\left(1,1\right),\quad F=1,\quad F_{\wdl}=0,\quad F_{\wdr}=1,\nonumber \\
\left(p,q\right) & =\left(2,1\right),\quad F=2,\quad F_{\wdl}=2,\quad F_{\wdr}=0.
\end{align}

Later, we relate toric diagrams with (quantum) curves. Here we summarize notations related to this. We often use the same symbol $\lp$ for denoting a set of vertices of $\lp$, and we label the vertices with $\left(\mathfrak{m},\mathfrak{n}\right)$, where $\mathfrak{m}$ and $\mathfrak{n}$ increase in the up and right directions, respectively. The distance between the two vertices of the dual toric diagrams of $\left(1,0\right)$5-brane and $\left(0,1\right)$5-brane is normalized to be $1$. The label $\left(\mathfrak{m},\mathfrak{n}\right)$ is defined up to translations. For example, the vertices of the left toric diagram $\lp$ of figure \ref{fig:DualPoly-gen} can be labeled as
\begin{equation}
\lp=\left\{ \left(\frac{3}{2},-\frac{1}{2}\right),\left(\frac{1}{2},-\frac{1}{2}\right),\left(\frac{1}{2},\frac{1}{2}\right),\left(-\frac{1}{2},\frac{1}{2}\right),\left(-\frac{3}{2},\frac{1}{2}\right)\right\} .\label{eq:TDex1}
\end{equation}

Let $\mathsf{NP}$ be a set of $\left(\mathfrak{m},\mathfrak{n}\right)$ where relative values of $\mathfrak{m}$ (and $\mathfrak{n}$) are integers, and let ${\cal O}\left(x,y\right)$ be a function of the form
\begin{equation}
{\cal O}\left(x,y\right)=\sum_{\left(\mathfrak{m},\mathfrak{n}\right)\in\mathsf{NP}}c_{\mathfrak{m},\mathfrak{n}}e^{\mathfrak{m}x+\mathfrak{n}y},\label{eq:QC-NP}
\end{equation}
where $c_{\mathfrak{m},\mathfrak{n}}$ is a constant. In this paper, we call a function of this form a (quantum) curve and call $\mathsf{NP}$ Newton polygon. (If the variables are promoted to operators, we call it a quantum curve.) If $\mathsf{NP}=\lp$ with an appropriate representation of $\lp$, we say that the Newton polygon of a (quantum) curve is equal to the toric diagram. For example, the Newton polygon of the following curve is equal to the toric diagram in \eqref{eq:TDex1}
\begin{equation}
{\cal O}\left(x,y\right)=\left(\begin{array}{cc}
c_{\frac{3}{2},-\frac{1}{2}}e^{\frac{3}{2}x-\frac{1}{2}y} & +0\\
c_{\frac{1}{2},-\frac{1}{2}}e^{\frac{1}{2}x-\frac{1}{2}y} & c_{\frac{1}{2},\frac{1}{2}}e^{\frac{1}{2}x+\frac{1}{2}y}\\
+0 & c_{-\frac{1}{2},\frac{1}{2}}e^{-\frac{1}{2}x+\frac{1}{2}y}\\
+0 & c_{-\frac{3}{2},\frac{1}{2}}e^{-\frac{3}{2}x+\frac{1}{2}y}
\end{array}\right).
\end{equation}
Note that we often express a curve in this way to visualize the correspondence between $\left(p,q\right)$ webs, dual toric diagrams and Newton polygons.

In deep IR, because the direction 6 shrinks to zero, all of the $\left(p,q\right)$ webs $\mathsf{w}^{\left(r\right)}$ can be replaced with a single web in the projected 59 plane. We call this 5-brane web combined $\left(p,q\right)$ web and denote it by $\overline{\bw}$. Because $\overline{\bw}$ can be regarded as an ordinary $\left(p,q\right)$ web, we can obtain the dual toric diagram $\lp\left(\overline{\bw}\right)$ as well.

\begin{figure}[t]
\begin{centering}
\includegraphics[scale=0.5]{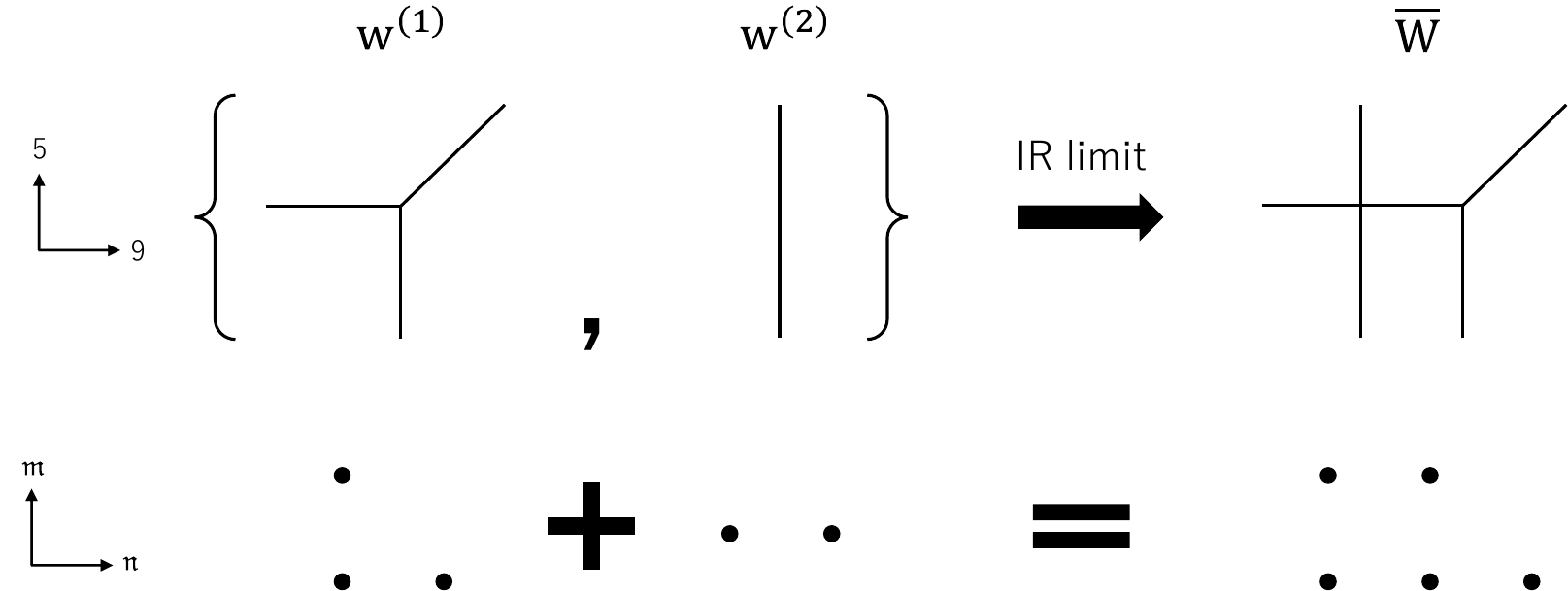}
\par\end{centering}
\caption{An example of $\protect\bw=\left\{ \mathsf{w}^{\left(1\right)},\mathsf{w}^{\left(2\right)}\right\} $ and the dual toric diagrams of them. The combined $\left(p,q\right)$ web $\overline{\protect\bw}$ and the dual toric diagram of it are also depicted on the right side.\label{fig:BW-LP}}
\end{figure}

This toric diagram can be obtained as a Minkowski sum of dual toric diagrams of the $\left(p,q\right)$ webs $\lp\left(\mathsf{w}^{\left(r\right)}\right)$. Let us see this operation in terms of associated curves. Consider two toric diagrams $\lp_{1},\lp_{2}$. We prepare curves whose Newton polygons are $\lp_{i}$
\begin{equation}
{\cal O}_{i}\left(x,y\right)=\sum_{\left(\mathfrak{m},\mathfrak{n}\right)\in\lp_{i}}e^{\mathfrak{m}x+\mathfrak{n}y}.
\end{equation}
The ambiguity of the overall power will not affect the conclusion. Then, the Minkowski sum of $\lp_{1}$ and $\lp_{2}$ is the Newton polygon of the product of the curves ${\cal O}_{1}{\cal O}_{2}$. Namely,
\begin{equation}
\lp_{1}+\lp_{2}=\left\{ \left(\mathfrak{m},\mathfrak{n}\right)|c_{\mathfrak{m},\mathfrak{n}}\neq0\right\} ,\quad\text{where}\quad{\cal O}_{1}\left(x,y\right){\cal O}_{2}\left(x,y\right)=\sum_{\left(\mathfrak{m},\mathfrak{n}\right)}c_{\mathfrak{m},\mathfrak{n}}e^{\mathfrak{m}x+\mathfrak{n}y}.
\end{equation}
The dual toric diagram of $\overline{\bw}$ is the Minkowski sum of $\lp\left(\mathsf{w}^{\left(r\right)}\right)$ 
\begin{equation}
\lp\left(\overline{\bw}\right)=\sum_{r=1}^{R}\lp\left(\mathsf{w}^{\left(r\right)}\right).\label{eq:CombLP}
\end{equation}
An example of $\lp\left(\overline{\bw}\right)$ is shown in figure \ref{fig:BW-LP}.
In this figure, $\mathsf{w}^{\left(1\right)}=\left[\left(1,0\right)+\dfl\right]$ and $\mathsf{w}^{\left(2\right)}=\left[\left(1,0\right)\right]$, and their dual toric diagrams are $\lp\left(\mathsf{w}^{\left(1\right)}\right)=\left\{ \left(\frac{1}{2},-\frac{1}{2}\right),\left(-\frac{1}{2},\frac{1}{2}\right),\left(-\frac{1}{2},-\frac{1}{2}\right)\right\} $ and $\lp\left(\mathsf{w}^{\left(2\right)}\right)=\left\{ \left(0,\frac{1}{2}\right),\left(0,-\frac{1}{2}\right)\right\} $. $\overline{\bw}$ is also depicted on the top right in the figure, and $\lp\left(\overline{\bw}\right)=\left\{ \left(\frac{1}{2},0\right),\left(\frac{1}{2},-1\right),\left(-\frac{1}{2},1\right),\left(-\frac{1}{2},0\right),\left(-\frac{1}{2},-1\right)\right\} $. One can easily see that $\lp\left(\overline{\bw}\right)$ is indeed the Minkowski sum of $\lp\left(\mathsf{w}^{\left(1\right)}\right)$ and $\lp\left(\mathsf{w}^{\left(2\right)}\right)$.

\subsection{3d ${\cal N}=2$ supersymmetric theories\label{subsec:3dTh}}

The worldvolume theory on D3-branes is a 3d supersymmetric gauge theory. In deep IR, in our setup it becomes a superconformal CS theory \cite{Kitao:1998mf,Bergman:1999na}. 5-branes break the supersymmetries, and the amount of the remaining supersymmetries depends on the types of the inserted 5-branes \cite{Kitao:1998mf}. When a brane configuration consists only of the NS5-branes, the number of preserved supersymmetries is 8, while when the brane configuration also includes $\left(p,q\right)$5-branes with $q\geq$1, the number of them is 4. This means that the worldvolume theory is a 3d $\mathcal{N}=4$ supersymmetric theory for the former case, while it is an $\mathcal{N}=2$ theory for the latter case. When the brane configuration consists of only $\left(p,q\right)$5-branes with $p=0,1$, the Lagrangian of the worldvolume theory is known. In this section, we review this case.

When an interval is bounded by two NS5-branes, the corresponding gauge group factor is an $\mathcal{N}=4$ $\mathrm{U}\left(N\right)$ group, where $N$ is the number of the D3-branes. It gives an $\mathcal{N}=4$ vector multiplet, which is decomposed into an $\mathcal{N}=2$ vector multiplet and a chiral multiplet with an adjoint representation. Since the direction 6 is compact, if $R$ NS5-branes are inserted, the gauge group is described by a $\mathrm{U}\left(N\right)^{R}$ circular quiver diagram. Note that because we restrict the number of the D3-branes spanning all of the intervals to be the same, the ranks of all the unitary groups are uniform. For an interval, a massless chiral multiplet $\phi$ with an adjoint representation appears. Because of the $\mathcal{N}=4$ supersymmetry, $\phi$ has R-charge $1$. In addition, each NS5-brane introduces an $\mathcal{N}=4$ hypermultiplet transforming in the bifundamental representation of $\mathrm{U}\left(N\right)\times\mathrm{U}\left(N\right)$. In terms of $\mathcal{N}=2$ supermultiplets, it is decomposed to a bifundamental and an anti-bifundamental chiral multiplet $A$ and $\tilde{A}$. Because the superpotential has a cubic term $W\sim\phi^{\left(1\right)}A\tilde{A}-\tilde{A}A\phi^{\left(2\right)}$ (see \eqref{eq:BCnot1} for the notation), the marginality of the superpotential implies that the R-charges of the (anti-)bifundamental matters are $1/2$.

For introducing fundamental matters while preserving $\mathcal{N}=4$ supersymmetry, we need to add D5'-branes which extend along the directions 012789 instead of the directions 012349. When $F$ D5'-branes are added to an interval, $F$ $\mathcal{N}=4$ fundamental hypermultiplets appear. In terms of $\mathcal{N}=2$ supermultiplet, they are $F$ fundamental chiral multiplets $Q_{f}$ and $F$ anti-fundamental chiral multiplets $\tilde{Q}_{f}$ ($f=1,2,\ldots,F$). (When $F=1$, we often omit the subscript $f$.) The superpotential demanded by $\mathcal{N}=4$ supersymmetry is $W\sim\phi Q_{f}\tilde{Q}_{f}$, and the global symmetry rotating the flavors is ${\rm U}\left(F\right)$. An important phenomenon is that when we rotate the D5'-branes in the 37 and 48 planes simultaneously with an angle $\theta\in\left[0,\pi/2\right]$, the cubic superpotential is continuously turned off, and when $\theta=\pi/2$, the cubic superpotential vanishes \cite{Elitzur:1997hc,Aharony:1997ju}. Thus, the flavor symmetry is enhanced to $\mathrm{U}\left(F\right)\times\mathrm{U}\left(F\right)$. With this rotation the D5'-branes become the D5-branes and they reduce the supersymmetry to ${\cal N}=2$. Notice that there is a quartic superpotential $W\sim\left(Q_{f}\tilde{Q}_{f}\right)^{2}$ \cite{Aharony:1997ju}, which implies that the R-charges of $Q_{f}$ and $\tilde{Q}_{f}$, which we denote by $\Delta_{f}$ and $\tilde{\Delta}_{f}$ respectively, satisfy
\begin{equation}
\Delta_{f}+\tilde{\Delta}_{f}=1\quad\text{(for matters from D5-branes at an interval)}.\label{eq:R-cond2}
\end{equation}

We can put the $F$ D5-branes on top of a $\left(1,q\right)$5-brane. In this case, D5-branes give fundamental and anti-fundamental chiral multiplets to both of the neighboring $\mathrm{U}\left(N\right)$ groups. This mechanism is called the flavor doubling \cite{Brodie:1997sz,Brunner:1998jr}. More preciously, a D5-brane extending in the $\pm x^{9}$ direction introduces $Q^{\left(1\right)}$ and $\tilde{Q}^{\left(2\right)}$ or $Q^{\left(2\right)}$ and $\tilde{Q}^{\left(1\right)}$, respectively. The $4F$ chiral multiplets couple via cubic superpotentials to the bifundamental hypermultiplet localized at the NS5-brane as
\begin{equation}
W=Q_{f}^{\left(1\right)}\tilde{A}\tilde{Q}_{f}^{\left(2\right)}+Q_{f}^{\left(2\right)}A\tilde{Q}_{f}^{\left(1\right)}.\label{eq:W-QAQ}
\end{equation}
The flavor symmetry is still ${\rm U}\left(F\right)\times{\rm U}\left(F\right)$. This means that we can choose the mass for $Q_{f}^{\left(r\right)}$ and $\tilde{Q}_{f}^{\left(r\right)}$ independently. On the other hand, the masses of $Q_{f}^{\left(1\right)}$ ($\tilde{Q}_{f}^{\left(1\right)}$) and $\tilde{Q}_{f}^{\left(2\right)}$ ($Q_{f}^{\left(2\right)}$) are not independent and actually have equal magnitudes and opposite signs
\begin{equation}
M_{f}^{\left(1\right)}=-\tilde{M}_{f}^{\left(2\right)},\quad\tilde{M}_{f}^{\left(1\right)}=-M_{f}^{\left(2\right)},\label{eq:Mass-cond}
\end{equation}
where $M_{f}^{\left(r\right)}$ and $\tilde{M}_{f}^{\left(r\right)}$ are the real masses of $Q_{f}^{\left(r\right)}$ and $\tilde{Q}_{f}^{\left(r\right)}$, respectively. Since the R-charges of $A$ and $\tilde{A}$ are $1/2$, the marginality of the superpotential implies that
\begin{equation}
\Delta_{f}^{\left(1\right)}+\tilde{\Delta}_{f}^{\left(2\right)}=\frac{3}{2},\quad\tilde{\Delta}_{f}^{\left(1\right)}+\Delta_{f}^{\left(2\right)}=\frac{3}{2},\label{eq:R-cond}
\end{equation}
where $\Delta_{f}^{\left(r\right)}$ and $\tilde{\Delta}_{f}^{\left(r\right)}$ are the R-charges of $Q_{f}^{\left(r\right)}$ and $\tilde{Q}_{f}^{\left(r\right)}$, respectively.

\subsubsection*{Mass, FI and brane configurations}

We now explain how the mass and FI parameters are related to the brane configuration. These parameters are indeed related to the position of 5-branes in the 59 plane, where the D3-branes sit on the origin \cite{Hanany:1996ie,Aharony:1997bx}. For example, a $\left(1,q\right)$5-brane can be moved in the direction 9, and this provides the FI term for the $\mathrm{U}\left(1\right)$ factors of the two $\mathrm{U}\left(N\right)$ groups in the both sides of the $\left(1,q\right)$5-brane. In other words, for the brane configuration
\begin{equation}
\left[\left(1,q^{\left(0\right)}\right)\right]-\left[\left(1,q^{\left(1\right)}\right)\right],
\end{equation}
the FI parameter on the interval is
\begin{equation}
\zeta=\eta^{\left(0\right)}-\eta^{\left(1\right)},\label{eq:BC-FI}
\end{equation}
where $\eta^{\left(0\right)}$ and $\eta^{\left(1\right)}$ denote the position of the $\left(1,q^{\left(0\right)}\right)$5-brane and the $\left(1,q^{\left(1\right)}\right)$5-brane along the direction 9, respectively.

When a $\left(p,q\right)$ web gets a finite web deformation (like the $\left(p,q\right)$ web in figure \ref{fig:ND_NDd_k} (ii)), however, a subtlety arises for the definition of the position. Moreover, when a $\left(p,q\right)$5-brane gets additional D5-brane charges (like the $\left(p,q\right)$ web in figure \ref{fig:ND_NDd_k} (iii) or (iv)), the definition becomes more obscure. To make it clear, let us focus on the $\left(p,q\right)$ webs depicted in figure \ref{fig:BW-gen}. In this case, we adopt a definition under which the position is invariant under the move of the D5-branes. For this purpose, we focus on the two external legs in the $\pm x^{5}$ directions, namely the $\left(p,q\right)$5-brane and the $\left(p,q+F_{\pm}\right)$5-brane. Then, the position $\eta$ can be defined as the position of the middle point of the two external legs. Here the middle point is measured along the direction 9. The FI parameter is again related to $\eta$ as \eqref{eq:BC-FI}.

A D5-brane can also be moved in the direction 5, and this gives a mass to the (anti-)fundamental chiral matters. For an isolated D5-brane, the mass parameter is related to the position in the direction 5 denoted by $m$ as
\begin{equation}
M=-\tilde{M}=m.
\end{equation}
For a D5-brane on a $\left(p,q\right)$ web, the location is again defined by the position of the external legs. Remember that a D5-brane extending in the $\pm x^{9}$ direction introduces $Q^{\left(1\right)}$ and $\tilde{Q}^{\left(2\right)}$ or $Q^{\left(2\right)}$ and $\tilde{Q}^{\left(1\right)}$, respectively. If $m$ (or $\tilde{m}$) denotes the position of a D5-brane extending in the $\pm x^{9}$ direction in the direction 5, the relation between the position and the mass parameters of the corresponding matters is
\begin{equation}
M^{\left(1\right)}=-\tilde{M}^{\left(2\right)}=m,\quad-\tilde{M}^{\left(1\right)}=M^{\left(2\right)}=\tilde{m}.\label{eq:BC-Mass}
\end{equation}
Note that this is consistent with \eqref{eq:Mass-cond}.

\begin{figure}[t]
\begin{centering}
\includegraphics[scale=0.55]{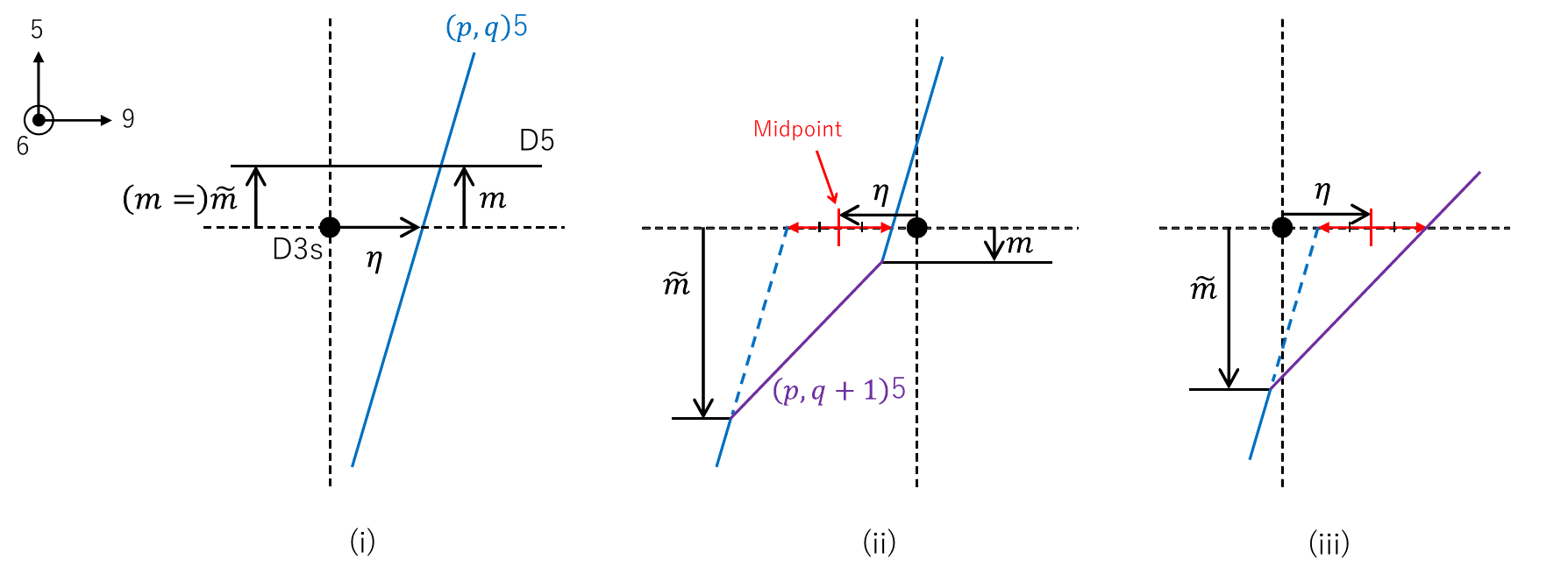}
\par\end{centering}
\caption{The position of 5-branes for three examples. The up direction is $+x^{5}$ and the right direction is $+x^{9}$. (Hence in (ii), $\eta<0$, $m<0$ and $\tilde{m}<0$, and in (iii), $\tilde{m}<0$.) $\eta$, $m$ and $\tilde{m}$ are related to the FI and mass parameters with \eqref{eq:BC-FI} and \eqref{eq:BC-Mass}.\label{fig:Web-MassFI}}
\end{figure}

Figure \ref{fig:Web-MassFI} shows three $\left(p,q\right)$ webs as examples.
Figure \ref{fig:Web-MassFI} (i) shows the simple case. $\eta$ is the position of the $\left(p,q\right)$5-brane in the direction 9, and $m$($=\tilde{m}$) is the position of the D5-brane in the 5-direction. Figure \ref{fig:Web-MassFI} (ii) shows the case when the $\left(p,q\right)$5-brane and the D5-brane get a finite web deformation. In this case $\eta$ is the position of the middle point of the upper and lower $\left(p,q\right)$5-brane. Figure \ref{fig:Web-MassFI} (iii) shows the junction of the $\left(p,q\right)$5-brane and the D5-brane merging into the $\left(p,q+1\right)$5-brane. In this case $\eta$ is the position of the middle point of the $\left(p,q\right)$5-brane and the $\left(p,q+1\right)$5-brane.

On the other hand, the R-charge is not directly related to the brane configuration. The correct R-charge would be determined via the $F$-maximization procedure with the restriction coming from the marginality of the superpotential \cite{Jafferis:2010un,Jafferis:2011zi,Willett:2011gp,Closset:2012vg}. However, in this paper we only impose the marginality conditions \eqref{eq:R-cond2} or \eqref{eq:R-cond}.

\subsubsection*{Large mass, CS level and web deformations}

In section \ref{subsec:BraneConfig}, we reviewed the web deformation. In terms of the 3d theory, the web deformation introduces the CS terms. It is known that if we give a large mass to an (anti-)chiral multiplet and integrate it out, the corresponding gauge field gets a CS term with the CS level $\pm1/2$ \cite{Redlich:1983dv}. In terms of the brane picture, this corresponds to moving a D5-brane far away in the $\pm x^{5}$ direction.

\begin{figure}[t]
\begin{centering}
\includegraphics[scale=0.55]{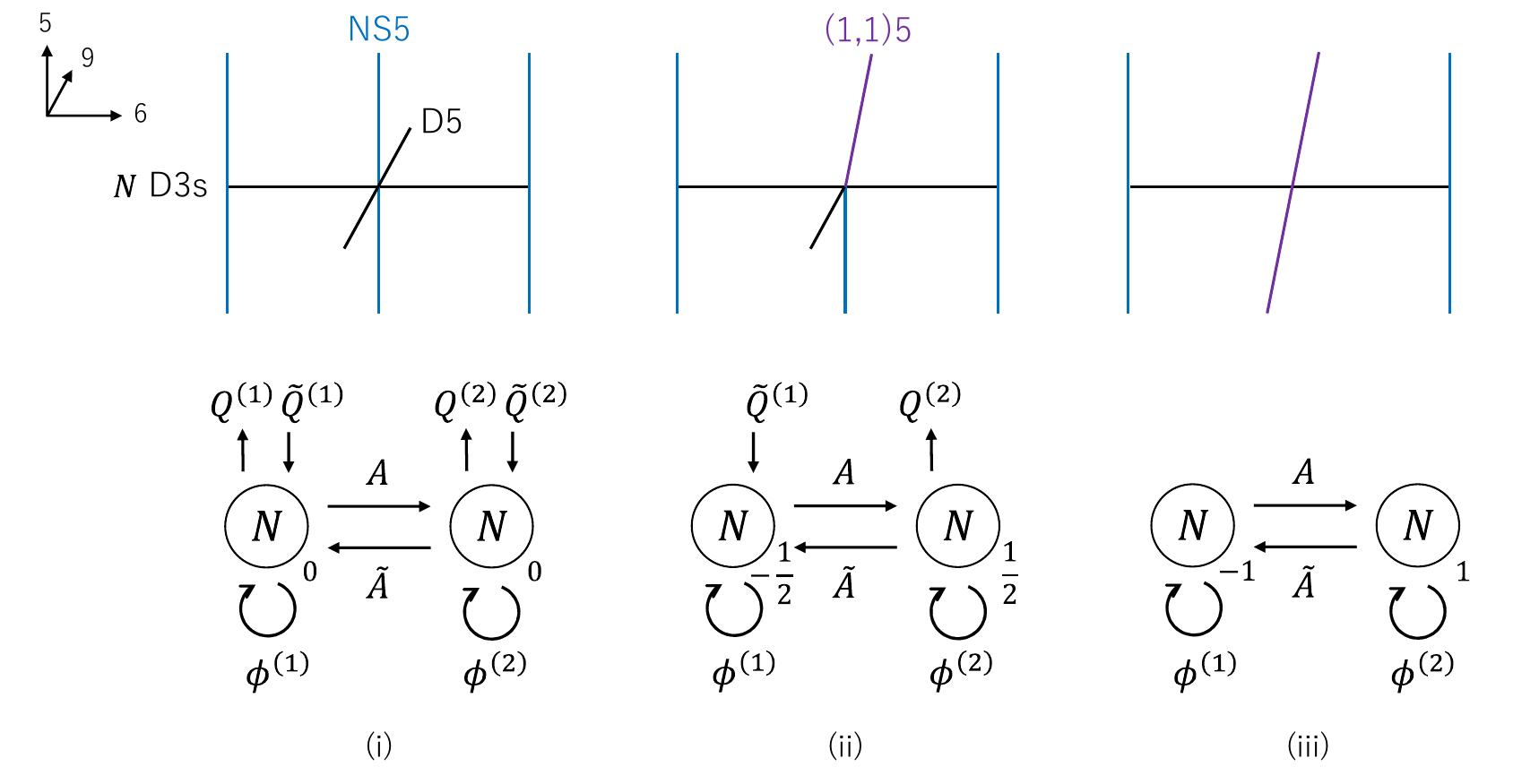}
\par\end{centering}
\caption{Three $\left(p,q\right)$ webs and the corresponding quiver diagrams. (ii) (or (iii)) is obtained from (i) (or (ii)) by the web deformation. The subscripts of the gauge nodes denote the CS levels.\label{fig:ND_NDd_k_Quiv}}
\end{figure}

Let us consider an example of a brane configuration
\begin{equation}
\left[\left(1,0\right)\right]-\left[\left(1,0\right)+\df\right]-\left[\left(1,0\right)\right].
\end{equation}
Figure \ref{fig:ND_NDd_k_Quiv} (i) shows this $\left(p,q\right)$ web and the corresponding quiver diagram.
If we perform the web deformation where the D5-brane extending in the $+x^{9}$ direction moves to far away in the $+x^{5}$ direction, the first and second nodes get the CS level $\pm1/2$ according to \eqref{eq:BC-Mass}, and we arrive at figure \ref{fig:ND_NDd_k_Quiv} (ii). We can further move the remaining D5-brane in the $-x^{5}$ direction. Then, the first and second nodes get the additional CS level $\pm1/2$, and we arrive at figure \ref{fig:ND_NDd_k_Quiv} (iii). Note that if we move the D5-brane in the $+x^{5}$ direction in the last step, the CS level cancels. This shows the importance of the web deformation for generating the CS terms.

From the above explanation, one can find that the CS level $k$ appearing on the following interval
\begin{equation}
\left[\left(1,q^{\left(0\right)}\right)+F_{\wdr}^{\left(0\right)}\dfr+F_{\wdl}^{\left(0\right)}\dfl\right]-\left[\left(1,q^{\left(1\right)}\right)+F_{\wdr}^{\left(1\right)}\dfr+F_{\wdl}^{\left(1\right)}\dfl\right],
\end{equation}
is \cite{Kitao:1998mf,Bergman:1999na}
\begin{equation}
k=q^{\left(0\right)}+\frac{1}{2}\left(F_{\wdr}^{\left(0\right)}+F_{\wdl}^{\left(0\right)}\right)-q^{\left(1\right)}-\frac{1}{2}\left(F_{\wdr}^{\left(1\right)}+F_{\wdl}^{\left(1\right)}\right).\label{eq:q-CSlevel}
\end{equation}

\subsection{Localization and matrix models\label{subsec:Localization}}

In this section we review the matrix model for 3d $\mathcal{N}=2$ CS matter theories with a gauge group written by a quiver diagram. We obtain the matrix model by applying the localization technique to the partition function on the round three sphere \cite{Kapustin:2009kz} (see also \cite{Willett:2016adv} for a review). We assume that all of the nodes are the unitary groups with the rank $N$, $\mathrm{U}\left(N\right)$.

After applying the localization technique, the partition function reduces to integrations over eigenvalues of scalars of unitary groups. The integrand of the matrix model comes from two kinds of contributions, namely, the classical contribution and the 1-loop contribution. In our case, each factor appearing in both the contributions is a function of the eigenvalues of either $\mathrm{U}\left(N\right)$ or $\mathrm{U}\left(N\right)\times\mathrm{U}\left(N\right)$, which is denoted by $\mu$ or $\mu,\nu$, respectively.

The classical contributions come from the CS terms and the FI terms of $\mathrm{U}\left(N\right)$. The corresponding factor is\footnote{The integration variables are rescaled by $2\pi$ for a later convenience.}
\begin{equation}
Z_{\text{CS+FI}}\left(\mu\right)=e^{-\frac{ik}{4\pi}\sum_{a}^{N}\mu_{a}^{2}-i\zeta\sum_{a}^{N}\mu_{a}},\label{eq:Zcl}
\end{equation}
where $k$ denotes the CS level and $\zeta$ denotes the FI parameter. In this paper the label for the integration variables (which is $a$ in this case) always starts from $1$. Next, we consider the 1-loop part. The contribution from a chiral multiplet with representation $\mathcal{R}$, mass $M$ and R-charge $\Delta$ is
\begin{equation}
Z_{\text{chiral},\mathcal{R}}\left(M,\Delta;\mu\right)=\prod_{\rho\in\mathcal{R}}s_{1}\left(M+i\left(1-\Delta\right)-\rho\left(\frac{\mu}{2\pi}\right)\right).
\end{equation}
Here one can see that this factor is a holomorphic function of the mass and R-charge. For example, an $\mathcal{N}=4$ $\mathrm{U}\left(N\right)\times\mathrm{U}\left(N\right)$ bi-fundamental hypermultiplet consists of two $\mathcal{N}=2$ (anti-)bi-fundamental chiral multiplets with R-charge $1/2$, and hence the corresponding factor is
\begin{equation}
Z_{\text{bi-fund}}\left(\mu,\nu\right)=\frac{1}{\prod_{a,b}^{N}2\cosh\frac{\mu_{a}-\nu_{b}}{2}}.\label{eq:Zhyp}
\end{equation}
The contribution from the $\mathrm{U}\left(N\right)$ (anti)-fundamental chiral multiplet with mass $M$ and R-charge $\Delta$ is
\begin{align}
Z_{\text{(anti-)fund}}\left(M,\Delta;\mu\right) & =\prod_{a}^{N}s_{1}\left(M+i\left(1-\Delta\right)\mp\frac{\mu_{a}}{2\pi}\right)\nonumber \\
 & =\begin{cases}
\prod_{a}^{N}s_{1}\left(\frac{\mu_{a}}{2\pi}-M-i\left(1-\Delta\right)\right)^{-1} & \text{fundamental}\\
\prod_{a}^{N}s_{1}\left(\frac{\mu_{a}}{2\pi}+M+i\left(1-\Delta\right)\right) & \text{anti-fundamental}
\end{cases}.\label{eq:Zchi}
\end{align}
Here we used a property of the double sine function in \eqref{eq:DS-id}. The contribution from the ${\cal N}=2$ $\mathrm{U}\left(N\right)$ vector multiplet is
\begin{equation}
Z_{\text{vec}}\left(\mu\right)=\prod_{a<b}^{N}\left(2\sinh\frac{\mu_{a}-\mu_{b}}{2}\right)^{2}.\label{eq:Zvec}
\end{equation}
For each gauge factor $\mathrm{U}\left(N\right)$ we also have adjoint chiral multiplet. However, since in this paper their R-charges are always $1$, they do not contribute. We can obtain the whole matrix model by gluing all the factors with integrations
\begin{equation}
\prod_{r=1}^{R}\left(\frac{1}{N!}\int_{-\infty}^{\infty}\prod_{a}^{N}\frac{d\mu_{a}^{\left(r\right)}}{2\pi}\right).\label{eq:MM_gluing}
\end{equation}

\subsubsection{Dictionary between $\left(p,q\right)$ webs to matrix factors\label{subsec:BC-MM}}

Combining the materials, we can obtain the dictionary between brane configuration to the matrix model. A key point is that there is one to one correspondence between $\left(p,q\right)$ webs and factors of the integrand of the matrix model. In other words, one can obtain the integrand of the matrix model by replacing each $\left(p,q\right)$ web to a factor which we explain below and multiplying all of them.

We start with an NS5-brane, which is $\left(1,0\right)$5-brane. In our setup, there are always two gauge group factors $\mathrm{U}\left(N\right)$ on both sides of the NS5-brane. Remember that a $\mathrm{U}\left(N\right)$ gauge factor includes an ${\cal N}=2$ vector multiplet and hence the integrand includes $Z_{\text{vec}}$. Because this factor is the square of $2\sinh\left(\mu_{a}-\mu_{b}\right)/2$, an NS5-brane can be replaced to the matrix factor
\begin{equation}
\mathcal{Z}^{\left(1,0\right)}\left(\eta;\mu,\nu\right)=\frac{1}{\left(2\pi\right)^{N}}e^{i\eta\sum_{a}^{N}\left(\mu_{a}-\nu_{a}\right)}\frac{\prod_{a<b}^{N}2\sinh\frac{\mu_{a}-\mu_{b}}{2}\prod_{a<b}^{N}2\sinh\frac{\nu_{a}-\nu_{b}}{2}}{\prod_{a,b}^{N}2\cosh\frac{\mu_{a}-\nu_{b}}{2}}.\label{eq:NS-MF}
\end{equation}
Here $\eta$ denotes the position of the NS5-brane in the direction 9. Because the FI parameter of a ${\rm U}\left(N\right)$ gauge factor is the difference of $\eta$ as \eqref{eq:BC-FI}, this is consistent with \eqref{eq:Zcl}.

If this 5-brane has $q$ D5-charges, say $\left(1,q\right)$5-brane, the matrix factor gets corresponding factor as well as $\eta$. Namely, according to \eqref{eq:q-CSlevel} and \eqref{eq:Zcl},
\begin{equation}
\mathcal{Z}^{\left(1,q\right)}\left(\eta;\mu,\nu\right)=e^{\frac{iq}{4\pi}\sum_{a}^{N}\left(\mu_{a}^{2}-\nu_{a}^{2}\right)}\mathcal{Z}^{\left(1,0\right)}\left(\eta;\mu,\nu\right).\label{eq:1q-MF}
\end{equation}

We can put a D5-brane on this $\left(1,q\right)$5-brane, and we denote the corresponding D5-brane factor by $\mathcal{Z}_{\df}$. Similarly, we can put a D5-brane which extends in the $\pm x_{9}$ direction and ends on the $\left(1,q\right)5$-brane, and we denote the corresponding D5-brane factor by $\mathcal{Z}_{\mathrm{D5}^{\pm}}$. The former D5-brane introduces four chiral multiplets, while the latter one introduces two chiral multiplets. The mass parameters for these matters are not completely independent and related to the position of the corresponding D5-brane as \eqref{eq:BC-Mass}. There is also a condition for the R-charge as in \eqref{eq:R-cond}. Under these conditions, the D5-brane factors are parameterized by the four parameters $\left(m,\tilde{m};D,\tilde{D}\right)$ (or the two of them) as \eqref{eq:BC-Mass} and (see \eqref{eq:BCnot1} for the notation)
\begin{equation}
\frac{3}{4}-\Delta^{\left(1\right)}=-\left(\frac{3}{4}-\tilde{\Delta}^{\left(2\right)}\right)=D,\quad-\left(\frac{3}{4}-\tilde{\Delta}^{\left(1\right)}\right)=\frac{3}{4}-\Delta^{\left(2\right)}=\tilde{D}.\label{eq:R-Para}
\end{equation}
Explicitly, the D5-brane factors are
\begin{align}
\mathcal{Z}_{\df}\left(z,\tilde{z};\mu,\nu\right) & =\frac{\prod_{a}^{N}s_{1}\left(\frac{\mu_{a}}{2\pi}-\tilde{z}+\frac{i}{4}\right)}{\prod_{a}^{N}s_{1}\left(\frac{\mu_{a}}{2\pi}-z-\frac{i}{4}\right)}\frac{\prod_{a}^{N}s_{1}\left(\frac{\nu_{a}}{2\pi}-z+\frac{i}{4}\right)}{\prod_{a}^{N}s_{1}\left(\frac{\nu_{a}}{2\pi}-\tilde{z}-\frac{i}{4}\right)},\nonumber \\
\mathcal{Z}_{\dfr}\left(z;\mu,\nu\right) & =e^{\frac{i}{8\pi}\sum_{a}^{N}\left(\mu_{a}^{2}-\nu_{a}^{2}\right)}\frac{\prod_{a}^{N}s_{1}\left(\frac{\nu_{a}}{2\pi}-z+\frac{i}{4}\right)}{\prod_{a}^{N}s_{1}\left(\frac{\mu_{a}}{2\pi}-z-\frac{i}{4}\right)},\nonumber \\
\mathcal{Z}_{\dfl}\left(\tilde{z};\mu,\nu\right) & =e^{\frac{i}{8\pi}\sum_{a}^{N}\left(\mu_{a}^{2}-\nu_{a}^{2}\right)}\frac{\prod_{a}^{N}s_{1}\left(\frac{\mu_{a}}{2\pi}-\tilde{z}+\frac{i}{4}\right)}{\prod_{a}^{N}s_{1}\left(\frac{\nu_{a}}{2\pi}-\tilde{z}-\frac{i}{4}\right)},\label{eq:D5wd-MF}
\end{align}
where
\begin{equation}
z=m+iD,\quad\tilde{z}=\tilde{m}+i\tilde{D}.\label{eq:z-Def}
\end{equation}
We remark that these factors always appear with a $\left(1,q\right)$5-brane, and we will define a matrix factor corresponding to an isolated D5-brane later. Therefore, the matrix factor corresponding to the $\left(p,q\right)$ web (see also figure \ref{fig:BW-gen})
\begin{equation}
\left[\left(1,q\right)+F\df+F_{\wdr}\dfr+F_{\wdl}\dfl\right],
\end{equation}
is
\begin{align}
{\cal Z}_{F,F_{\wdr},F_{\wdl}}^{\left(1,q\right)}\left(\eta,\mathbf{z};\mu,\nu\right) & =\begin{cases}
{\cal Z}^{\left(1,q\right)}\left(\zeta;\mu,\nu\right)\prod_{f=1}^{F}\mathcal{Z}_{\mathrm{D5}}\left(z_{f},\tilde{z}_{f};\mu,\nu\right)\prod_{f=1}^{F_{\wdr}}\mathcal{Z}_{\mathrm{D5}^{\wdr}}\left(z_{F+f};\mu,\nu\right) & \left(F_{\wdl}=0\right)\\
{\cal Z}^{\left(1,q\right)}\left(\zeta;\mu,\nu\right)\prod_{f=1}^{F}\mathcal{Z}_{\mathrm{D5}}\left(z_{f},\tilde{z}_{f};\mu,\nu\right)\prod_{f=1}^{F_{\wdl}}\mathcal{Z}_{\mathrm{D5}^{\wdl}}\left(\tilde{z}_{F+f};\mu,\nu\right) & \left(F_{\wdr}=0\right)
\end{cases},\label{eq:1qD5-MF}
\end{align}
where
\begin{align}
 & z_{f}=m_{f}+iD_{f},\quad\tilde{z}_{f}=\tilde{m}_{f}+i\tilde{D}_{f},\nonumber \\
 & \mathbf{z}=\begin{cases}
\left(z_{1},\tilde{z}_{1},z_{2},\tilde{z}_{2},\ldots,z_{F},\tilde{z}_{F}|z_{F+1},z_{F+2},\ldots,z_{F+F_{\wdr}}\right) & \left(F_{\wdl}=0\right)\\
\left(z_{1},\tilde{z}_{1},z_{2},\tilde{z}_{2},\ldots,z_{F},\tilde{z}_{F}|\tilde{z}_{F+1},\tilde{z}_{F+2},\ldots,\tilde{z}_{F+F_{\wdl}}\right) & \left(F_{\wdr}=0\right)
\end{cases}.\label{eq:1qD5-MFpara}
\end{align}

Multiplying $\mathcal{Z}^{\left(1,q\right)}$ by $\mathcal{Z}_{\mathrm{D5}^{\pm}}$ means adding a D5-brane extending in the $\pm x^{9}$ direction and adding the unit D5-charge to the lower (upper) half of the $\left(1,q\right)$5-brane (see figure \ref{fig:BW-gen}). Therefore, if $\mathcal{Z}^{\left(1,q\right)}$ is multiplied by both $\mathcal{Z}_{\dfl}$ and $\mathcal{Z}_{\dfr}$, one must obtain the matrix factor for the $\left(1,q+1\right)$5-brane with one D5-brane factor. Actually, the following identity holds
\begin{equation}
\mathcal{Z}^{\left(1,q\right)}\left(\eta;\mu,\nu\right)\mathcal{Z}_{\dfr}\left(z;\mu,\nu\right)\mathcal{Z}_{\dfl}\left(\tilde{z};\mu,\nu\right)=\mathcal{Z}^{\left(1,q+1\right)}\left(\eta;\mu,\nu\right)\mathcal{Z}_{\df}\left(z,\tilde{z};\mu,\nu\right).
\end{equation}

Next, we consider a factor corresponding to a D5-brane on an interval. For a D5-brane, a fundamental chiral multiplet and an anti-fundamental chiral multiplet appear. Because the R-charges of the (anti-)fundamental matters satisfy \eqref{eq:R-cond2}, the contributions from them \eqref{eq:Zchi} simplify thanks to \eqref{eq:DS-Trig}. We introduce
\begin{equation}
\frac{1}{2}-\Delta=-\left(\frac{1}{2}-\tilde{\Delta}\right)=D.
\end{equation}
Then, an isolated D5-brane on an interval can be replaced to
\begin{equation}
\mathcal{Z}^{\left(0,1\right)}\left(z;\mu,\nu\right)=\frac{N!}{\prod_{a}^{N}2\cosh\frac{\mu_{a}-2\pi z}{2}}\prod_{a}^{N}\delta\left(\mu_{a}-\nu_{b}\right),\label{eq:D5-MF}
\end{equation}
where
\begin{equation}
z=m+iD,\label{eq:D5-MFpara}
\end{equation}
and $m$ is the position of the D5-brane in the direction 5.

We can obtain the whole matrix model of a brane configuration $\bw=\left\{ \mathsf{w}^{\left(1\right)},\mathsf{w}^{\left(2\right)},\ldots,\mathsf{w}^{\left(R\right)}\right\} $ by gluing the matrix factors with integrations \eqref{eq:MM_gluing} as
\begin{equation}
Z^{\bw}=\frac{1}{\left(N!\right)^{R}}\int\prod_{r=1}^{R}\prod_{a}^{N}d\mu_{a}^{\left(r\right)}\prod_{r=1}^{R}\mathcal{Z}^{\left(\mathsf{w}^{\left(r\right)}\right)}\left(\mu^{\left(r\right)},\mu^{\left(r+1\right)}\right),\label{eq:pqMF-MM}
\end{equation}
where we omitted the parameters. Since the direction 6 is periodic (and hence the quiver diagram is the circular one), $\mu^{\left(R+1\right)}=\mu^{\left(1\right)}$.

By using the matrix factors, one can check that the relations between the FI or mass parameters and the position of the 5-branes, \eqref{eq:BC-FI} and \eqref{eq:BC-Mass}, are consistent with the definition of the position. Especially, for a $\left(p,q\right)$ web with (finite) web deformations, the definition is non-trivial. For the check, we use the fact that the values of the integration variables corresponds to the Cartans of the adjoint chiral multiples, and hence they correspond to the position of the D3-branes in the direction 5. This allows us to check the relative normalization of the mass and FI parameters by observing the effect of a constant shift of the integration variables to the parameters. Let us consider the $\left(p,q\right)$webs in figure \ref{fig:Web-MassFI} as examples. First, we consider the $\left(p,q\right)$ web in figure \ref{fig:Web-MassFI} (ii). The corresponding matrix factor is \eqref{eq:1qD5-MF} with $F=1$, $F_{\wdl}=F_{\wdr}=0$. We shift the integration variables as $\left(\mu,\nu\right)\rightarrow\left(\mu+c,\nu+c\right)$. In figure \ref{fig:Web-MassFI}, this shift corresponds to moving the D3-branes to the $+x^{5}$ direction by $c/\left(2\pi\right)$. The normalization $\left(2\pi\right)^{-1}$ can be found from the observation that under the shift the mass parameters are shifted as $m\rightarrow m-c/\left(2\pi\right)$, $\tilde{m}\rightarrow\tilde{m}-c/\left(2\pi\right)$. On the other hand, the shift of the FI parameter comes from the CS term, which is $\eta\rightarrow\eta+cq/\left(2\pi\right)$. This is consistent with the definition of the position and the relation between $q$ and the angle of the $\left(1,q\right)$5-brane $\theta$, $\tan\theta=q$. Second, we consider the $\left(p,q\right)$ web in figure \ref{fig:Web-MassFI} (iii). In this case, the angle $\theta$ for measuring the position satisfies $\tan\theta=q+1/2$. This is consistent with the fact that the web deformation gives $\pm1/2$ CS levels as \eqref{eq:D5wd-MF} and thus the effect of the shift of the integration variables form the CS level is added by $\pm1/2$.

\section{From $\left(p,q\right)$ webs to quantum curves\label{sec:BW-QC}}

As discussed in the introduction, we claim that given an $\mathcal{N}=2$ brane configuration, the Newton polygon of a quantum curve arising from the $S^{3}$ partition function is equal to the toric diagram which is dual to the combined $\left(p,q\right)$ web. In section \ref{subsec:Conjecture}, we explain the conjecture in detail. We also discuss a local version of the conjecture. The remaining sections are for giving various evidences for the conjecture. In section \ref{subsec:BW-QC-FGF}, we give a proof for the Lagrangian theories. In section \ref{subsec:WebDef}, we see that the web deformations are consistent with the conjecture. In section \ref{subsec:SL2Ztrans}, we see that the $\mathrm{SL}\left(2,\mathbb{Z}\right)$ transformations are consistent with the conjecture.

\subsection{Conjecture\label{subsec:Conjecture}}

As discussed in section \ref{subsec:BraneConfig}, we consider 3d ${\cal N}=2$ brane configurations with $\left(p,q\right)$ webs $\mathsf{w}^{\left(r\right)}$
\begin{equation}
\bw=\left\{ \mathsf{w}^{\left(1\right)},\mathsf{w}^{\left(2\right)},\ldots,\mathsf{w}^{\left(R\right)}\right\} .
\end{equation}
If all of the $\left(p,q\right)$ webs are ones in figure \ref{fig:BW-gen} with $p=1$ and isolated D5-branes, the Lagrangian of the 3d theory is known.\footnote{A 3d theory on a interval between an NS5-brane and a $\left(p,q\right)$5-brane with general $p$ has a dual Lagrangian description with $T\left(\mathrm{U}\left(N\right)\right)$ theories \cite{Gaiotto:2008ak}.} However, we also consider brane configurations which include general $\left(p,q\right)$ webs and thus the 3d theories do not have Lagrangian description. Those are not only the $\left(p,q\right)$ webs in figure \ref{fig:BW-gen} with $p\geq2$, but also more general $\left(p,q\right)$ webs. Even in these cases, the low energy dynamics on the D3-branes would be described by 3d $\mathcal{N}=2$ theories.

In this paper we propose two conjectures. The first conjecture focuses on the whole brane configuration $\bw$, and especially the combined $\left(p,q\right)$ web $\overline{\bw}$, while the second conjecture focuses on each $\left(p,q\right)$ web $\mathsf{w}^{\left(r\right)}$. In this paper we only consider the case when all the ranks are uniform, and in this case the first conjecture is obtained from the second conjecture. On the other hand, we expect that the first conjecture holds also for non-uniform ranks cases as we will remark later. 

\subsubsection{Conjecture for whole brane configuration\label{subsec:Conjecture1}}

We start with the first conjecture, which focuses on the whole brane configuration $\bw$. The notation is summarized in section \ref{sec:Reviews}.

We conjecture that the $S^{3}$ partition function of the worldvolume theory on $\bw$ can be written as
\begin{equation}
Z^{\bw}=\int\prod_{a}^{N}d\mu_{a}\det\left(\left[\braket{\mu_{a}|\hat{{\cal O}}\left(\hat{x},\hat{y}\right)^{-1}|\mu_{b}}\right]_{a,b}^{N\times N}\right),\label{eq:Conjecture1-MM}
\end{equation}
where $\hat{{\cal O}}$ is an operator of the quantum mechanics and satisfies the following two properties. First, $\hat{{\cal O}}$ is a quantum curve as a function of $\left(\hat{x},\hat{y}\right)$ where $\hat{x}$ and $\hat{y}$ are the position and momentum operators satisfying $\left[\hat{x},\hat{y}\right]=2\pi i$,\footnote{The notation of the quantum mechanics is summarized in appendix \ref{sec:QMnotation}.} and the Newton polygon of $\hat{{\cal O}}$ is equal to the dual toric diagram of the combined $\left(p,q\right)$ web $\overline{\bw}$. Second, the real parts of the asymptotic values of the classical curve $\mathcal{O}\left(x,y\right)=0$ correspond to the asymptotic positions of the external legs of $\overline{\bw}$.

We explain the second property in detail. We can obtain a classical curve $\mathcal{O}\left(x,y\right)$ by replacing the operators $\left(\hat{x},\hat{y}\right)$ by the coordinates $\left(x,y\right)$. We then identify the $\left(x,y\right)$ plane with the 59 plane for $\overline{\bw}$ under an appropriate normalization, where the $x$-axis and $y$-axis are identified with the directions 5 and 9, respectively. Under the identification, the first property ensures that the asymptotic behaviors of the curve are in one-to-one correspondence with the external legs of $\overline{\bw}$. This is because the directions of them are in the transverse directions of the edges of the toric diagram. Furthermore, we claim that the real parts of the asymptotic behavior of the curves correspond to the asymptotic positions of the external legs. Some concrete examples will appear in sections \ref{subsec:BW-QC-FGF} and \ref{subsec:pqQC}.

We give some remarks. First, as discussed around \eqref{eq:QC-NP}, since the equality between the toric diagram and the Newton polygon is determined up to translations, the conjecture only determines the relative powers of the quantum curve. Second, the coefficients of the quantum curve are functions of the mass parameters, the FI parameters, the R-charges and the order of the $\left(p,q\right)$ webs. Especially, the mass and FI parameter dependence of the coefficients of terms on the boundary of the Newton polygon is determined by the second property up to an overall factor. Third, although in this paper we assume that all the numbers of D3-branes on intervals are the same, we expect that this conjecture holds for general numbers.\footnote{We do not consider pathological cases. For example, when the brane configuration breaks s-rule, the partition function becomes zero.\label{fn:Conj-rd}} In this case, the relative ranks would also affect to the coefficients. An additional $N$-independent factor would also appear in \eqref{eq:Conjecture1-MM}. Explicit relations between the circular quiver gauge theories with rank deformations and the quantum curves are known for some examples \cite{Honda:2013pea,Kashaev:2015wia,Kubo:2019ejc,Kubo:2020qed}. These examples would, however, not satisfy the second conjecture. Fourth, we emphasize that our conjecture includes not only the $\left(p,q\right)$ webs depicted in figure \ref{fig:BW-gen} but also general $\left(p,q\right)$ webs.
\begin{figure}[t]
\begin{centering}
\includegraphics[scale=0.3]{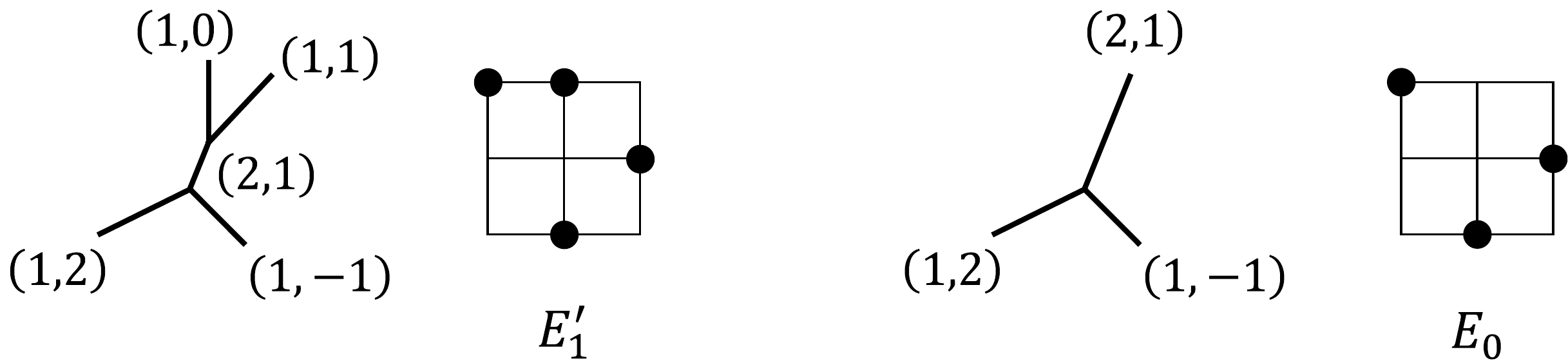}
\par\end{centering}
\caption{Two $\left(p,q\right)$ webs and their dual toric diagrams.\label{fig:BW-LP-E1}}
\end{figure}
For example, we conjecture that the partition functions of brane configurations consisting of only one $\left(p,q\right)$ web in figure \ref{fig:BW-LP-E1} are written in the form of \eqref{eq:Conjecture1-MM} with the following quantum curves up to an overall factor
\begin{align}
\hat{\mathcal{O}}^{\left(E_{1}'\right)}\left(\hat{x},\hat{y}\right) & \propto\left(\begin{array}{ccc}
+c_{1,-1}e^{\hat{x}-\hat{y}} & +c_{1,0}e^{\hat{x}} & +0\\
+0 & +0 & +c_{0,1}e^{\hat{y}}\\
+0 & +c_{-1,0}e^{-\hat{x}} & +0
\end{array}\right),\nonumber \\
\hat{\mathcal{O}}^{\left(E_{0}\right)}\left(\hat{x},\hat{y}\right) & \propto\left(\begin{array}{ccc}
+c_{1,-1}e^{\hat{x}-\hat{y}} & +0 & +0\\
+0 & +0 & +c_{0,1}e^{\hat{y}}\\
+0 & +c_{-1,0}e^{-\hat{x}} & +0
\end{array}\right).\label{eq:E1E0}
\end{align}
The Newton polygons of these curves are indeed equal to the toric diagrams in figure \ref{fig:BW-LP-E1}. $E_{1}'$ and $E_{0}$ are the symmetries of the curves, which we will explain later.

\subsubsection{Conjecture for each $\left(p,q\right)$ web\label{subsec:Conjecture2}}

Next, we explain the second conjecture. We emphasize that, unlike the first conjecture, which we expect to hold even if the numbers of D3-branes on the intervals are not the same, we assume that all the numbers of the D3-branes on all the intervals are the same $N$ in this conjecture. The claim is almost the same with the first conjecture, but this conjecture focuses on each $\left(p,q\right)$ web.

We conjecture that the $S^{3}$ partition function of the worldvolume theory on $\bw$ can be written as
\begin{equation}
Z^{\bw}=\frac{1}{\left(N!\right)^{R}}\int\prod_{r=1}^{R}\prod_{a}^{N}d\mu_{a}^{\left(r\right)}\prod_{r=1}^{R}\det\left(\left[\braket{\mu_{a}^{\left(r\right)}|\hat{{\cal O}}^{\left(r\right)}\left(\hat{x},\hat{y}\right)^{-1}|\mu_{b}^{\left(r+1\right)}}\right]_{a,b}^{N\times N}\right),\label{eq:Conjecture2-MM}
\end{equation}
where $\hat{{\cal O}}^{\left(r\right)}$ is an operator of the quantum mechanics and satisfies the following two properties. First, $\hat{{\cal O}}^{\left(r\right)}$ is a quantum curve as a function of $\left(\hat{x},\hat{y}\right)$ where $\hat{x}$ and $\hat{y}$ are the position and momentum operators satisfying $\left[\hat{x},\hat{y}\right]=2\pi i$, and the Newton polygon of $\hat{{\cal O}}^{\left(r\right)}$ is equal to the dual toric diagram of the $\left(p,q\right)$ web $\mathsf{w}^{\left(r\right)}$. Second, the real parts of the asymptotic values of the classical curve $\mathcal{O}^{\left(r\right)}\left(x,y\right)=0$ correspond to the asymptotic positions of the external legs of $\mathsf{w}^{\left(r\right)}$. (Note that $\mu_{a}^{\left(R+1\right)}=\mu_{a}^{\left(1\right)}$.)

The idea of the above conjecture is that we can associate a ``matrix factor'' not only for the Lagrangian theory as discussed in section \ref{subsec:BC-MM} but also for an arbitrary $\left(p,q\right)$ web $\mathsf{w}$. Namely, we expect that the matrix factor associated with $\mathsf{w}$ can be written as
\begin{equation}
\mathcal{Z}^{\left(\mathsf{w}\right)}\left(\mu,\nu\right)=\det\left(\left[\braket{\mu_{a}|\hat{{\cal O}}^{\left(\mathsf{w}\right)}\left(\hat{x},\hat{y}\right)^{-1}|\nu_{b}}\right]_{a,b}^{N\times N}\right),\label{eq:MF-QC}
\end{equation}
where the Newton polygon of $\hat{{\cal O}}^{\left(\mathsf{w}\right)}$ is $\lp\left(\mathsf{w}\right)$, and we can obtain the whole matrix model by gluing the matrix factors as \eqref{eq:pqMF-MM}.

One can derive the first conjecture from the second conjecture when all the ranks are uniform. Thanks to a formula for arbitrary operators $\hat{A},\hat{B}$
\begin{align}
 & \frac{1}{N!}\int\prod_{a}^{N}d\alpha_{a}\det\left(\left[\braket{\mu_{a}|\hat{A}|\alpha_{b}}\right]_{a,b}^{N\times N}\right)\det\left(\left[\braket{\alpha_{a}|\hat{B}|\nu_{b}}\right]_{a,b}^{N\times N}\right)\nonumber \\
 & =\det\left(\left[\braket{\mu_{a}|\hat{A}\hat{B}|\nu_{b}}\right]_{a,b}^{N\times N}\right),\label{eq:GlueForm}
\end{align}
the whole matrix model \eqref{eq:Conjecture2-MM} becomes
\begin{equation}
Z^{\bw}=\int\prod_{a}^{N}d\mu_{a}\det\left(\left[\braket{\mu_{a}|\prod_{r=1}^{R}\hat{{\cal O}}^{\left(r\right)}\left(\hat{x},\hat{y}\right)^{-1}|\mu_{b}}\right]_{a,b}^{N\times N}\right).
\end{equation}
Because the conjecture claims that the Newton polygon of $\hat{{\cal O}}^{\left(r\right)}$ is equal to the dual toric diagram of $\mathsf{w}^{\left(r\right)}$, thanks to the formula \eqref{eq:CombLP} the Newton polygon of $\prod_{r=1}^{R}\hat{{\cal O}}^{\left(r\right)}$ is $\lp\left(\overline{\bw}\right)$. Furthermore, the real part of the asymptotic values of the product of the classical curves $\prod_{r=1}^{R}{\cal O}^{\left(r\right)}\left(x,y\right)$ are the collection of the real part of the asymptotic values of ${\cal O}^{\left(r\right)}\left(x,y\right)$. Note that the non-commutativity of the operators generates additional phases, which, however, do not affect to the real part. Thus, the second conjecture leads to the first conjecture.

\subsection{Explicit derivation for Lagrangian theories\label{subsec:BW-QC-FGF}}

In this section we concretely confirm our conjecture described in section \ref{subsec:Conjecture} for a class of brane configurations whose worldvolume theories are Lagrangian theories. Namely, we focus on the $\left(p,q\right)$ webs in figure \ref{fig:BW-gen} with $p=0,1$. Furthermore, as discussed in section \ref{subsec:Conjecture}, since we focus on uniform ranks case, it is enough to check the second conjecture which is for each $\left(p,q\right)$ web discussed in section \ref{subsec:Conjecture2}.

For Lagrangian theories, as discussed in section \ref{subsec:Localization}, after applying the supersymmetric localization, the $S^{3}$ partition function reduces to the matrix model. As discussed in section \ref{subsec:BC-MM}, there is the one-to-one correspondence between the $\left(p,q\right)$ webs and the matrix factors, and the whole matrix model can be obtained as \eqref{eq:pqMF-MM}. An important point is that the form of \eqref{eq:pqMF-MM} is the same with the whole matrix model appeared in the second conjecture \eqref{eq:Conjecture2-MM}. Thus, our task is to show that the matrix factor appearing in section \ref{subsec:BC-MM} can be written in an operator formalism as \eqref{eq:MF-QC} and the operator $\hat{{\cal O}}^{\left(r\right)}$ satisfies the conditions.

To obtain the right hand side of \eqref{eq:MF-QC}, we adopt the Fermi gas formalism \cite{Marino:2011eh}. There are three types of the matrix factors, namely $\mathcal{Z}^{\left(1,q\right)}$ ,$\mathcal{Z}_{F,F_{\wdl},F_{\wdr}}^{\left(1,q\right)}$ and $\mathcal{Z}^{\left(0,1\right)}$. (Note that the first factor is the special case of the second factor.) We see that we can apply the Fermi gas formalism for each factor, and as a result the matrix factor for a $\left(p,q\right)$ web $\mathsf{w}$ can be written as a factor appearing in the partition function of an ideal Fermi gas system
\begin{equation}
\mathcal{Z}^{\left(\mathsf{w}\right)}\left(\mu,\nu\right)=\det\left(\left[\braket{\mu_{a}|\hat{\rho}^{\left(\mathsf{w}\right)}\left(\hat{x},\hat{y}\right)|\nu_{b}}\right]_{a,b}^{N\times N}\right).\label{eq:MF-DM}
\end{equation}
In terms of the Fermi gas system, $\hat{\rho}^{\left(\mathsf{w}\right)}$ is a one-particle density matrix. Then, following the conjecture, we introduce an operator $\hat{\mathcal{O}}^{\left(\mathsf{w}\right)}$ as the inverse of the density matrix $\hat{\rho}^{\left(\mathsf{w}\right)}$,
\begin{equation}
\hat{\mathcal{O}}^{\left(\mathsf{w}\right)}\left(\hat{x},\hat{y}\right)=\hat{\rho}^{\left(\mathsf{w}\right)}\left(\hat{x},\hat{y}\right)^{-1}.
\end{equation}
We show that the operator $\hat{\mathcal{O}}^{\left(\mathsf{w}\right)}$ is the form of a curve and its Newton polygon is equal to the dual toric diagram of $\mathsf{w}$, $\lp\left(\mathsf{w}\right)$. After proving these for the matrix factors, we finally show that the real part of the asymptotic behavior of the classical curve $\mathcal{O}^{\left(\mathsf{w}\right)}$ corresponds to the position of the external legs of $\mathsf{w}$.

First, we compute the $\left(1,q\right)$5-brane factor defined in \eqref{eq:1q-MF}. The first step to apply the Fermi gas formalism is to use the Cauchy determinant formula \cite{Kapustin:2010xq,Marino:2011eh}
\begin{equation}
\frac{\prod_{a<b}^{N}2\sinh\frac{\mu_{a}-\mu_{b}}{2}\prod_{a<b}^{N}2\sinh\frac{\nu_{a}-\nu_{b}}{2}}{\prod_{a,b}^{N}2\cosh\frac{\mu_{a}-\nu_{b}}{2}}=\det\left(\left[\frac{1}{2\cosh\frac{\mu_{a}-\nu_{b}}{2}}\right]_{a,b}^{N\times N}\right).\label{eq:1Loop-Det}
\end{equation}
After putting the remaining factors into the matrix, we obtain
\begin{equation}
\mathcal{Z}^{\left(1,q\right)}\left(\eta;\mu,\nu\right)=\frac{1}{\left(2\pi\right)^{N}}\det\left(\left[\frac{e^{\frac{iq}{4\pi}\left(\mu_{a}^{2}-\nu_{b}^{2}\right)+i\eta\left(\mu_{a}-\nu_{b}\right)}}{2\cosh\frac{\mu_{a}-\nu_{b}}{2}}\right]_{a,b}^{N\times N}\right).
\end{equation}
We then introduce the quantum mechanical system by using \eqref{eq:Cosh-op}
\begin{equation}
\mathcal{Z}^{\left(1,q\right)}\left(\eta;\mu,\nu\right)=\det\left(\left[\braket{\mu_{a}|e^{\frac{iq}{4\pi}\hat{x}^{2}+i\eta\hat{x}}\frac{1}{2\cosh\frac{\hat{y}}{2}}e^{-\frac{iq}{4\pi}\hat{x}^{2}-i\eta\hat{x}}|\nu_{b}}\right]_{a,b}^{N\times N}\right).\label{eq:1q-FGF}
\end{equation}
This is the form of \eqref{eq:MF-DM} with the density matrix
\begin{equation}
\hat{\rho}^{\left(1,q\right)}\left(\eta;\hat{x},\hat{y}\right)=\frac{1}{2\cosh\frac{\hat{y}-q\hat{x}-2\pi\eta}{2}},\label{eq:1q-DM}
\end{equation}
where we used \eqref{eq:OpSim}. For this density matrix, one can easily find that the inverse of $\hat{\rho}^{\left(1,q\right)}$ is
\begin{equation}
\hat{\mathcal{O}}^{\left(1,q\right)}\left(\eta;\hat{x},\hat{y}\right)=e^{\pi\eta}e^{\frac{q}{2}\hat{x}-\frac{1}{2}\hat{y}}+e^{-\pi\eta}e^{-\frac{q}{2}\hat{x}+\frac{1}{2}\hat{y}}.
\end{equation}
This is clearly the form of the quantum curve, and the Newton polygon of this quantum curve is indeed equal to $\lp\left(\left[\left(1,q\right)\right]\right)$.

Second, we consider the matrix factor $\mathcal{Z}_{F,F_{\wdr},F_{\wdl}}^{\left(1,q\right)}$ defined in \eqref{eq:1qD5-MF}, which is associated with the $\left(p,q\right)$ web
\begin{equation}
\mathsf{w}_{F,F_{\wdr},F_{\wdl}}^{\left(1,q\right)}=\left[\left(1,q\right)+F\df+F_{\wdr}\dfr+F_{\wdl}\dfl\right].
\end{equation}
One can perform the computation in the same way with $\left[\left(1,q\right)\right]$ and again rewrite $\mathcal{Z}^{\left(1,q\right)}$ in $\mathcal{Z}_{F,F_{\wdr},F_{\wdl}}^{\left(1,q\right)}$ as \eqref{eq:1q-FGF}. After putting the remaining D5-brane factors $\mathcal{Z}_{\mathrm{D5}}$ and $\mathcal{Z}_{\mathrm{D5}^{\pm}}$ into the matrix, we again obtain the form of \eqref{eq:MF-DM}. In this case the density matrix is
\begin{align}
 & \hat{\rho}_{F,F_{\wdr},F_{\wdl}}^{\left(1,q\right)}\left(\eta,\mathbf{z};\hat{x},\hat{y}\right)\nonumber \\
 & =\frac{\prod_{f=1}^{F+F_{\wdl}}s_{1}\left(\frac{\hat{x}}{2\pi}-\tilde{z}_{f}+\frac{i}{4}\right)}{\prod_{f=1}^{F+F_{\wdr}}s_{1}\left(\frac{\hat{x}}{2\pi}-z_{f}-\frac{i}{4}\right)}\frac{1}{2\cosh\frac{\hat{y}-\left(q+\frac{1}{2}F_{\wdr}+\frac{1}{2}F_{\wdl}\right)\hat{x}-2\pi\eta}{2}}\frac{\prod_{f=1}^{F+F_{\wdr}}s_{1}\left(\frac{\hat{x}}{2\pi}-z_{f}+\frac{i}{4}\right)}{\prod_{f=1}^{F+F_{\wdl}}s_{1}\left(\frac{\hat{x}}{2\pi}-\tilde{z}_{f}-\frac{i}{4}\right)},\label{eq:1qD5-DM}
\end{align}
where $\mathbf{z}$ is defined in \eqref{eq:1qD5-MFpara}. For this density matrix, we can again show that the inverse is the form of the quantum curve in a following way. By using \eqref{eq:OpSim} and \eqref{eq:DS-Trig}, we obtain identities
\begin{align}
\frac{1}{s_{1}\left(\frac{\hat{x}}{2\pi}-z+\frac{i}{4}\right)}\left(e^{\frac{1}{2}\hat{y}}+e^{-\frac{1}{2}\hat{y}}\right)s_{1}\left(\frac{\hat{x}}{2\pi}-z-\frac{i}{4}\right) & =e^{\frac{1}{4}\hat{y}}2\cosh\left(\frac{\hat{x}}{2}-\pi z\right)e^{\frac{1}{4}\hat{y}}+e^{-\frac{1}{2}\hat{y}},\nonumber \\
s_{1}\left(\frac{\hat{x}}{2\pi}-\tilde{z}-\frac{i}{4}\right)\left(e^{\frac{1}{2}\hat{y}}+e^{-\frac{1}{2}\hat{y}}\right)\frac{1}{s_{1}\left(\frac{\hat{x}}{2\pi}-\tilde{z}+\frac{i}{4}\right)} & =e^{\frac{1}{2}\hat{y}}+e^{-\frac{1}{4}\hat{y}}2\cosh\left(\frac{\hat{x}}{2}-\pi\tilde{z}\right)e^{-\frac{1}{4}\hat{y}}.\label{eq:DS-sim}
\end{align}
Notice that the formula \eqref{eq:DS-Trig} cannot be applied if the two $z$ ($\tilde{z}$) appearing on the left-hand side of the first (second) line are different as $s_{1}\left(\frac{\hat{x}}{2\pi}-z_{1}+\frac{i}{4}\right)^{-1}\left(e^{\frac{1}{2}\hat{y}}+e^{-\frac{1}{2}\hat{y}}\right)s_{1}\left(\frac{\hat{x}}{2\pi}-z_{2}-\frac{i}{4}\right)$. The equality $z_{1}=z_{2}=z$ comes from \eqref{eq:Mass-cond} and \eqref{eq:R-cond}, which come from the $\left(p,q\right)$ web setup. In this sense, the curve form reflects the brane setup behind the 3d theories. By using this identity, we obtain
\begin{align}
\hat{\mathcal{O}}_{F,F_{\wdr},F_{\wdl}}^{\left(1,q\right)}\left(\eta,\mathbf{z};\hat{x},\hat{y}\right) & =e^{-\pi\eta}e^{\frac{1}{4}\hat{y}}\left\{ e^{-\frac{1}{2}\left(q+\frac{1}{2}F_{\wdr}+\frac{1}{2}F_{\wdl}\right)\hat{x}}\prod_{f=1}^{F+F_{\wdr}}\left(e^{\frac{1}{2}\hat{x}-\pi z_{f}}+e^{-\frac{1}{2}\hat{x}+\pi z_{f}}\right)\right\} e^{\frac{1}{4}\hat{y}}\nonumber \\
 & \quad+e^{\pi\eta}e^{-\frac{1}{4}\hat{y}}\left\{ e^{\frac{1}{2}\left(q+\frac{1}{2}F_{\wdr}+\frac{1}{2}F_{\wdl}\right)\hat{x}}\prod_{f=1}^{F+F_{\wdl}}\left(e^{\frac{1}{2}\hat{x}-\pi\tilde{z}_{f}}+e^{-\frac{1}{2}\hat{x}+\pi\tilde{z}_{f}}\right)\right\} e^{-\frac{1}{4}\hat{y}}.\label{eq:1qD5-QC}
\end{align}
This is clearly the form of the quantum curve, and the Newton polygon of it is indeed $\lp\left(\mathsf{w}_{F,F_{\wdr},F_{\wdl}}^{\left(1,q\right)}\right)$. Note that this expression clarifies that for each term the position and momentum operators $\hat{x}$ and $\hat{y}$ are on single exponential factor thanks to the identity
\begin{equation}
e^{\frac{c}{2}\hat{y}}e^{f\left(\hat{x}\right)}e^{\frac{c}{2}\hat{y}}=e^{f\left(\hat{x}\right)+c\hat{y}}.
\end{equation}
This identity comes from \eqref{eq:CBHform}.

Third, we consider the matrix factor $\mathcal{Z}^{\left(0,1\right)}$ defined in \eqref{eq:D5-MF}. Notice that $\mathcal{Z}^{\left(0,1\right)}$ is always glued with another matrix factor by the integrations as \eqref{eq:pqMF-MM}. If the matrix factor is an anti-symmetric function, we can rewrite $\mathcal{Z}^{\left(0,1\right)}$ into a determinant form by using the following formula for an anti-symmetric function $f\left(\nu\right)$
\begin{equation}
N!\int\prod_{a}^{N}d\nu_{a}\prod_{a}^{N}\delta\left(\mu_{a}-\nu_{a}\right)f\left(\nu\right)=\int\prod_{a}^{N}d\nu_{a}\det\left(\left[\delta\left(\mu_{a}-\nu_{b}\right)\right]_{a,b}^{N\times N}\right)f\left(\nu\right).
\end{equation}
Indeed, the matrix factor $\mathcal{Z}_{F,F_{\wdr},F_{\wdl}}^{\left(1,q\right)}$ is the anti-symmetric function, and $\mathcal{Z}^{\left(0,1\right)}$ can also be anti-symmetrized recursively with the above formula. (Note that the conjectured form of a matrix factor \eqref{eq:MF-QC} is also an anti-symmetric function.) Then, by using \eqref{eq:Normalization} and putting the factors $\left(2\cosh\right)^{-1}$ into the determinant, we obtain the form of \eqref{eq:MF-DM} with the density matrix
\begin{equation}
\hat{\rho}^{\left(0,1\right)}\left(z;\hat{x}\right)=\frac{1}{2\cosh\frac{\hat{x}-2\pi z}{2}},\label{eq:D5-DM}
\end{equation}
where $z$ is defined in \eqref{eq:D5-MFpara}. Therefore, the inverse of the density matrix $\hat{\rho}^{\left(0,1\right)}$ is
\begin{equation}
\hat{\mathcal{O}}^{\left(0,1\right)}\left(z;\hat{x}\right)=e^{-\pi z}e^{\frac{1}{2}\hat{x}}+e^{\pi z}e^{-\frac{1}{2}\hat{x}}.\label{eq:D5-QC}
\end{equation}
This is clearly the form of the quantum curve, and the Newton polygon of it is indeed $\lp\left(\left[\left(0,1\right)\right]\right)$.

Finally, let us consider the asymptotic behavior of the curves. We identify the $\left(x,y\right)$ plane with the 59 plane. Because $\left[\left(1,q\right)\right]$ is a special case of $\mathsf{w}_{F,F_{\wdr},F_{\wdl}}^{\left(1,q\right)}$, we only consider the latter $\left(p,q\right)$ web and $\left[\left(0,1\right)\right]$. For $\mathsf{w}_{F,F_{\wdr},F_{\wdl}}^{\left(1,q\right)}$, we have four directions where the external legs extend. Namely, for example when $F_{\wdr}=0$, there are a $\left(1,q+F_{\wdl}\right)$5-brane extending in the $+x$ direction, a $\left(1,q\right)$5-brane extending in the $-x$ direction, $F$ D5-branes extending in the $+y$ direction and $F+F_{\wdl}$ D5-branes extending in the $-y$ direction (see right side of figure \ref{fig:BW-gen}). For the $+x$ direction, we introduce new variables $\left(u,v\right)$ by $x=u$, $y=\left(q+F_{\wdl}\right)u+v$. Then, in the limit $u\rightarrow\infty$ only two terms remain
\begin{equation}
\mathcal{O}_{F,0,F_{\wdl}}^{\left(1,q\right)}\left(\eta,\mathbf{z};u,v\right)\rightarrow\left(e^{\pi\eta-\pi\sum_{f=1}^{F+F_{\wdl}}\tilde{z}_{f}}e^{-\frac{1}{2}v}+e^{-\pi\eta-\pi\sum_{f=1}^{F}z_{f}}e^{\frac{1}{2}v}\right)e^{\left(\frac{1}{2}F+\frac{1}{4}F_{\wdl}\right)u}.
\end{equation}
Hence the real part of the solution of $\lim_{u\rightarrow\infty}\mathcal{O}_{F,0,F_{\wdl}}^{\left(1,q\right)}=0$ is 
\begin{equation}
\left(2\pi\right)^{-1}\Re\left(v\right)=\eta+\frac{1}{2}\sum_{f=1}^{F}m_{f}-\frac{1}{2}\sum_{f=1}^{F+F_{\wdl}}\tilde{m}_{f}.
\end{equation}
(The normalization $\left(2\pi\right)^{-1}$ is expected from the observation in the last paragraph of section \ref{subsec:BC-MM}.) This solution actually matches the position of the $\left(1,q+F_{\wdl}\right)$5-brane. The effect of $\eta$ agrees with this interpretation by definition. The effects from $m_{f}$, $\tilde{m}_{f}$ also correspond because the move of a D5-brane extending in the $\pm x^{9}$ direction by $m$ (or $\tilde{m}$) moves the $\left(1,q+F_{\wdl}\right)$5-brane by $m/2$ (or $-\tilde{m}/2$). (We will consider the same move in a context of the web deformations in section \ref{subsec:WebDef}, see e.g. figure \ref{fig:WebDef-1q}.) For the $-y$ direction, in the limit $y\rightarrow-\infty$ the remaining terms are
\begin{equation}
\mathcal{O}_{F,0,F_{\wdl}}^{\left(1,q\right)}\left(\eta,\mathbf{z};x,y\right)\rightarrow e^{\pi\eta}\left\{ e^{\frac{1}{2}\left(q+\frac{1}{2}F_{\wdr}+\frac{1}{2}F_{\wdl}\right)x}\prod_{f=1}^{F+F_{\wdl}}\left(e^{\frac{1}{2}x-\pi\tilde{z}_{f}}+e^{-\frac{1}{2}x+\pi\tilde{z}_{f}}\right)\right\} e^{-\frac{1}{2}y}.
\end{equation}
Hence the real part of the solutions of $\lim_{y\rightarrow-\infty}\mathcal{O}_{F,0,F_{\wdl}}^{\left(1,q\right)}=0$ are 
\begin{equation}
\left(2\pi\right)^{-1}\Re\left(x\right)=\tilde{m}_{f},\quad\left(f=1,2,\ldots,F+F_{\wdl}\right).
\end{equation}
Since we have $F+F_{\wdl}$ D5-branes in the $-x^{9}$ direction and their positions are $\tilde{m}_{f}$, this is again an expected result. The other directions or the $F_{\wdl}=0$ case can be checked in a similar way. For $\left[\left(0,1\right)\right]$, the real part of the solution of $\mathcal{O}^{\left(0,1\right)}\left(z;x\right)=0$ is $x/\left(2\pi\right)=m$, which is again an expected result.

\subsection{Web deformations\label{subsec:WebDef}}

In this section we show how the web deformations are realized in matrix models and quantum curves. We first consider the case when we have an explicit computation, and then we discuss more general cases.

\subsubsection{From $\left[\left(1,0\right)+\protect\df\right]$ to $\left[\left(1,1\right)\right]$\label{subsec:WebDefEx}}

In this section we consider two web deformations. The first one is from $\left[\left(1,0\right)+\df\right]$ to $\left[\left(1,0\right)+\dfl\right]$, and the second one is from it to $\left[\left(1,1\right)\right]$.

The matrix factor for $\left[\left(1,0\right)+\df\right]$ cab be read off from \eqref{eq:1qD5-MF} as
\begin{align}
\mathcal{Z}_{1,0,0}^{\left(1,0\right)}\left(\eta,z,\tilde{z};\mu,\nu\right) & =\frac{1}{\left(2\pi\right)^{N}}e^{i\eta\sum_{a}^{N}\left(\mu_{a}-\nu_{a}\right)}\frac{\prod_{a<b}^{N}2\sinh\frac{\mu_{a}-\mu_{b}}{2}\prod_{a<b}^{N}2\sinh\frac{\nu_{a}-\nu_{b}}{2}}{\prod_{a,b}^{N}2\cosh\frac{\mu_{a}-\nu_{b}}{2}}\nonumber \\
 & \quad\times\frac{\prod_{a}^{N}s_{1}\left(\frac{\mu_{a}}{2\pi}-\tilde{z}+\frac{i}{4}\right)}{\prod_{a}^{N}s_{1}\left(\frac{\mu_{a}}{2\pi}-z-\frac{i}{4}\right)}\frac{\prod_{a}^{N}s_{1}\left(\frac{\nu_{a}}{2\pi}-z+\frac{i}{4}\right)}{\prod_{a}^{N}s_{1}\left(\frac{\nu_{a}}{2\pi}-\tilde{z}-\frac{i}{4}\right)},
\end{align}
where
\begin{equation}
z=m+iD,\quad\tilde{z}=\tilde{m}+i\tilde{D}.
\end{equation}
We consider the web deformation where the right D5-brane goes far away in the $+x^{5}$ direction. At first sight, taking $m\rightarrow+\infty$ limit would be enough for realizing this deformation. However, this simple limit moves the remaining junction far away in the $-x^{9}$ direction. Note that this effect is obtained purely in the gauge theory \cite{Aharony:1997bx,Willett:2016adv}. We can keep the junction by adjusting the FI parameter by hand.
\begin{figure}[t]
\begin{centering}
\includegraphics[scale=0.5]{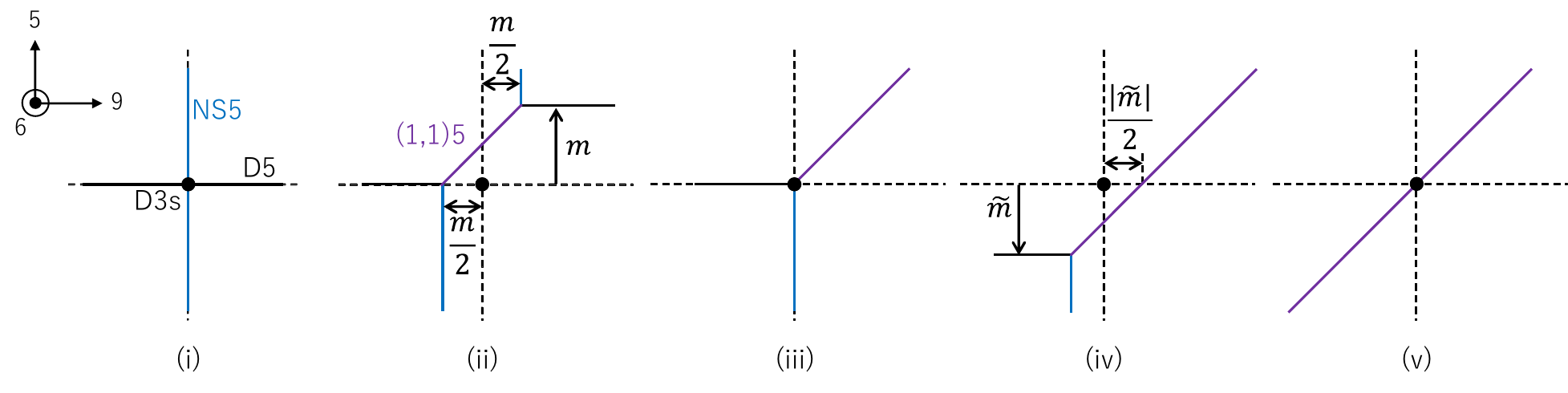}
\par\end{centering}
\caption{Two examples of web deformations. Starting from (i), we move the right D5-brane in the $+x^{5}$ direction while keeping the position of the left D5-brane. The NS5-brane gets the unexpected move as (ii). The amount of the move of the upper and lower NS5-branes is $\pm\tilde{m}/2$. For cancelling this effect we move the $\left(p,q\right)$ web to the $+x^{9}$ direction by $+\tilde{m}/2$. After moving the right D5-brane far away, we arrive at (iii). Once again, we start from (iii) and move the left D5-brane in the $-x^{5}$ direction. The $\left(1,1\right)$5-brane gets the unexpected move as (iv), thus we move the $\left(p,q\right)$ web in the $-x^{9}$ direction. After moving the left D5-brane far away, we arrive at (v).\label{fig:pq_WebDef1}}
\end{figure}
Figure \ref{fig:pq_WebDef1} (i), (ii) and (iii) show this situation and the amount of the shift of the FI parameter compared with the mass parameter. According to the figure, we take the limit
\begin{equation}
\eta=\frac{1}{2}\Lambda+\eta',\quad m=\Lambda,\quad\Lambda\rightarrow+\infty.\label{eq:WD-Lim1}
\end{equation}
By using \eqref{eq:DSasym}, we obtain
\begin{align}
 & e^{i\eta\left(\mu-\nu\right)}\frac{s_{1}\left(\frac{\nu}{2\pi}-z+\frac{i}{4}\right)}{s_{1}\left(\frac{\mu}{2\pi}-z-\frac{i}{4}\right)}\nonumber \\
 & \rightarrow\exp\left[\frac{i}{8\pi}\left(\mu^{2}-\nu^{2}\right)+i\left(\eta'-\frac{i}{2}D\right)\left(\mu-\nu\right)+\frac{1}{8}\left(\mu+\nu\right)-\frac{\pi}{2}\Lambda-\frac{i\pi}{2}D\right].
\end{align}
By using this asymptotic behavior, after rescaling the matrix factor and cancelling the phase factor we find
\begin{align}
\lim_{\Lambda\rightarrow\infty}e^{\frac{\pi}{2}N\Lambda+\frac{i\pi}{2}ND}{\cal Z}_{1,0,0}^{\left(1,0\right)}\left(\eta,z,\tilde{z};\mu,\nu\right) & =e^{\frac{i}{8\pi}\sum_{a}^{N}\left(\mu_{a}^{2}-\nu_{a}^{2}\right)}e^{i\left(\eta'-\frac{i}{2}D\right)\sum_{a}^{N}\left(\mu_{a}-\nu_{a}\right)}e^{\frac{1}{8}\sum_{a}^{N}\left(\mu_{a}+\nu_{a}\right)}\nonumber \\
 & \quad\times\frac{\prod_{a<b}^{N}2\sinh\frac{\mu_{a}-\mu_{b}}{2}\prod_{a<b}^{N}2\sinh\frac{\nu_{a}-\nu_{b}}{2}}{\prod_{a,b}^{N}2\cosh\frac{\mu_{a}-\nu_{b}}{2}}\frac{\prod_{a}^{N}s_{1}\left(\frac{\mu_{a}}{2\pi}-\tilde{z}+\frac{i}{4}\right)}{\prod_{a}^{N}s_{1}\left(\frac{\nu_{a}}{2\pi}-\tilde{z}-\frac{i}{4}\right)}.
\end{align}
This is the same with the matrix factor with $F=0$, $F_{\wdl}=1$. Therefore, we find the limit \eqref{eq:WD-Lim1}
\begin{equation}
\lim_{\Lambda\rightarrow\infty}e^{\frac{\pi}{2}N\Lambda+\frac{i\pi}{2}ND}{\cal Z}_{1,0,0}^{\left(1,0\right)}\left(\eta,z,\tilde{z};\mu,\nu\right)=e^{\frac{1}{8}\sum_{a}^{N}\left(\mu_{a}+\nu_{a}\right)}{\cal Z}_{0,0,1}^{\left(1,0\right)}\left(\eta'-\frac{i}{2}D,\tilde{z};\mu,\nu\right).\label{eq:PF-NDtoND11}
\end{equation}
The factor $\exp\left[\frac{1}{8}\sum_{a}^{N}\left(\mu_{a}+\nu_{a}\right)\right]$ would come from monopole operators with non-zero R-charges \cite{Jafferis:2010un}.\footnote{In general, if we add factors of this type (namely, FI terms with pure imaginably FI parameters) to the matrix model, the resulting quantum curve is multiplied by $\exp\left(c\hat{x}\right)$ where $c\in\mathbb{R}$ and $-ic$ is the total value of the FI parameters. (An additional phase also appears in each term.) However, since the equality between Newton polygons and toric diagrams is insensitive to $c$, we do not care about such an overall factor.\label{fn:CurveShift}} 

Next, we consider the web deformation from (iii) to (v) in figure \ref{fig:pq_WebDef1}. The matrix factor for $\left[\left(1,0\right)+\dfl\right]$ can be read off from \eqref{eq:1qD5-MF} as
\begin{align}
 & \mathcal{Z}_{0,0,1}^{\left(1,0\right)}\left(\eta,\tilde{z};\mu,\nu\right)\nonumber \\
 & =\frac{1}{\left(2\pi\right)^{N}}e^{\frac{i}{8\pi}\sum_{a}^{N}\left(\mu_{a}^{2}-\nu_{a}^{2}\right)}e^{i\eta\sum_{a}^{N}\left(\mu_{a}-\nu_{a}\right)}\frac{\prod_{a<b}^{N}2\sinh\frac{\mu_{a}-\mu_{b}}{2}\prod_{a<b}^{N}2\sinh\frac{\nu_{a}-\nu_{b}}{2}}{\prod_{a,b}^{N}2\cosh\frac{\mu_{a}-\nu_{b}}{2}}\frac{\prod_{a}^{N}s_{1}\left(\frac{\mu_{a}}{2\pi}-\tilde{z}+\frac{i}{4}\right)}{\prod_{a}^{N}s_{1}\left(\frac{\nu_{a}}{2\pi}-\tilde{z}-\frac{i}{4}\right)},
\end{align}
where
\begin{equation}
\tilde{z}=\tilde{m}+i\tilde{D}.
\end{equation}
We again have to move the $\left(p,q\right)$ web by shifting the FI parameter. Figures \ref{fig:pq_WebDef1} (iii), (iv) and (v) show this situation and the amount of the shift of the FI parameter compared with the mass parameter. According to the figure, we take the limit
\begin{equation}
\eta=-\frac{1}{2}\Lambda+\eta',\quad\tilde{m}=-\Lambda,\quad\Lambda\rightarrow+\infty.\label{eq:WD-Lim2}
\end{equation}
By using \eqref{eq:DSasym}, we obtain
\begin{align}
 & e^{i\eta\left(\mu-\nu\right)}\frac{s_{1}\left(\frac{\mu}{2\pi}-\tilde{z}+\frac{i}{4}\right)}{s_{1}\left(\frac{\nu}{2\pi}-\tilde{z}-\frac{i}{4}\right)}\nonumber \\
 & \rightarrow\exp\left[\frac{i}{8\pi}\left(\mu^{2}-\nu^{2}\right)+i\left(\eta'-\frac{i}{2}\tilde{D}\right)\left(\mu-\nu\right)-\frac{1}{8}\left(\mu+\nu\right)-\frac{\pi}{2}\Lambda+\frac{i\pi}{2}\tilde{D}\right].
\end{align}
By using this asymptotic behavior, we find
\begin{align}
 & \lim_{\Lambda\rightarrow\infty}e^{\frac{\pi}{2}N\Lambda-\frac{i\pi}{2}N\tilde{D}}\mathcal{Z}_{0,0,1}^{\left(1,0\right)}\left(\eta,\tilde{z};\mu,\nu\right)\nonumber \\
 & =e^{\frac{i}{4\pi}\sum_{a}^{N}\left(\mu_{a}^{2}-\nu_{a}^{2}\right)}e^{i\left(\eta'-\frac{i}{2}\tilde{D}\right)\sum_{a}^{N}\left(\mu_{a}-\nu_{a}\right)}e^{-\frac{1}{8}\sum_{a}^{N}\left(\mu_{a}+\nu_{a}\right)}\frac{\prod_{a<b}^{N}2\sinh\frac{\mu_{a}-\mu_{b}}{2}\prod_{a<b}^{N}2\sinh\frac{\nu_{a}-\nu_{b}}{2}}{\prod_{a,b}^{N}2\cosh\frac{\mu_{a}-\nu_{b}}{2}}.
\end{align}
This is the same with the matrix factor with $F=0$, $F_{\wdl}=0$, $q=1$. Therefore, we obtain the limit \eqref{eq:WD-Lim2}
\begin{equation}
\lim_{\Lambda\rightarrow\infty}e^{\frac{\pi}{2}N\Lambda-\frac{i\pi}{2}N\tilde{D}}\mathcal{Z}_{0,0,1}^{\left(1,0\right)}\left(\eta,\tilde{z};\mu,\nu\right)=e^{-\frac{1}{8}\sum_{a}^{N}\left(\mu_{a}+\nu_{a}\right)}\mathcal{Z}_{0,0,0}^{\left(1,1\right)}\left(\eta'-\frac{i}{2}\tilde{D};\mu,\nu\right).\label{eq:PF-ND11to11}
\end{equation}

Next, we consider the quantum curves. The quantum curve \eqref{eq:1qD5-QC} with $F=1$, $F_{\wdl}=F_{\wdr}=0$, $q=0$ is
\begin{equation}
\hat{\mathcal{O}}_{1,0,0}^{\left(1,0\right)}\left(\eta,z,\tilde{z};\hat{x},\hat{y}\right)=\left(\begin{array}{cc}
+e^{\pi\eta-\pi\tilde{z}}e^{\frac{1}{2}\hat{x}-\frac{1}{2}\hat{y}} & +e^{-\pi\eta-\pi z}e^{\frac{1}{2}\hat{x}+\frac{1}{2}\hat{y}}\\
+e^{\pi\eta+\pi\tilde{z}}e^{-\frac{1}{2}\hat{x}-\frac{1}{2}\hat{y}} & +e^{-\pi\eta+\pi z}e^{-\frac{1}{2}\hat{x}+\frac{1}{2}\hat{y}}
\end{array}\right).
\end{equation}
We first take the limit \eqref{eq:WD-Lim1}. One can easily find that
\begin{align}
\lim_{\Lambda\rightarrow\infty}e^{-\frac{\pi}{2}\Lambda-\frac{i\pi}{2}D}\hat{\mathcal{O}}_{1,0,0}^{\left(1,0\right)}\left(\eta,z,\tilde{z};\hat{x},\hat{y}\right) & =e^{-\frac{i\pi}{2}D}\left(\begin{array}{cc}
+e^{\pi\eta'-\pi\tilde{z}}e^{\frac{1}{2}\hat{x}-\frac{1}{2}\hat{y}} & +0\\
+e^{\pi\eta'+\pi\tilde{z}}e^{-\frac{1}{2}\hat{x}-\frac{1}{2}\hat{y}} & +e^{-\pi\eta'+\pi iD}e^{-\frac{1}{2}\hat{x}+\frac{1}{2}\hat{y}}
\end{array}\right),\nonumber \\
 & =e^{-\frac{1}{8}\hat{x}}\hat{\mathcal{O}}_{0,0,1}^{\left(1,0\right)}\left(\eta'-\frac{i}{2}D,\tilde{z};\hat{x},\hat{y}\right)e^{-\frac{1}{8}\hat{x}}.\label{eq:QC-NDtoND11}
\end{align}
This is consistent with the result of the matrix factor \eqref{eq:PF-NDtoND11}. Next, we take the limit \eqref{eq:WD-Lim2} for $\hat{\mathcal{O}}_{0,0,1}^{\left(1,0\right)}\left(\eta,\tilde{z}\right)$, which is
\begin{align}
\lim_{\Lambda\rightarrow\infty}e^{-\frac{\pi}{2}\Lambda+\frac{i\pi}{2}\tilde{D}}\hat{\mathcal{O}}_{0,0,1}^{\left(1,0\right)}\left(\eta,\tilde{z};\hat{x},\hat{y}\right) & =e^{\frac{i\pi}{2}\tilde{D}}e^{\frac{1}{8}\hat{x}}\left(\begin{array}{cc}
+e^{\pi\eta'-\pi i\tilde{D}}e^{\frac{1}{2}\hat{x}-\frac{1}{2}\hat{y}} & +0\\
+0 & +e^{-\pi\eta'}e^{-\frac{1}{2}\hat{x}+\frac{1}{2}\hat{y}}
\end{array}\right)e^{\frac{1}{8}\hat{x}},\nonumber \\
 & =e^{\frac{1}{8}\hat{x}}\hat{\mathcal{O}}_{0,0,0}^{\left(1,1\right)}\left(\eta'-\frac{i}{2}\tilde{D};\hat{x},\hat{y}\right)e^{\frac{1}{8}\hat{x}}.\label{eq:QC-ND11to11}
\end{align}
This is again consistent with the result of the matrix factor \eqref{eq:PF-ND11to11}.

\subsubsection{General cases\label{subsec:WebDefGen}}

Next, we consider more general cases. First, we consider the $\left(p,q\right)$ webs in figure \ref{fig:BW-gen} with $p=1$. The way of the computation in terms of the matrix factor is clearly the same as the previous examples since a web deformation affects only one D5-brane factor even if there are many D5-brane factors. Correspondingly, the computation in terms of the quantum curve is the same. Notice that the computation is clearly independent of $q$. On the other hand, in the brane picture it would be unclear whether the amount of move of the $\left(p,q\right)$ web is independent of $q$ or not. However, by carefully seeing the definition of the FI and mass parameter discussed in section \ref{subsec:3dTh}, one can find that it is indeed independent.
\begin{figure}[t]
\begin{centering}
\includegraphics[scale=0.5]{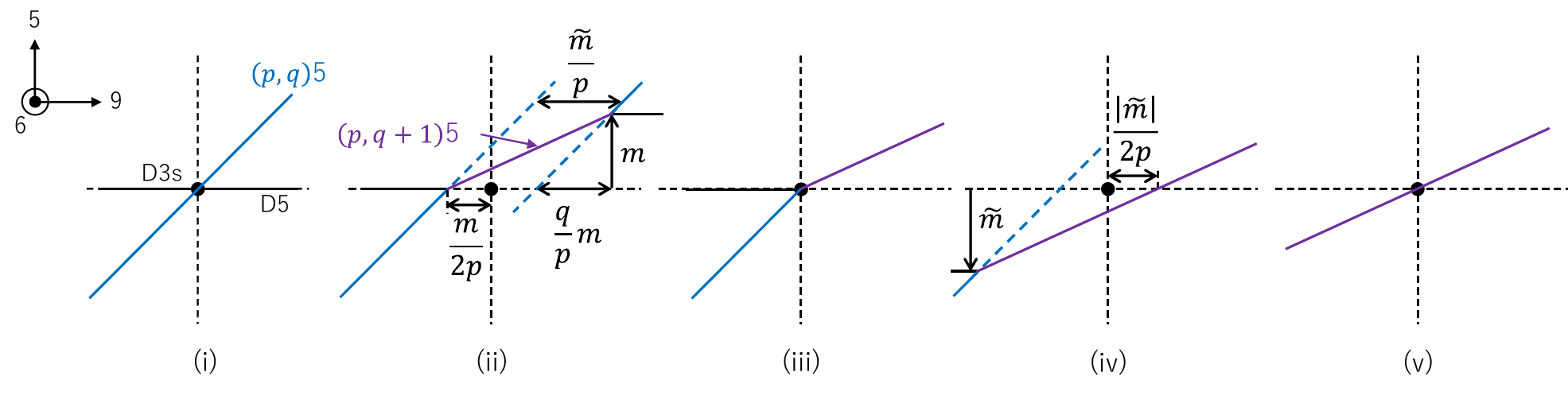}
\par\end{centering}
\caption{Two examples of web deformations. In each step, we need to move the $\left(p,q\right)$ web by adjusting the FI parameter as well as the $\left(p,q\right)=\left(1,0\right)$ case in figure \ref{fig:pq_WebDef1}. The amount of the move depends on $p$ but is independent of $q$.\label{fig:WebDef-1q}}
\end{figure}
Figure \ref{fig:WebDef-1q} shows two examples, which are generalized versions of the two examples discussed in the previous section. Note that in this figure we consider a general $p$ for later convenience. We discuss web deformations for $p\geq2$ cases in section \ref{sec:pqMF-QC}.

\begin{figure}[t]
\begin{centering}
\includegraphics[scale=0.6]{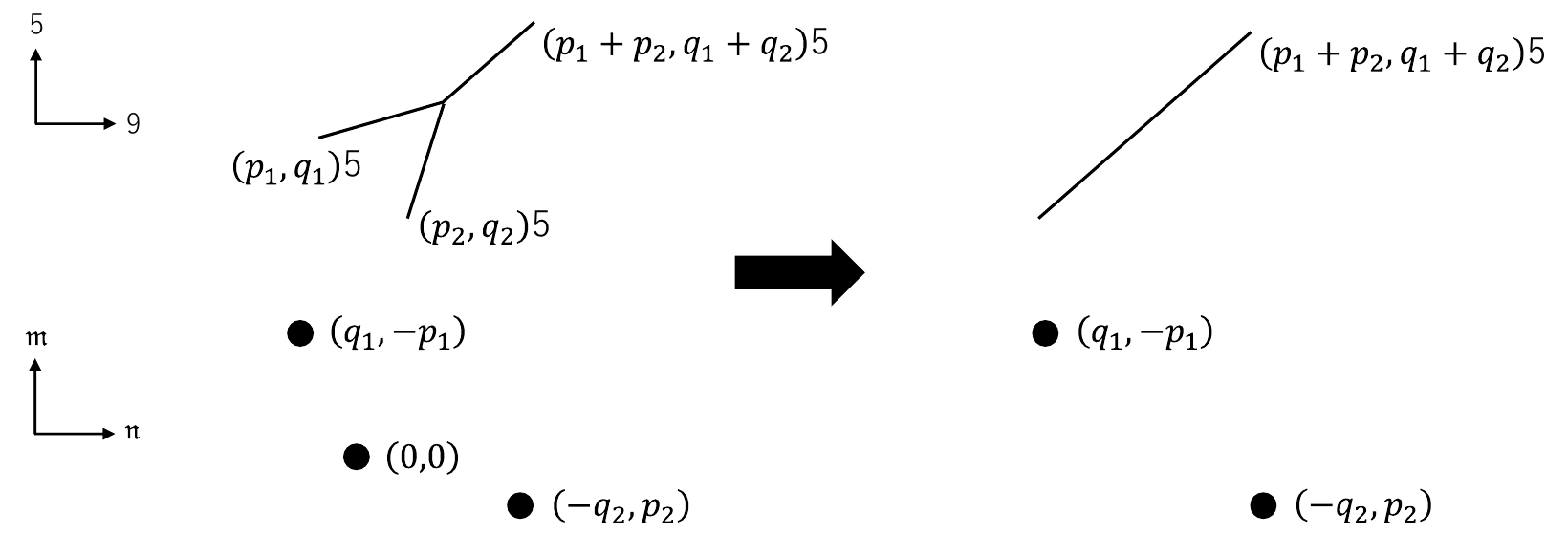}
\par\end{centering}
\caption{A web deformation and it in terms of the dual toric diagrams. The web deformation corresponds to the degeneration of the corresponding quantum curve.\label{fig:WD-QCgen}}
\end{figure}

As we saw, the web deformations are realized as degenerations of the quantum curve. Namely, in terms of the Newton polygon, after a web deformation the corresponding vertex vanishes. We expect that this holds for more general cases. This is consistent with the conjecture in the following sense. Let us consider the web deformation in figure \ref{fig:WD-QCgen}. We start with a $\left(p,q\right)$ web where a $\left(p_{1},q_{1}\right)$5-brane and a $\left(p_{2},q_{2}\right)$5-brane merge into a $\left(p_{1}+p_{2},q_{1}+q_{2}\right)$5-brane. Note that we can form any $\left(p,q\right)$ web by gluing this $\left(p,q\right)$ web, and thus this is not merely a simple case. After the web deformation where the $\left(p_{1},q_{1}\right)$5-brane goes to $x^{5}\rightarrow-\infty$, it becomes the single $\left(p_{1}+p_{2},q_{1}+q_{2}\right)$5-brane. On the other hand, the lower side of figure \ref{fig:WD-QCgen} shows the corresponding toric diagrams. Then, if we assume the conjecture, namely if we assume that the Newton polygons of the corresponding curves are equal to the toric diagrams, we find that the web deformation corresponds to the degeneration of the quantum curve.

\subsection{$\mathrm{SL}\left(2,\mathbb{Z}\right)$ transformations\label{subsec:SL2Ztrans}}

In this section we see that the conjecture is consistent with the $\mathrm{SL}\left(2,\mathbb{Z}\right)$ transformations. It is known that type IIB string theory enjoys the $\mathrm{SL}\left(2,\mathbb{Z}\right)$ transformations. An element $A\in\mathrm{SL}\left(2,\mathbb{Z}\right)$ changes a $\left(p,q\right)$5-brane as
\begin{equation}
\left(\begin{array}{c}
p'\\
q'
\end{array}\right)=A\left(\begin{array}{c}
p\\
q
\end{array}\right).
\end{equation}
Here we also rotate in the 59 plane appropriately so that the angles of all $\left(p,q\right)$5-branes keep $\tan\theta=q/p$. The $\mathrm{SL}\left(2,\mathbb{Z}\right)$ transformations are generated by following two elements
\begin{equation}
S=\left(\begin{array}{cc}
0 & -1\\
1 & 0
\end{array}\right),\quad T=\left(\begin{array}{cc}
1 & 0\\
1 & 1
\end{array}\right).\label{eq:ST}
\end{equation}
Namely, they change a $\left(p,q\right)$5-brane as $S:\left(p,q\right)\rightarrow\left(-q,p\right)$ and $T:\left(p,q\right)\rightarrow\left(p,p+q\right)$. The action $S$ in a $\left(p,q\right)$ web picture is simple. Namely, it is a clockwise rotation by $\pi/2$.

In terms of the gauge theory, the $\mathrm{SL}\left(2,\mathbb{Z}\right)$ transformations are interpreted as dualities. This implies that the $S^{3}$ partition functions related by the $\mathrm{SL}\left(2,\mathbb{Z}\right)$ transformations are equal. On the other hand, our conjecture implies that the quantum curves of dual brane configurations are not equal in general. However, an important property of the Fermi gas expression \eqref{eq:Conjecture1-MM} is that the value of the partition function is invariant under similarity transformations of the quantum curve. We see that the $\mathrm{SL}\left(2,\mathbb{Z}\right)$ transformations are part of similarity transformations, and thus our conjecture is consistent with the $\mathrm{SL}\left(2,\mathbb{Z}\right)$ transformations. Note that the relation between $S\in\mathrm{SL}\left(2,\mathbb{Z}\right)$ and the similarity transformation has been already pointed out in \cite{Drukker:2015awa}.

We check the above claim. Since the $\mathrm{SL}\left(2,\mathbb{Z}\right)$ transformation is generated by $S$ and $T$, it is sufficient to consider these two elements. The action of the elements $S,T$ on the $\left(p,q\right)$ web is interpreted as the following actions on quantum curves via the conjecture
\begin{equation}
S^{\mathrm{QC}}:\left(\begin{array}{c}
\hat{x}\\
\hat{y}
\end{array}\right)\rightarrow\left(\begin{array}{c}
\hat{y}\\
-\hat{x}
\end{array}\right),\quad T^{\mathrm{QC}}:\left(\begin{array}{c}
\hat{x}\\
\hat{y}
\end{array}\right)\rightarrow\left(\begin{array}{c}
\hat{x}\\
-\hat{x}+\hat{y}
\end{array}\right).\label{eq:ST-xy}
\end{equation}
The action $S^{\mathrm{QC}}$ is a clockwise rotation in the $\left(x,y\right)$ plane by $\pi/2$, which is consistent with $S$. The action $T^{\mathrm{QC}}$ also corresponds to $T$. For example, the curve in \eqref{eq:1qD5-QC} corresponding to a $\left(1,q\right)$5-brane is $\hat{\mathcal{O}}_{0,0,0}^{\left(1,q\right)}\left(0\right)=e^{\frac{q}{2}\hat{x}-\frac{1}{2}\hat{y}}+e^{-\frac{q}{2}\hat{x}+\frac{1}{2}\hat{y}}$. (For the simplicity we set the all parameters to be zero.) After a transformation by $T$, the $\left(1,q\right)$5-brane becomes the $\left(1,q+1\right)$5-brane and the corresponding quantum curve is $\hat{\mathcal{O}}_{0,0,0}^{\left(1,q+1\right)}\left(0\right)=e^{\frac{q+1}{2}\hat{x}-\frac{1}{2}\hat{y}}+e^{-\frac{q+1}{2}\hat{x}+\frac{1}{2}\hat{y}}$. This is consistent with $T^{\mathrm{QC}}$. Indeed, these transformations can be realized by the similarity transformations generated by the following operators
\begin{equation}
\hat{S}=e^{\frac{i}{2\hbar}\hat{x}^{2}}e^{\frac{i}{2\hbar}\hat{y}^{2}}e^{\frac{i}{2\hbar}\hat{x}^{2}},\quad\hat{T}=e^{\frac{i}{2\hbar}\hat{x}^{2}},\label{eq:SThat}
\end{equation}
with the similarity transformation $\hat{{\cal O}}\rightarrow\hat{A}\hat{{\cal O}}\hat{A}^{-1}$.

The above realization \eqref{eq:SThat} has an ambiguity which comes from the fact that the conjecture does not fully mention the coefficients of quantum curves (although \eqref{eq:SThat} is consistent with the second property of the quantum curves in the conjecture). For improving this situation, we consider the local $\mathrm{SL}\left(2,\mathbb{Z}\right)$ transformations. It is known that in the brane setup,one can locally apply the $\mathrm{SL}\left(2,\mathbb{Z}\right)$ transformations by introducing duality walls \cite{Gaiotto:2008ak,Gulotta:2011si,Nishioka:2011dq}. Conceptually, one can apply the $\mathrm{SL}\left(2,\mathbb{Z}\right)$ transformations locally by sandwiching a $\left(p,q\right)$ web by the duality wall $W_{A}$ and its inverse $W_{A}^{-1}$ as
\begin{equation}
-\left[A\left(\mathsf{w}\right)\right]-=-W_{A}\left[\mathsf{w}\right]W_{A}^{-1}-.
\end{equation}
Especially, the $S$ duality wall $W_{S}$ is realized as an interpolating $T\left(\mathrm{U}\left(N\right)\right)$ coupling, whose matrix model has been computed in \cite{Benvenuti:2011ga,Gulotta:2011si,Nishioka:2011dq}. In terms of the matrix model, at least for a $\left(p,q\right)$ web including only a single $\left(p,q\right)$5-brane, the effect of the duality wall for $S$ and $T$ has been found to be \cite{Gulotta:2011si,Assel:2014awa}
\begin{align}
\mathcal{Z}^{\left(-q,p\right)}\left(\mu,\nu\right) & =\int\prod_{a}^{N}\frac{d\alpha_{a}}{2\pi}\frac{d\beta_{a}}{2\pi}e^{\frac{i}{\hbar}\sum_{a}^{N}\mu_{a}\alpha_{a}}\mathcal{Z}^{\left(p,q\right)}\left(\alpha,\beta\right)e^{-\frac{i}{\hbar}\sum_{a}^{N}\beta_{a}\nu_{a}},\nonumber \\
\mathcal{Z}^{\left(p,q+p\right)}\left(\mu,\nu\right) & =e^{\frac{i}{2\hbar}\sum_{a}^{N}\mu_{a}^{2}}\mathcal{Z}^{\left(p,q\right)}\left(\mu,\nu\right)e^{-\frac{i}{2\hbar}\sum_{a}^{N}\nu_{a}^{2}}.\label{eq:DualityWall}
\end{align}
The concept of the duality wall, however, is independent of the detail of the $\left(p,q\right)$ web. Hence we expect that this holds for general $\left(p,q\right)$ webs. Let us check that this is consistent with \eqref{eq:SThat}. We assume our conjecture \eqref{eq:MF-QC}, so that
\begin{equation}
\mathcal{Z}^{\left(A\left(\mathsf{w}\right)\right)}\left(\mu,\nu\right)=\det\left(\left[\braket{\mu_{a}|\hat{A}\hat{{\cal O}}^{\left(\mathsf{w}\right)}\left(\hat{x},\hat{y}\right)^{-1}\hat{A}^{-1}|\nu_{b}}\right]_{a,b}^{N\times N}\right).
\end{equation}
For $S$, by inserting $1=\int d\alpha\ket{\alpha}\bra{\alpha}$, we obtain
\begin{align}
 & \mathcal{Z}^{\left(S\left(\mathsf{w}\right)\right)}\left(\mu,\nu\right)\nonumber \\
 & =\int\prod_{a}^{N}d\alpha_{a}d\beta_{a}\prod_{a}^{N}\braket{\mu_{a}|e^{\frac{i}{2\hbar}\hat{x}^{2}}e^{\frac{i}{2\hbar}\hat{y}^{2}}e^{\frac{i}{2\hbar}\hat{x}^{2}}|\alpha_{a}}\mathcal{Z}^{\left(\mathsf{w}\right)}\left(\alpha,\beta\right)\prod_{a}^{N}\braket{\beta_{a}|e^{+\frac{i}{2\hbar}\hat{x}^{2}}e^{+\frac{i}{2\hbar}\hat{y}^{2}}e^{+\frac{i}{2\hbar}\hat{x}^{2}}|\nu_{a}}\nonumber \\
 & =\int\prod_{a}^{N}\frac{d\alpha_{a}}{2\pi}\frac{d\beta_{a}}{2\pi}e^{\frac{i}{\hbar}\sum_{a}^{N}\mu_{a}\alpha_{a}}\mathcal{Z}^{\left(\mathsf{w}\right)}\left(\alpha,\beta\right)e^{-\frac{i}{\hbar}\sum_{a}^{N}\beta_{a}\nu_{a}},
\end{align}
where we used \eqref{eq:VecSim} and \eqref{eq:Normalization}. This is the same with \eqref{eq:DualityWall}. For $T$, it is easy to see that
\begin{equation}
\mathcal{Z}^{\left(T\left(\mathsf{w}\right)\right)}\left(\mu,\nu\right)=e^{\frac{i}{2\hbar}\sum_{a}^{N}\mu_{a}^{2}}\mathcal{Z}^{\left(\mathsf{w}\right)}\left(\mu,\nu\right)e^{-\frac{i}{2\hbar}\sum_{a}^{N}\nu_{a}^{2}}.
\end{equation}
This is again the same with \eqref{eq:DualityWall}.

\begin{figure}[t]
\begin{centering}
\includegraphics[scale=0.6]{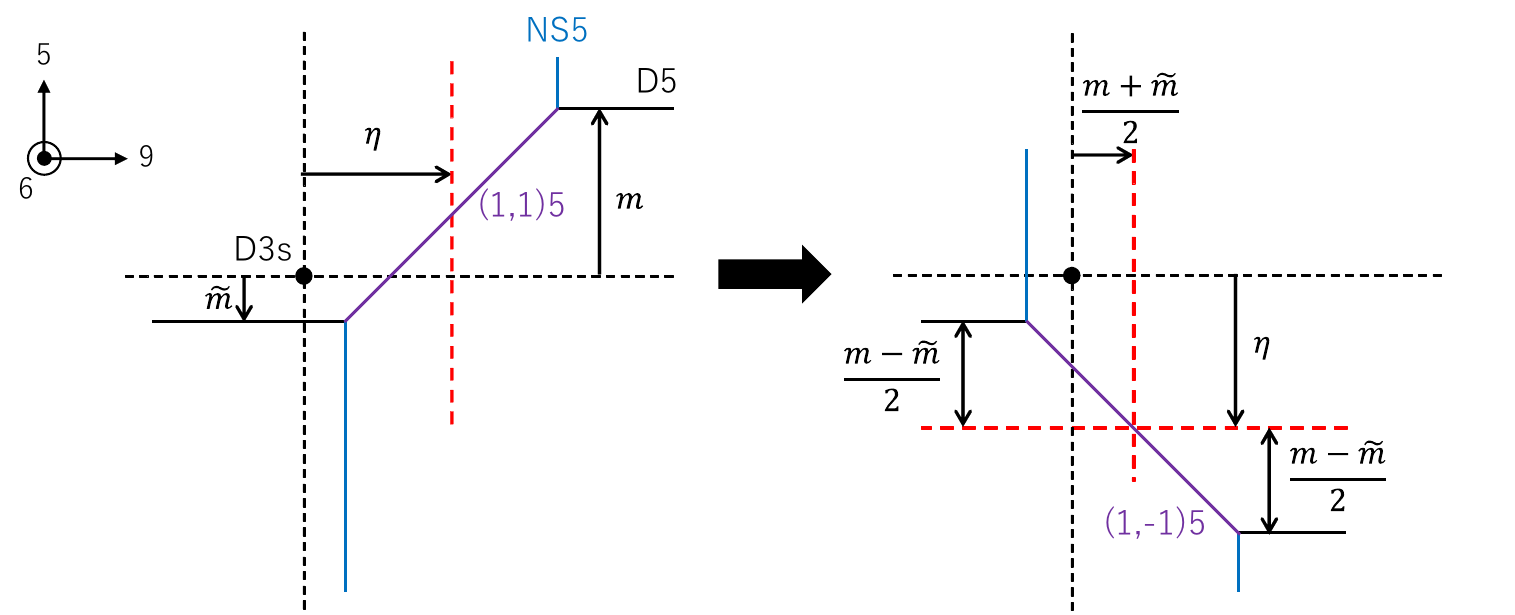}
\par\end{centering}
\caption{The relation of the deformation parameters under the $S$ transformation \eqref{eq:S-Trans}. In the brane picture, the $\left(p,q\right)$ web is rotated clockwise in the 59 plane. The dotted red lines are the middle of the NS5-branes or the D5-branes.\label{fig:BW-SL2Z}}
\end{figure}

Finally, we see the local $S$ transformation for the $\left(p,q\right)$ web $\left[\left(1,0\right)+\left(0,1\right)\right]$. This is self-dual as\footnote{The last relation shows the invariance of the matrix model, which has been discussed in \cite{Assel:2014awa}.}
\begin{equation}
S:\left[\left(1,0\right)+\left(0,1\right)\right]\rightarrow\left[\left(-1,0\right)+\left(0,1\right)\right]\sim\left[\left(1,0\right)+\left(0,1\right)\right].\label{eq:S-Trans}
\end{equation}
We check this transformation in the quantum curve and the brane picture and especially focus on the parameter correspondence. The quantum curve for $\left[\left(1,0\right)+\left(0,1\right)\right]$ is
\begin{equation}
\hat{\mathcal{O}}_{1,0,0}^{\left(1,0\right)}\left(\eta,m,\tilde{m};\hat{x},\hat{y}\right)=\left(\begin{array}{cc}
+e^{\pi\eta-\pi\tilde{m}}e^{\frac{1}{2}\hat{x}-\frac{1}{2}\hat{y}} & +e^{-\pi\eta-\pi m}e^{\frac{1}{2}\hat{x}+\frac{1}{2}\hat{y}}\\
+e^{\pi\eta+\pi\tilde{m}}e^{-\frac{1}{2}\hat{x}-\frac{1}{2}\hat{y}} & +e^{-\pi\eta+\pi m}e^{-\frac{1}{2}\hat{x}+\frac{1}{2}\hat{y}}
\end{array}\right).
\end{equation}
Here we set $D$ and $\tilde{D}$ to be zero since they do not have a brane interpretation. After the similarity transformation by $\hat{S}$, one obtains
\begin{equation}
\hat{S}\hat{\mathcal{O}}_{1,0,0}^{\left(1,0\right)}\left(\eta,m,\tilde{m};\hat{x},\hat{y}\right)\hat{S}^{-1}=\left(\begin{array}{cc}
+e^{\pi\eta+\pi\tilde{m}}e^{\frac{1}{2}\hat{x}-\frac{1}{2}\hat{y}} & +e^{\pi\eta-\pi\tilde{m}}e^{\frac{1}{2}\hat{x}+\frac{1}{2}\hat{y}}\\
+e^{-\pi\eta+\pi m}e^{-\frac{1}{2}\hat{x}-\frac{1}{2}\hat{y}} & +e^{-\pi\eta-\pi m}e^{-\frac{1}{2}\hat{x}+\frac{1}{2}\hat{y}}
\end{array}\right).
\end{equation}
This is indeed the quantum curve for $\left[\left(1,0\right)+\left(0,1\right)\right]$. The relation of the parameters between the original system $\left(\eta,m,\tilde{m}\right)$ and the dual system $\left(\eta',m',\tilde{m}'\right)$ can be read off (from $\hat{S}\hat{\mathcal{O}}_{1,0,0}^{\left(1,0\right)}\left(\eta,m,\tilde{m}\right)\hat{S}^{-1}=\hat{\mathcal{O}}_{1,0,0}^{\left(1,0\right)}\left(\eta',m',\tilde{m}'\right)$) as
\begin{equation}
\eta'=\frac{1}{2}\left(m+\tilde{m}\right),\quad m'=-\eta-\frac{1}{2}\left(m-\tilde{m}\right),\quad\tilde{m}'=-\eta+\frac{1}{2}\left(m-\tilde{m}\right).
\end{equation}
This is consistent with the brane picture, see figure \ref{fig:BW-SL2Z}.

Notice that the above argument shows the systematic way of the proof of the equality of the $S^{3}$ partition functions between the dual theories. Namely, once the Fermi gas expressions with the quantum curves \eqref{eq:Conjecture1-MM} are given for dual theories related by $A\in\mathrm{SL}\left(2,\mathbb{Z}\right)$, the quantum curves must be related by the similarity transformation by $\hat{A}$. Because $S$ and $T$ are the generators of the $\mathrm{SL}\left(2,\mathbb{Z}\right)$, $\hat{A}$ is also generated by $\hat{S}$ and $\hat{T}$ in \eqref{eq:SThat}.

\section{Matrix factors for $\left(p,q\right)$ webs with $\left(p,q\right)$5-brane\label{sec:pqMF-QC}}

In section \ref{subsec:Conjecture} we gave a conjecture which relates $\left(p,q\right)$ webs to quantum curves. In this section, we show that the conjecture gives matrix factors for the $\left(p,q\right)$ webs in figure \ref{fig:BW-gen} with $p\geq2$.

\subsection{Suggestion of matrix factors\label{subsec:pqMF}}

As discussed in section \ref{subsec:3dTh}, if a brane configuration includes a $\left(p,q\right)$ web containing a $\left(p,q\right)$5-brane with $p\geq2$, the Lagrangian of an associated 3d theory is not known. Nevertheless, it was conjectured that the matrix factor for a $\left(p,q\right)$5-brane is \cite{Gulotta:2011si,Assel:2014awa}
\begin{align}
\mathcal{Z}^{\left(p,q\right)}\left(\eta;\mu,\nu\right) & =\frac{1}{\left(2\pi p\right)^{N}}e^{\frac{iq}{4\pi p}\sum_{a}^{N}\left(\mu_{a}^{2}-\nu_{a}^{2}\right)}e^{i\eta\sum_{a}^{N}\left(\mu_{a}-\nu_{a}\right)}\frac{\prod_{a<b}^{N}2\sinh\frac{\mu_{a}-\mu_{b}}{2p}\prod_{a<b}^{N}2\sinh\frac{\nu_{a}-\nu_{b}}{2p}}{\prod_{a,b}^{N}2\cosh\frac{\mu_{a}-\nu_{b}}{2p}}.\label{eq:pq-MF}
\end{align}
Here $\eta$ corresponds to the position of the $\left(p,q\right)$5-brane in the direction 9, see figure \ref{fig:Web-MassFI}. We will discuss the normalization of $\eta$ soon later. In this section we study a matrix factor $\mathcal{Z}_{F,F_{\wdr},F_{\wdl}}^{\left(p,q\right)}$ corresponding to
\begin{equation}
\mathsf{w}_{F,F_{\wdr},F_{\wdl}}^{\left(p,q\right)}=\left[\left(p,q\right)+F\df+F_{\wdr}\dfr+F_{\wdl}\dfl\right].\label{eq:pqWeb-Gen}
\end{equation}

At first sight, the D5-brane factors $\mathcal{Z}_{\df}$ and $\mathcal{Z}_{\mathrm{D5}^{\pm}}$ defined in \eqref{eq:D5wd-MF} seem to sufficient for constructing $\mathcal{Z}_{F,F_{\wdr},F_{\wdl}}^{\left(p,q\right)}$. However, these D5-brane factors do not work. A simple reason is that the CS factor (the Fresnel factor) is now divided by $p$. Since a web deformation where a D5-brane goes far away changes a $\left(p,q\right)$5-brane into a $\left(p,q+1\right)$5-brane (like the one discussed in section \ref{subsec:WebDefEx}), it changes the coefficient of the CS factor $iq/\left(4\pi p\right)$ into $i\left(q+1\right)/\left(4\pi p\right)$. Namely, after the web deformation the coefficient of the CS factor is shifted by not $i/\left(4\pi\right)$ but $i/\left(4\pi p\right)$. Thus we need to modify the D5-brane factors.

As a first step for obtaining the correct D5-brane factors, we focus on our conjecture, which claims that the operator appearing in \eqref{eq:MF-QC} must be (the inverse of) a quantum curve. Remember that the property of the double sine function \eqref{eq:DS-Trig} played a crucial role in obtaining the curve form in the Fermi gas formalism as one can see in \eqref{eq:DS-sim}. Thus we assume that the D5-brane factors for general $\left(p,q\right)$5-branes can be also written in terms of the double sine functions.

Next, we consider the arguments of the double sine functions. As we mentioned above, the integration variables $\mu/\left(2\pi\right)$ in \eqref{eq:D5wd-MF} do not reproduce the correct CS factors under the web deformations. A simple modification for solving this discrepancy is to rescale the integration variables as $\mu\rightarrow\mu/\sqrt{p}$. This, however, causes an unwanted rescaling of the coefficient of $\hat{x}$ in the quantum curve as $\exp\left(c\hat{x}\right)\rightarrow\exp\left(c\hat{x}/\sqrt{p}\right)$. We can solve this problem by replacing the subscription of the double sine function from $1$ to $\sqrt{p}$ (see \eqref{eq:DS-Trig}). Notice that the asymptotic behavior of the double sine function \eqref{eq:DSasym} does not depend on the subscription (up to a trivial phase), and thus the web deformation does not receive any corrections.

In the D5-brane factors $\mathcal{Z}_{\df}$ and $\mathcal{Z}_{\mathrm{D5}^{\pm}}$, the parameters $m$ and $D$ also enter in the arguments. We assume that these parameters also enter into the arguments of the D5-brane factors for general $\left(p,q\right)$5-branes. Remark that $m$ denotes the position of an associated D5-brane in the direction 5. On the other hand, a physical interpretation of $D$ for the $p\geq2$ is unclear since the Lagrangian is not known. Here we use $D\in\mathbb{R}$ as an imaginary part of a holomorphic function $z=m+iD$. (We use $\tilde{D}\in\mathbb{R}$ in the same way as $\tilde{z}=\tilde{m}+i\tilde{D}$.) The coefficient of $z$ (or $\tilde{z}$) has not been determined yet, which we denote by $c_{m}$. In the D5-brane factors $\mathcal{Z}_{\df}$ and $\mathcal{Z}_{\mathrm{D5}^{\pm}}$, there are additional imaginary numbers $\pm i/4$. We introduce a corresponding constant $c_{i}$ ($\in\mathbb{R}$). Now, we arrive at the following form of the D5-brane factor
\begin{equation}
\frac{\prod_{a}^{N}s_{\sqrt{p}}\left(\frac{\mu_{a}}{2\pi\sqrt{p}}-c_{m}\tilde{z}+\frac{c_{i}}{4}i\right)}{\prod_{a}^{N}s_{\sqrt{p}}\left(\frac{\mu_{a}}{2\pi\sqrt{p}}-c_{m}z-\frac{c_{i}}{4}i\right)}\frac{\prod_{a}^{N}s_{\sqrt{p}}\left(\frac{\nu_{a}}{2\pi\sqrt{p}}-c_{m}z+\frac{c_{i}}{4}i\right)}{\prod_{a}^{N}s_{\sqrt{p}}\left(\frac{\nu_{a}}{2\pi\sqrt{p}}-c_{m}\tilde{z}-\frac{c_{i}}{4}i\right)},\label{eq:D5wdp-MF0}
\end{equation}
where
\begin{equation}
z=m+iD,\quad\tilde{z}=\tilde{m}+i\tilde{D}.
\end{equation}
Our remaining task is to determine $c_{m}$ and $c_{i}$.

$c_{i}$ can be determined from the conjecture. Namely, we again demand that (the inverse of) the operator becomes a quantum curve. Let us consider a computation for obtaining the quantum curve corresponding to \eqref{eq:DS-sim}. Here we have to care about the terms $\exp\left(\pm\hat{y}/2\right)$. In the current case this is modified. By using \eqref{eq:1Loop-Det} and \eqref{eq:Cosh-op}, we can rewrite the 1-loop determinant part of ${\cal Z}^{\left(p,q\right)}$ in \eqref{eq:pq-MF} as
\begin{equation}
\frac{\prod_{a<b}^{N}2\sinh\frac{\mu_{a}-\mu_{b}}{2p}\prod_{a<b}^{N}2\sinh\frac{\nu_{a}-\nu_{b}}{2p}}{\prod_{a,b}^{N}2\cosh\frac{\mu_{a}-\nu_{b}}{2p}}=\left(2\pi p\right)^{N}\det\left(\left[\braket{\mu_{a}|\frac{1}{2\cosh\frac{p\hat{y}}{2}}|\nu_{b}}\right]_{a,b}^{N\times N}\right).\label{eq:pq-Fermi1}
\end{equation}
This means that we have to replace them with $\exp\left(\pm p\hat{y}/2\right)$. After the similarity transformation by $\exp\left(\pm p\hat{y}/4\right)$, one obtains
\begin{equation}
\frac{s_{\sqrt{p}}\left(\frac{\hat{x}}{2\pi\sqrt{p}}-c_{m}z-\frac{c_{i}}{4}i\mp\frac{\sqrt{p}}{4}i\right)}{s_{\sqrt{p}}\left(\frac{\hat{x}}{2\pi\sqrt{p}}-c_{m}z+\frac{c_{i}}{4}i\pm\frac{\sqrt{p}}{4}i\right)},\quad\frac{s_{\sqrt{p}}\left(\frac{\hat{x}}{2\pi\sqrt{p}}-c_{m}\tilde{z}-\frac{c_{i}}{4}i\pm\frac{\sqrt{p}}{4}i\right)}{s_{\sqrt{p}}\left(\frac{\hat{x}}{2\pi\sqrt{p}}-c_{m}\tilde{z}+\frac{c_{i}}{4}i\mp\frac{\sqrt{p}}{4}i\right)}.
\end{equation}
According to the formula \eqref{eq:DS-Trig}, this expression implies
\begin{equation}
c_{i}=\sqrt{p}.
\end{equation}

We can obtain $c_{m}$ by considering a constant shift of the integration variables as we did in section \ref{subsec:BC-MM} for a $\left(p,q\right)$ web including a $\left(1,q\right)$5-brane. As we discussed, the constant shift of the integration variables $\mu\rightarrow\mu+2\pi c$, $\nu\rightarrow\nu+2\pi c$ corresponds to moving the D3-branes in the $+x^{5}$ direction by $c$. On the other hand, the shift affects $m$ in \eqref{eq:D5wdp-MF0} as $m\rightarrow m-c/\left(\sqrt{p}c_{m}\right)$ (and the same with $\tilde{m}$). Thus we find that
\begin{equation}
c_{m}=\frac{1}{\sqrt{p}}.
\end{equation}
Note that we can also check the normalization of $\eta$ in \eqref{eq:pq-MF} by using the above constant shift. The shift for $\eta$ comes from the Chern-Simons term in \eqref{eq:pq-MF}, which is $\eta\rightarrow\eta+cq/p$. This is consistent with the angle of the $\left(p,q\right)$5-brane, $\tan\theta=q/p$.

Therefore, we expect that the D5-brane factors for general $\left(p,q\right)$5-branes are
\begin{align}
\mathcal{Z}_{\df}^{p}\left(z,\tilde{z};\mu,\nu\right) & =\frac{\prod_{a}^{N}s_{\sqrt{p}}\left(\frac{\mu_{a}}{2\pi\sqrt{p}}-\frac{1}{\sqrt{p}}\tilde{z}+\frac{\sqrt{p}}{4}i\right)}{\prod_{a}^{N}s_{\sqrt{p}}\left(\frac{\mu_{a}}{2\pi\sqrt{p}}-\frac{1}{\sqrt{p}}z-\frac{\sqrt{p}}{4}i\right)}\frac{\prod_{a}^{N}s_{\sqrt{p}}\left(\frac{\nu_{a}}{2\pi\sqrt{p}}-\frac{1}{\sqrt{p}}z+\frac{\sqrt{p}}{4}i\right)}{\prod_{a}^{N}s_{\sqrt{p}}\left(\frac{\nu_{a}}{2\pi\sqrt{p}}-\frac{1}{\sqrt{p}}\tilde{z}-\frac{\sqrt{p}}{4}i\right)},\nonumber \\
\mathcal{Z}_{\dfr}^{p}\left(z;\mu,\nu\right) & =e^{\frac{i}{8\pi p}\sum_{a}\left(\mu_{a}^{2}-\nu_{a}^{2}\right)}\frac{\prod_{a}^{N}s_{\sqrt{p}}\left(\frac{\nu_{a}}{2\pi\sqrt{p}}-\frac{1}{\sqrt{p}}z+\frac{\sqrt{p}}{4}i\right)}{\prod_{a}^{N}s_{\sqrt{p}}\left(\frac{\mu_{a}}{2\pi\sqrt{p}}-\frac{1}{\sqrt{p}}z-\frac{\sqrt{p}}{4}i\right)},\nonumber \\
\mathcal{Z}_{\dfl}^{p}\left(\tilde{z};\mu,\nu\right) & =e^{\frac{i}{8\pi p}\sum_{a}\left(\mu_{a}^{2}-\nu_{a}^{2}\right)}\frac{\prod_{a}^{N}s_{\sqrt{p}}\left(\frac{\mu_{a}}{2\pi\sqrt{p}}-\frac{1}{\sqrt{p}}\tilde{z}+\frac{\sqrt{p}}{4}i\right)}{\prod_{a}^{N}s_{\sqrt{p}}\left(\frac{\nu_{a}}{2\pi\sqrt{p}}-\frac{1}{\sqrt{p}}\tilde{z}-\frac{\sqrt{p}}{4}i\right)}.\label{eq:D5wdp-MF}
\end{align}
We can obtain the matrix factor corresponding to the $\left(p,q\right)$ web \eqref{eq:pqWeb-Gen} by multiplying the matrix factor for a $\left(p,q\right)$5-brane by these factors as
\begin{align}
 & \mathcal{Z}_{F,F_{\wdr},F_{\wdl}}^{\left(p,q\right)}\left(\eta,\mathbf{z};\mu,\nu\right)\nonumber \\
 & =\begin{cases}
\mathcal{Z}^{\left(p,q\right)}\left(\eta;\mu,\nu\right)\prod_{f=1}^{F}\mathcal{Z}_{\df}^{p}\left(z_{f},\tilde{z}_{f};\mu,\nu\right)\prod_{f=1}^{F_{\wdr}}\mathcal{Z}_{\dfr}^{p}\left(z_{F+f};\mu,\nu\right) & \left(F_{\wdl}=0\right)\\
\mathcal{Z}^{\left(p,q\right)}\left(\eta;\mu,\nu\right)\prod_{f=1}^{F}\mathcal{Z}_{\df}^{p}\left(z_{f},\tilde{z}_{f};\mu,\nu\right)\prod_{f=1}^{F_{\wdl}}\mathcal{Z}_{\dfl}^{p}\left(\tilde{z}_{F+f};\mu,\nu\right) & \left(F_{\wdr}=0\right)
\end{cases},\label{eq:pqD5-MF}
\end{align}
where
\begin{align}
 & z_{f}=m_{f}+iD_{f},\quad\tilde{z}_{f}=\tilde{m}_{f}+i\tilde{D}_{f},\nonumber \\
 & \mathbf{z}=\begin{cases}
\left(z_{1},\tilde{z}_{1},z_{2},\tilde{z}_{2},\ldots,z_{F},\tilde{z}_{F}|z_{F+1},z_{F+2},\ldots,z_{F+F_{\wdr}}\right) & \left(F_{\wdl}=0\right)\\
\left(z_{1},\tilde{z}_{1},z_{2},\tilde{z}_{2},\ldots,z_{F},\tilde{z}_{F}|\tilde{z}_{F+1},\tilde{z}_{F+2},\ldots,\tilde{z}_{F+F_{\wdl}}\right) & \left(F_{\wdr}=0\right)
\end{cases}.\label{eq:pqD5-MFpara}
\end{align}
We remark that $m_{f}$ and $\tilde{m}_{f}$ denote the position of the corresponding D5-branes in the direction 5 as well as \eqref{eq:1qD5-MFpara}.

Now we see that this expectation is consistent with the web deformation. We have already demanded for determining the form of the D5-brane factors that in the large mass limit the corresponding D5-brane factor provides the appropriate CS factor. Therefore, our remaining task is to check whether the shift of the FI parameter in the matrix factor is consistent with the brane picture as we did for the $p=1$ case. Let us consider the web deformation in figure \ref{fig:WebDef-1q} (i), (ii) and (iii). According to this brane picture, we need to shift the FI parameter by $m/\left(2p\right)$ or $-\tilde{m}/\left(2p\right)$. Let us check this in the matrix factor. We take the large $m$ limit for the D5-brane factor \eqref{eq:D5wdp-MF}. In this step, we obtain
\begin{align}
 & e^{i\eta\left(\mu-\nu\right)}\frac{s_{\sqrt{p}}\left(\frac{\nu}{2\pi\sqrt{p}}-\frac{1}{\sqrt{p}}\left(m+iD\right)+\frac{\sqrt{p}}{4}i\right)}{s_{\sqrt{p}}\left(\frac{\mu}{2\pi\sqrt{p}}-\frac{1}{\sqrt{p}}\left(m+iD\right)-\frac{\sqrt{p}}{4}i\right)}\nonumber \\
 & \rightarrow\exp\left[\frac{i}{8\pi p}\left(\mu^{2}-\nu^{2}\right)+i\left(\eta-\frac{m}{2p}-\frac{i}{2p}D\right)\left(\mu-\nu\right)+\frac{1}{8}\left(\mu+\nu\right)-\frac{\pi}{2}\Lambda-\frac{i\pi}{2}D\right].
\end{align}
Therefore, we have to scale $\eta$ by $m/\left(2p\right)$ as expected.

\subsection{Quantum curves\label{subsec:pqQC}}

In this section we explicitly check that the proposed matrix factor $\mathcal{Z}_{F,F_{\wdr},F_{\wdl}}^{\left(p,q\right)}$ gives the quantum curve which is expected from the conjecture suggested in section \ref{subsec:Conjecture2}. The computation follows the one in section \ref{subsec:BW-QC-FGF}.

We start with the matrix factor for the $\left(p,q\right)$5-brane defined in \eqref{eq:pq-MF} as a warm-up. By using formulas \eqref{eq:1Loop-Det}, \eqref{eq:Cosh-op} (as \eqref{eq:pq-Fermi1}) and \eqref{eq:OpSim}, we obtain the form of \eqref{eq:MF-DM} with the density matrix
\begin{equation}
\hat{\rho}^{\left(p,q\right)}\left(\eta;\hat{x},\hat{y}\right)=\frac{1}{2\cosh\frac{p\hat{y}-q\hat{x}-2\pi p\eta}{2}}.\label{eq:1q-DM-1}
\end{equation}
Therefore, the inverse of this operator is
\begin{equation}
\hat{\mathcal{O}}^{\left(p,q\right)}\left(\eta;\hat{x},\hat{y}\right)=e^{\pi p\eta}e^{\frac{q}{2}\hat{x}-\frac{p}{2}\hat{y}}+e^{-\pi p\eta}e^{-\frac{q}{2}\hat{x}+\frac{p}{2}\hat{y}}.
\end{equation}
This is clearly the form of the quantum curve, and the Newton polygon of this curve is equal to $\lp\left(\left[\left(p,q\right)\right]\right)$.

Second, we consider the matrix factor $\mathcal{Z}_{F,F_{\wdr},F_{\wdl}}^{\left(p,q\right)}$. After a short computation which is the same as the one performed for the $p=1$ case, we find that the density matrix is
\begin{align}
\hat{\rho}_{F,F_{\wdr},F_{\wdl}}^{\left(p,q\right)}\left(\eta,\mathbf{z};\hat{x},\hat{y}\right) & =\frac{\prod_{f=1}^{F+F_{\wdl}}s_{\sqrt{p}}\left(\frac{\hat{x}}{2\pi\sqrt{p}}-\frac{1}{\sqrt{p}}\tilde{z}_{f}+\frac{\sqrt{p}}{4}i\right)}{\prod_{f=1}^{F+F_{\wdr}}s_{\sqrt{p}}\left(\frac{\hat{x}}{2\pi\sqrt{p}}-\frac{1}{\sqrt{p}}z_{f}-\frac{\sqrt{p}}{4}i\right)}\nonumber \\
 & \quad\times\frac{1}{2\cosh\frac{p\hat{y}-\left(q+\frac{1}{2}F_{\wdr}+\frac{1}{2}F_{\wdl}\right)\hat{x}-2\pi p\eta}{2}}\frac{\prod_{f=1}^{F+F_{\wdr}}s_{\sqrt{p}}\left(\frac{\hat{x}}{2\pi\sqrt{p}}-\frac{1}{\sqrt{p}}z_{f}+\frac{\sqrt{p}}{4}i\right)}{\prod_{f=1}^{F+F_{\wdl}}s_{\sqrt{p}}\left(\frac{\hat{x}}{2\pi\sqrt{p}}-\frac{1}{\sqrt{p}}\tilde{z}_{f}-\frac{\sqrt{p}}{4}i\right)},\label{eq:pqD5-DM}
\end{align}
where $\mathbf{z}$ is defined in \eqref{eq:pqD5-MFpara}. By using \eqref{eq:OpSim} and \eqref{eq:DS-Trig}, we obtain an identity
\begin{align}
 & \frac{s_{\sqrt{p}}\left(\frac{\hat{x}}{2\pi\sqrt{p}}-\frac{1}{\sqrt{p}}\tilde{z}-\frac{\sqrt{p}}{4}i\right)}{s_{\sqrt{p}}\left(\frac{\hat{x}}{2\pi\sqrt{p}}-\frac{1}{\sqrt{p}}z+\frac{\sqrt{p}}{4}i\right)}\left(e^{\frac{p}{2}\hat{y}}+e^{-\frac{p}{2}\hat{y}}\right)\frac{s_{\sqrt{p}}\left(\frac{\hat{x}}{2\pi\sqrt{p}}-\frac{1}{\sqrt{p}}z-\frac{\sqrt{p}}{4}i\right)}{s_{\sqrt{p}}\left(\frac{\hat{x}}{2\pi\sqrt{p}}-\frac{1}{\sqrt{p}}\tilde{z}+\frac{\sqrt{p}}{4}i\right)}\nonumber \\
 & =e^{\frac{p}{4}\hat{y}}2\cosh\left(\frac{\hat{x}}{2}-\pi z\right)e^{\frac{p}{4}\hat{y}}+e^{-\frac{p}{4}\hat{y}}2\cosh\left(\frac{\hat{x}}{2}-\pi\tilde{z}\right)e^{-\frac{p}{4}\hat{y}}.
\end{align}
By using this identity, we find that the inverse of the density matrix becomes
\begin{align}
\hat{\mathcal{O}}_{F,F_{\wdr},F_{\wdl}}^{\left(p,q\right)}\left(\eta,\mathbf{z};\hat{x},\hat{y}\right) & =e^{-\pi p\eta}e^{\frac{p}{4}\hat{y}}\left\{ e^{-\frac{1}{2}\left(q+\frac{1}{2}F_{\wdr}+\frac{1}{2}F_{\wdl}\right)\hat{x}}\prod_{f=1}^{F+F_{\wdr}}\left(e^{\frac{1}{2}\hat{x}-\pi z_{f}}+e^{-\frac{1}{2}\hat{x}+\pi z_{f}}\right)\right\} e^{\frac{p}{4}\hat{y}}\nonumber \\
 & \quad+e^{\pi p\eta}e^{-\frac{p}{4}\hat{y}}\left\{ e^{\frac{1}{2}\left(q+\frac{1}{2}F_{\wdr}+\frac{1}{2}F_{\wdl}\right)\hat{x}}\prod_{f=1}^{F+F_{\wdl}}\left(e^{\frac{1}{2}\hat{x}-\pi\tilde{z}_{f}}+e^{-\frac{1}{2}\hat{x}+\pi\tilde{z}_{f}}\right)\right\} e^{-\frac{p}{4}\hat{y}}.\label{eq:pqD5-QC}
\end{align}
This is clearly the form of the quantum curve, and the Newton polygon of this curve is equal to $\lp\left(\mathsf{w}_{F,F_{\wdr},F_{\wdl}}^{\left(p,q\right)}\right)$.

Finally, let us consider the asymptotic behavior of the classical curve $\mathcal{O}_{F,F_{\wdr},F_{\wdl}}^{\left(p,q\right)}$ We identify the $\left(x,y\right)$ plane with the 59 plane. We have four directions where the external legs extend. Namely, for example when $F_{\wdr}=0$, there are a $\left(1,q+F_{\wdl}\right)$5-brane extending in the $+x$ direction, a $\left(1,q\right)$5-brane extending in the $-x$ direction, $F$ D5-branes extending in the $+y$ direction and $F+F_{\wdl}$ D5-branes extending in the $-y$ direction (see right side of figure \ref{fig:BW-gen}). For the $+x$ direction, we introduce new variables $\left(u,v\right)$ by $x=pu$, $y=\left(q+F_{\wdl}\right)u+v$. Then, in the limit $u\rightarrow\infty$ only two terms remain
\begin{equation}
\mathcal{O}_{F,0,F_{\wdl}}^{\left(p,q\right)}\left(\eta,\mathbf{z};u,v\right)\rightarrow\left(e^{\pi p\eta-\pi\sum_{f=1}^{F+F_{\wdl}}\tilde{z}_{f}}e^{-\frac{1}{2}pv}+e^{-\pi p\eta-\pi\sum_{f=1}^{F}z_{f}}e^{\frac{1}{2}pv}\right)e^{\left(\frac{1}{2}F+\frac{1}{4}F_{\wdl}\right)pu}.
\end{equation}
Hence the real part of the solution of $\lim_{u\rightarrow\infty}\mathcal{O}_{F,0,F_{\wdl}}^{\left(p,q\right)}=0$ is 
\begin{equation}
\left(2\pi\right)^{-1}\Re\left(v\right)=\eta+\frac{1}{2p}\sum_{f=1}^{F}m_{f}-\frac{1}{2p}\sum_{f=1}^{F+F_{\wdl}}\tilde{m}_{f}.
\end{equation}
This value actually matches the position of the $\left(p,q+F_{\wdl}\right)$5-brane. The effect of $\eta$ agrees on this interpretation by definition. The effects from $m_{f}$, $\tilde{m}_{f}$ also correspond because the move of a D5-brane extending in the $\pm x^{9}$ direction by $m$ (or $\tilde{m}$) moves the $\left(p,q+F_{\wdl}\right)$5-brane by $m/\left(2p\right)$ (or $-\tilde{m}/\left(2p\right)$) (see figure \ref{fig:WebDef-1q}). For the $-y$ direction, in the limit $y\rightarrow-\infty$ the remaining terms are
\begin{equation}
\mathcal{O}_{F,0,F_{\wdl}}^{\left(p,q\right)}\left(\eta,\mathbf{z};x,y\right)\rightarrow e^{\pi p\eta}\left\{ e^{\frac{1}{2}\left(q+\frac{1}{2}F_{\wdr}+\frac{1}{2}F_{\wdl}\right)x}\prod_{f=1}^{F+F_{\wdl}}\left(e^{\frac{1}{2}x-\pi\tilde{z}_{f}}+e^{-\frac{1}{2}x+\pi\tilde{z}_{f}}\right)\right\} e^{-\frac{1}{2}py}.
\end{equation}
Hence the real part of the solutions of $\lim_{y\rightarrow-\infty}\mathcal{O}_{F,0,F_{\wdl}}^{\left(p,q\right)}=0$ are 
\begin{equation}
\left(2\pi\right)^{-1}\Re\left(x\right)=\tilde{m}_{f},\quad\left(f=1,2,\ldots,F+F_{\wdl}\right).
\end{equation}
Since we have $F+F_{\wdl}$ D5-branes in the $-x^{9}$ direction and their positions are $\tilde{m}_{f}$, this is again an expected result. The other directions or the $F_{\wdl}=0$ case can be checked in a similar way.

\section{Quantum curves of general $\hbar$\label{sec:QC-genh}}

So far we considered the case when $\hbar=2\pi$. However, from our setup we can obtain more general $\hbar$, and in this section we show this. One motivation is that the quantum curves with arbitrary $\hbar$ are related to the topological strings and the integrable systems, where the parameter $\hbar$ plays an important role. In these theories, the genus one cases have been especially well-studied. Another motivation is that $\mathcal{N}=4$ supersymmetric CS theories corresponding to genus one curves have already been studied. A simple but important example is the ABJM theory. In this case, $\hbar=2\pi k$ where $k$ is the CS level. In section \ref{subsec:Genus1} we study the genus one curves, and in section \ref{subsec:p1D-ABJM} we study the relation between our setup and the ABJM theory through the quantum curve.

\subsection{Genus one curves\label{subsec:Genus1}}

Our conjecture suggests brane configurations for arbitrary quantum curves with a toric diagram. However, it is non-trivial whether one can obtain the matrix models for those brane configurations. If the worldvolume theory of the brane configuration is a Lagrangian theory, one can obtain the matrix model representation systematically by using the supersymmetric localization. We also conjectured the matrix models for more general $\left(p,q\right)$ webs in section \ref{sec:pqMF-QC}. In this section we suggest brane configurations and matrix models for the genus one quantum curves. For this purpose the conjectured matrix models play a crucial role.

We especially focus on the genus one curves which are related to the $q$-Painlev\'e equations. In the context of the integrable systems, the $q$-Painlev\'e equations are classified by their symmetries, which are the affine Weyl groups of $E_{n}$ \cite{2001CMaPh.220..165S}. Correspondingly, the asymptotic behaviors of the genus one classical curves are classified by them. These symmetries are uplifted to quantum curves without any modifications (up to the affine direction), where the symmetries are defined with similarity transformations \cite{Kubo:2018cqw,Moriyama:2020lyk}. We hence label both of the classical and quantum genus one curves by $E_{n}$.\footnote{The symmetry of a quantum curve plays an important role for connecting the $E_{5}$ curve with the $q$-Painlev\'e $\mathrm{VI}$ \cite{Bonelli:2022dse}. On the other hand, the overall powers of a curve played an important role in the correspondence between the mass deformed ABJM theory and the $\mathrm{SU}\left(N\right)$ $q$-Toda equations \cite{Nosaka:2020tyv}. The overall powers can be shifted in a way commented in footnote \ref{fn:CurveShift}.}
\begin{figure}[t]
\begin{centering}
\includegraphics[scale=0.3]{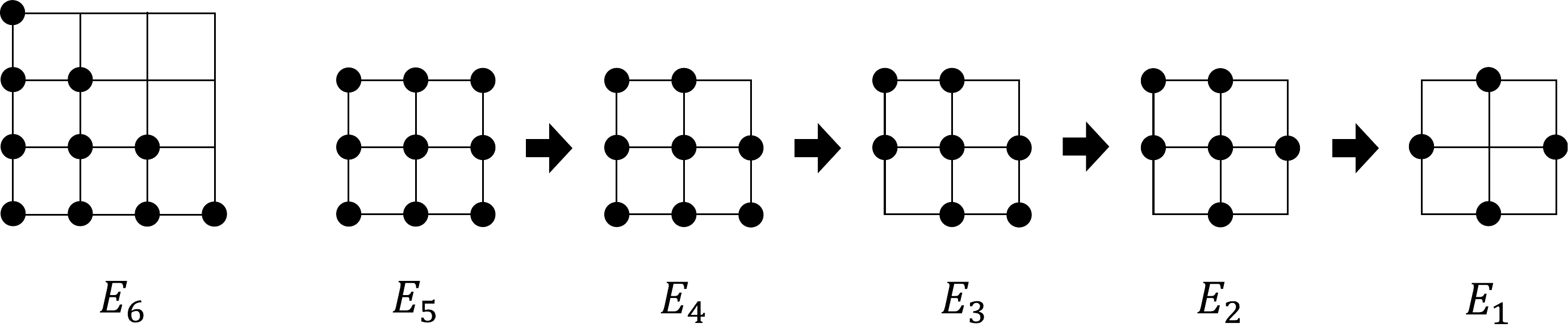}
\par\end{centering}
\caption{Genus one curves in terms of the Newton polygons (or equivalently the toric diagrams). They are classified by their symmetries. The arrows represent coalescences.\label{fig:G1Polygon}}
\end{figure}
Figure \ref{fig:G1Polygon} shows these curves in terms of the Newton polygons.

In section \ref{subsec:Genus1-Gen} we try to generalize $\hbar=2\pi$ to $\hbar=2\pi\ell$ with positive integer $\ell$. The meaning of $\hbar=c$ is that quantum curves are written by $\hat{x}$ and $\hat{y}$ whose commutation relation is $\left[\hat{x},\hat{y}\right]=ci$. The quantum curves appearing in our conjecture always have $\ell=1$, which we write $\left(\hat{x},\hat{y}\right)$ as before. On the other hand, in section \ref{subsec:Genus1-Gen} we obtain the quantum curves whose Newton polygons are depicted in figure \ref{fig:G1Polygon} and which are written by $c=\ell$ operators $\left(\hat{X},\hat{Y}\right)$. In this section $\left(\hat{x},\hat{y}\right)$ and $\left(\hat{X},\hat{Y}\right)$ always satisfy $\left[\hat{x},\hat{y}\right]=2\pi i$ and $\left[\hat{X},\hat{Y}\right]=2\pi\ell i$.

\subsubsection{$\hbar=2\pi$ case\label{subsec:Genus1-k1}}

Although our main interest is the $\hbar>2\pi$ case, in this section we start with the $\hbar=2\pi$ case. Because now we can suggest brane configurations from the Newton polygons by using the conjecture, for each Newton polygon we start with a brane configuration, and then we show a corresponding matrix model. We also give a quantum curve associated with the matrix model and check the Newton polygon of the curve. Note that there is an arbitrariness in the choice of the brane configuration. We chose one which has the maximum number of parameters under the condition that it has a matrix model representation.

In this section the following quantum curves appear (see \eqref{eq:1qD5-QC})
\begin{align}
\hat{\mathcal{O}}_{1,0,0}^{\left(1,0\right)}\left(\eta,z,\tilde{z};\hat{x},\hat{y}\right) & =\left(\begin{array}{cc}
+e^{\pi\eta-\pi\tilde{z}}e^{\frac{1}{2}\hat{x}-\frac{1}{2}\hat{y}} & +e^{-\pi\eta-\pi z}e^{\frac{1}{2}\hat{x}+\frac{1}{2}\hat{y}}\\
+e^{\pi\eta+\pi\tilde{z}}e^{-\frac{1}{2}\hat{x}-\frac{1}{2}\hat{y}} & +e^{-\pi\eta+\pi z}e^{-\frac{1}{2}\hat{x}+\frac{1}{2}\hat{y}}
\end{array}\right),\nonumber \\
\hat{\mathcal{O}}_{0,0,1}^{\left(1,0\right)}\left(\eta,\tilde{z};\hat{x},\hat{y}\right) & =\left(\begin{array}{cc}
+e^{\pi\eta-\pi\tilde{z}}e^{\frac{3}{4}\hat{x}-\frac{1}{2}\hat{y}} & +0\\
+e^{\pi\eta+\pi\tilde{z}}e^{-\frac{1}{4}\hat{x}-\frac{1}{2}\hat{y}} & +e^{-\pi\eta}e^{-\frac{1}{4}\hat{x}+\frac{1}{2}\hat{y}}
\end{array}\right),\nonumber \\
\hat{\mathcal{O}}_{0,0,1}^{\left(1,-1\right)}\left(\eta,\tilde{z};\hat{x},\hat{y}\right) & =\left(\begin{array}{cc}
+e^{\pi\eta-\pi\tilde{z}}e^{\frac{1}{4}\hat{x}-\frac{1}{2}\hat{y}} & +e^{-\pi\eta}e^{\frac{1}{4}\hat{x}+\frac{1}{2}\hat{y}}\\
+e^{\pi\eta+\pi\tilde{z}}e^{-\frac{3}{4}\hat{x}-\frac{1}{2}\hat{y}} & +0
\end{array}\right),\nonumber \\
\hat{\mathcal{O}}_{0,0,0}^{\left(1,1\right)}\left(\eta;\hat{x},\hat{y}\right) & =\left(\begin{array}{cc}
+e^{\pi\eta}e^{\frac{1}{2}\hat{x}-\frac{1}{2}\hat{y}} & +0\\
+0 & +e^{-\pi\eta}e^{-\frac{1}{2}\hat{x}+\frac{1}{2}\hat{y}}
\end{array}\right),\nonumber \\
\hat{\mathcal{O}}_{0,0,0}^{\left(1,-1\right)}\left(\eta;\hat{x},\hat{y}\right) & =\left(\begin{array}{cc}
+0 & +e^{-\pi\eta}e^{\frac{1}{2}\hat{x}+\frac{1}{2}\hat{y}}\\
+e^{\pi\eta}e^{-\frac{1}{2}\hat{x}-\frac{1}{2}\hat{y}} & +0
\end{array}\right).
\end{align}
We also write the toric diagrams corresponding to $\hat{\mathcal{O}}_{F,F_{\wdr},F_{\wdl}}^{\left(p,q\right)}$ as $\lp_{F,F_{\wdr},F_{\wdl}}^{\left(p,q\right)}=\lp\left(\mathsf{w}_{F,F_{\wdr},F_{\wdl}}^{\left(p,q\right)}\right)$. One can see that the Newton polygon of each curve expressed above is indeed the corresponding $\lp_{F,F_{\wdr},F_{\wdl}}^{\left(p,q\right)}$.

Note that when we consider the integrable systems, the overall factor of the curve is also important. Especially, for relating the curves to the $q$-Painlev\'e systems, we would need to use $e^{-\frac{1}{8}\hat{x}}\hat{\mathcal{O}}_{0,0,1}^{\left(1,q\right)}e^{-\frac{1}{8}\hat{x}}$ rather than $\hat{\mathcal{O}}_{0,0,1}^{\left(1,q\right)}$ itself. Interestingly, the former curve appears in the web deformation, see \eqref{eq:QC-NDtoND11}. Hereafter we do not care about this point.

We start studying each curve. The $E_{6}$ toric diagram depicted in figure \ref{fig:G1Polygon} can be regarded as a sum of three $\lp_{0,0,1}^{\left(1,0\right)}$. Namely, the $E_{6}$ toric diagram is realized by the following brane configuration with three $\left(p,q\right)$ webs
\begin{equation}
\left[\left(1,0\right)+\dfl\right]-\left[\left(1,0\right)+\dfl\right]-\left[\left(1,0\right)+\dfl\right]-_{\mathrm{p}}.
\end{equation}
The matrix model of this brane configuration is
\begin{align}
 & Z^{\left(E_{6}\right)}\left(\eta^{\left(r\right)},\tilde{z}^{\left(r\right)}\right)\nonumber \\
 & =\frac{1}{\left(N!\right)^{3}}\int\prod_{a}^{N}\frac{d\mu_{a}}{2\pi}\frac{d\nu_{a}}{2\pi}\frac{d\rho_{a}}{2\pi}{\cal Z}_{0,0,1}^{\left(1,0\right)}\left(\eta^{\left(1\right)},\tilde{z}^{\left(1\right)};\mu,\nu\right){\cal Z}_{0,0,1}^{\left(1,0\right)}\left(\eta^{\left(2\right)},\tilde{z}^{\left(2\right)};\nu,\rho\right){\cal Z}_{0,0,1}^{\left(1,0\right)}\left(\eta^{\left(3\right)},\tilde{z}^{\left(3\right)};\rho,\mu\right).
\end{align}
The associated quantum curve is
\begin{equation}
\hat{\mathcal{O}}^{\left(E_{6}\right)}\left(\eta^{\left(r\right)},\tilde{z}^{\left(r\right)};\hat{x},\hat{y}\right)=\hat{\mathcal{O}}_{0,0,1}^{\left(1,0\right)}\left(\eta^{\left(3\right)},\tilde{z}^{\left(3\right)};\hat{x},\hat{y}\right)\hat{\mathcal{O}}_{0,0,1}^{\left(1,0\right)}\left(\eta^{\left(2\right)},\tilde{z}^{\left(2\right)};\hat{x},\hat{y}\right)\hat{\mathcal{O}}_{0,0,1}^{\left(1,0\right)}\left(\eta^{\left(1\right)},\tilde{z}^{\left(1\right)};\hat{x},\hat{y}\right).\label{eq:E6QC}
\end{equation}
One can easily check that the Newton polygon of $\hat{\mathcal{O}}^{\left(E_{6}\right)}$ is the $E_{6}$ one.

The $E_{5}$ toric diagram can be regarded as a sum of two $\lp_{1,0,0}^{\left(1,0\right)}$. Namely, the $E_{5}$ toric diagram is realized by the following brane configuration with two $\left(p,q\right)$ webs
\begin{equation}
\left[\left(1,0\right)+\df\right]-\left[\left(1,0\right)+\df\right]-_{\mathrm{p}}.
\end{equation}
The matrix model of this brane configuration is
\begin{align}
Z^{\left(E_{5}\right)}\left(\eta^{\left(r\right)},\mathbf{z}^{\left(r\right)}\right) & =\frac{1}{\left(N!\right)^{2}}\int\prod_{a}^{N}\frac{d\mu_{a}}{2\pi}\frac{d\nu_{a}}{2\pi}{\cal Z}_{1,0,0}^{\left(1,0\right)}\left(\eta^{\left(1\right)},z^{\left(1\right)},\tilde{z}^{\left(1\right)};\mu,\nu\right){\cal Z}_{1,0,0}^{\left(1,0\right)}\left(\eta^{\left(2\right)},z^{\left(2\right)},\tilde{z}^{\left(2\right)};\nu,\mu\right).
\end{align}
The associated quantum curve is
\begin{equation}
\hat{\mathcal{O}}^{\left(E_{5}\right)}\left(\eta^{\left(r\right)},\mathbf{z}^{\left(r\right)};\hat{x},\hat{y}\right)=\hat{\mathcal{O}}_{1,0,0}^{\left(1,0\right)}\left(\eta^{\left(2\right)},z^{\left(2\right)},\tilde{z}^{\left(2\right)};\hat{x},\hat{y}\right)\hat{\mathcal{O}}_{1,0,0}^{\left(1,0\right)}\left(\eta^{\left(1\right)},z^{\left(1\right)},\tilde{z}^{\left(1\right)};\hat{x},\hat{y}\right).\label{eq:E5QC}
\end{equation}
One can easily check that the Newton polygon of $\hat{\mathcal{O}}^{\left(E_{5}\right)}$ is the $E_{5}$ one.

The $E_{4}$ toric diagram can be regarded as a sum of $\lp_{1,0,0}^{\left(1,0\right)}$ and $\lp_{0,0,1}^{\left(1,0\right)}$. Namely, the $E_{4}$ toric diagram is realized by the following brane configuration with two $\left(p,q\right)$ webs
\begin{equation}
\left[\left(1,0\right)+\df\right]-\left[\left(1,0\right)+\dfl\right]-_{\mathrm{p}}.
\end{equation}
The matrix model of this brane configuration is
\begin{align}
Z^{\left(E_{4}\right)}\left(\eta^{\left(r\right)},\mathbf{z}^{\left(r\right)}\right) & =\frac{1}{\left(N!\right)^{2}}\int\prod_{a}^{N}\frac{d\mu_{a}}{2\pi}\frac{d\nu_{a}}{2\pi}{\cal Z}_{1,0,0}^{\left(1,0\right)}\left(\eta^{\left(1\right)},z^{\left(1\right)},\tilde{z}^{\left(1\right)};\mu,\nu\right){\cal Z}_{0,0,1}^{\left(1,0\right)}\left(\eta^{\left(2\right)},\tilde{z}^{\left(2\right)};\nu,\mu\right).
\end{align}
The associated quantum curve is
\begin{equation}
\hat{\mathcal{O}}^{\left(E_{4}\right)}\left(\eta^{\left(r\right)},\mathbf{z}^{\left(r\right)};\hat{x},\hat{y}\right)=\hat{\mathcal{O}}_{0,0,1}^{\left(1,0\right)}\left(\eta^{\left(2\right)},\tilde{z}^{\left(2\right)};\hat{x},\hat{y}\right)\hat{\mathcal{O}}_{1,0,0}^{\left(1,0\right)}\left(\eta^{\left(1\right)},z^{\left(1\right)},\tilde{z}^{\left(1\right)};\hat{x},\hat{y}\right).\label{eq:E4QC}
\end{equation}
One can easily check that the Newton polygon of $\hat{\mathcal{O}}^{\left(E_{4}\right)}$ is the $E_{4}$ one.

The $E_{3}$ toric diagram can be regarded as a sum of $\lp_{1,0,0}^{\left(1,0\right)}$ and $\lp_{0,0,0}^{\left(1,1\right)}$. Namely, the $E_{3}$ toric diagram is realized by the following brane configuration with two $\left(p,q\right)$ webs
\begin{equation}
\left[\left(1,0\right)+\df\right]-\left[\left(1,1\right)\right]-_{\mathrm{p}}.
\end{equation}
The matrix model of this brane configuration is
\begin{align}
Z^{\left(E_{3}\right)}\left(\eta^{\left(r\right)},\mathbf{z}^{\left(1\right)}\right) & =\frac{1}{\left(N!\right)^{2}}\int\prod_{a}^{N}\frac{d\mu_{a}}{2\pi}\frac{d\nu_{a}}{2\pi}{\cal Z}_{1,0,0}^{\left(1,0\right)}\left(\eta^{\left(1\right)},z^{\left(1\right)},\tilde{z}^{\left(1\right)};\mu,\nu\right){\cal Z}_{0,0,0}^{\left(1,1\right)}\left(\eta^{\left(2\right)};\nu,\mu\right).
\end{align}
The associated quantum curve is
\begin{equation}
\hat{\mathcal{O}}^{\left(E_{3}\right)}\left(\eta^{\left(r\right)},\mathbf{z}^{\left(1\right)};\hat{x},\hat{y}\right)=\hat{\mathcal{O}}_{0,0,0}^{\left(1,1\right)}\left(\eta^{\left(2\right)};\hat{x},\hat{y}\right)\hat{\mathcal{O}}_{1,0,0}^{\left(1,0\right)}\left(\eta^{\left(1\right)},z^{\left(1\right)},\tilde{z}^{\left(1\right)};\hat{x},\hat{y}\right).\label{eq:E3QC}
\end{equation}
One can easily check that the Newton polygon of $\hat{\mathcal{O}}^{\left(E_{3}\right)}$ is the $E_{3}$ one.

The $E_{2}$ toric diagram can be regarded as a sum of $\lp_{0,1,0}^{\left(1,-1\right)}$ and $\lp_{0,0,0}^{\left(1,1\right)}$. Namely, the $E_{2}$ toric diagram is realized by the following brane configuration with two $\left(p,q\right)$ webs
\begin{equation}
\left[\left(1,-1\right)+\dfl\right]-\left[\left(1,1\right)\right]-_{\mathrm{p}}.
\end{equation}
The matrix model of this brane configuration is
\begin{align}
Z^{\left(E_{2}\right)}\left(\eta^{\left(r\right)},\tilde{z}^{\left(1\right)}\right) & =\frac{1}{\left(N!\right)^{2}}\int\prod_{a}^{N}\frac{d\mu_{a}}{2\pi}\frac{d\nu_{a}}{2\pi}{\cal Z}_{0,0,1}^{\left(1,-1\right)}\left(\eta^{\left(1\right)},\tilde{z}^{\left(1\right)};\mu,\nu\right){\cal Z}_{0,0,0}^{\left(1,1\right)}\left(\eta^{\left(2\right)};\nu,\mu\right).
\end{align}
The associated quantum curve is
\begin{equation}
\hat{\mathcal{O}}^{\left(E_{2}\right)}\left(\eta^{\left(r\right)},\tilde{z}^{\left(1\right)};\hat{x},\hat{y}\right)=\hat{\mathcal{O}}_{0,0,0}^{\left(1,1\right)}\left(\eta^{\left(2\right)};\hat{x},\hat{y}\right)\hat{\mathcal{O}}_{0,0,1}^{\left(1,-1\right)}\left(\eta^{\left(1\right)},\tilde{z}^{\left(1\right)};\hat{x},\hat{y}\right).\label{eq:E2QC}
\end{equation}
One can easily check that the Newton polygon of $\hat{\mathcal{O}}^{\left(E_{2}\right)}$ is the $E_{2}$ one.

The $E_{1}$ toric diagram can be regarded as a sum of $\lp_{0,0,0}^{\left(1,-1\right)}$ and $\lp_{0,0,0}^{\left(1,1\right)}$. Namely, the $E_{1}$ toric diagram is realized by the following brane configuration with two $\left(p,q\right)$ webs
\begin{equation}
\left[\left(1,-1\right)\right]-\left[\left(1,1\right)\right]-_{\mathrm{p}}.
\end{equation}
The matrix model of this brane configuration is
\begin{align}
Z^{\left(E_{1}\right)}\left(\eta^{\left(r\right)}\right) & =\frac{1}{\left(N!\right)^{2}}\int\prod_{a}^{N}\frac{d\mu_{a}}{2\pi}\frac{d\nu_{a}}{2\pi}{\cal Z}_{0,0,0}^{\left(1,-1\right)}\left(\eta^{\left(1\right)};\mu,\nu\right){\cal Z}_{0,0,0}^{\left(1,1\right)}\left(\eta^{\left(2\right)};\nu,\mu\right).
\end{align}
The associated quantum curve is
\begin{equation}
\hat{\mathcal{O}}^{\left(E_{1}\right)}\left(\eta^{\left(r\right)};\hat{x},\hat{y}\right)=\hat{\mathcal{O}}_{0,0,0}^{\left(1,1\right)}\left(\eta^{\left(2\right)};\hat{x},\hat{y}\right)\hat{\mathcal{O}}_{0,0,0}^{\left(1,-1\right)}\left(\eta^{\left(1\right)};\hat{x},\hat{y}\right).\label{eq:E1QC}
\end{equation}
One can easily check that the Newton polygon of $\hat{\mathcal{O}}^{\left(E_{1}\right)}$ is the $E_{1}$ one.

Before going to the $\hbar>2\pi$ case, let us give some comments. First, we comment on the number of parameters. It is known that the $E_{n}$ curve possesses $n$ parameters. This means that the asymptotic behavior of the $E_{n}$ (classical and quantum) curve is controlled by $n$ parameters up to the shift of the coordinates. Let us see the number of the parameters of the quantum curves we obtained. For example, although it looks that the $E_{6}$ quantum curve \eqref{eq:E6QC} has six parameters $\eta^{\left(r\right)},z^{\left(r\right)}$ ($r=1,2,3$), we have two degrees of freedom to shift $\hat{x}$ and $\hat{y}$ without changing the commutation relation, and thus the $E_{6}$ quantum curve effectively have only four parameters. What are the remaining two parameters? In the brane configuration, we have three $\left(p,q\right)$ webs. Correspondingly, we can turn on two rank deformations. (For example, the rank deformation in the ABJM theory becomes a parameter of the ABJM quantum curve \cite{Kashaev:2015wia}.) Hence we expect that these two rank deformations are the remaining parameters. Another possibility of introducing two additional parameters would be to combine the three $\left(p,q\right)$ webs and consider a single $\left(p,q\right)$ web which realizes the toric diagram of $E_{6}$. In that case, the NS5-branes can break by ending on the D5-branes, and thus they can have two additional FI parameters. In this case, however, either a Lagrangian or a matrix model is not known.
\begin{table}[t]
\begin{centering}
\begin{tabular}{|c|c|c|c|c|c|c|c|}
\hline 
 & $E_{6}$ & $E_{5}$ & $E_{4}$ & $E_{3}$ & $E_{2}$ & $E_{1}$ & $E_{0}$\tabularnewline
\hline 
\hline 
Number of parameters of the $E_{n}$ curve & $6$ & $5$ & $4$ & $3$ & $2$ & $1$ & $0$\tabularnewline
\hline 
Number of parameters of $\hat{\mathcal{O}}^{\left(E_{n}\right)}$ & $4$ & $4$ & $3$ & $2$ & $1$ & $0$ & $0$\tabularnewline
\hline 
Number of $\left(p,q\right)$ webs in our setup & $3$ & $2$ & $2$ & $2$ & $2$ & $2$ & $1$\tabularnewline
\hline 
\end{tabular}
\par\end{centering}
\caption{The number of the parameters of the $E_{n}$ curves, the number of parameters of the quantum curves we obtained and the number of $\left(p,q\right)$ webs in our setup.\label{tab:Paras}}
\end{table}
The similar story holds for the other curves, see table \ref{tab:Paras}. Actually, one can find that for each $E_{n}$ the sum of the number of the parameters of $\hat{\mathcal{O}}^{\left(E_{n}\right)}$ and $j_{n}-1$ is equal to $n$, where $j_{n}$ is the number of the $\left(p,q\right)$ webs in our setup. Because $j$ $\left(p,q\right)$ webs provide $j-1$ parameters and the number of parameters of the $E_{n}$ curve is $n$, this is the expected result. Note that the $E_{1}$ curve can be realized in a different way, and in that case the number of the parameters becomes one. See section \ref{subsec:p1D-ABJM}.

The brane configurations of the $E_{n}$ ($4\geq n\geq1$) types are obtained from the $E_{5}$ one by the web deformations. This can be checked in the direct computation as we performed in section \eqref{subsec:WebDef}. Alternatively, one can easily see in figure \ref{fig:G1Polygon} that from $E_{5}$ to $E_{1}$ each curve receives one degeneration. On the other hand, in the context of the $q$-Painlev\'e equations, the arrows in figure \ref{fig:G1Polygon} are known to be the coalescence. Especially, the coalescence has been studied in terms of the matrix models and the quantum curves \cite{Bonelli:2022dse}. Therefore, our setup gives a physical interpretation to the coalescence as the web deformations.

Note that there are also $q$-Painlev\'e equations possessing the $E_{1}'$ and $E_{0}$ symmetries. The corresponding Newton polygons are in figure \ref{fig:BW-LP-E1}, and we discussed the corresponding $\left(p,q\right)$ webs and quantum curves in the figure and \eqref{eq:E1E0}. We also discuss the $E_{0}$ curve in the next section.

\subsubsection{$\hbar>2\pi$ case\label{subsec:Genus1-Gen}}

In the previous section we studied the genus one curves with $\hbar=2\pi$. In this section we try to generalize $\hbar$ to $\hbar=2\pi\ell$ with positive integer $\ell$.

The technique to obtain such a general $\hbar$ is to ``stretch'' the Newton polygons in the horizontal direction. Then, even if the corresponding curves are written by $\left(\hat{x},\text{\ensuremath{\hat{y}}}\right)$ with $\left[\hat{x},\text{\ensuremath{\hat{y}}}\right]=2\pi i$, by ``compressing'' the Newton polygons we can obtain the $E_{n}$ curves depicted in figure \ref{fig:G1Polygon}. More concretely, if we change the variables as $\left(\hat{X},\hat{Y}\right)\sim\left(\hat{x},\ell\text{\ensuremath{\hat{y}}}\right)$, the Newton polygons in terms of $\left(\hat{X},\hat{Y}\right)$ becomes the $E_{n}$ ones and the commutation relation becomes $\left[\hat{X},\hat{Y}\right]=2\pi\ell i$. Here the detail of the change of the variables depends on $E_{n}$.

However, if we stretch the to the Newton polygons in figure \ref{fig:G1Polygon}, additional vertices always appear. In terms of the $\left(p,q\right)$ web, this corresponds to appearance of $\left(p,q\right)$5-branes where $\mathrm{gcd}\left(p,q\right)>1$. Because all the $\left(p,q\right)$5-branes must satisfy $\mathrm{gcd}\left(p,q\right)=1$, this situation is unacceptable. To avoid that, we first perform the $\mathrm{SL}\left(2,\mathbb{Z}\right)$ transformations appropriately. Explicitly, we perform $T$ for the $E_{n}$ ($6\geq n\geq3$), $T^{2}$ for the $E_{n}$ ($2\geq n\geq1$) and $S^{-1}T^{-1}S$ for the $E_{0}$. We then perform the stretching in the horizontal direction by $\ell$.
\begin{figure}[t]
\begin{centering}
\includegraphics[scale=0.25]{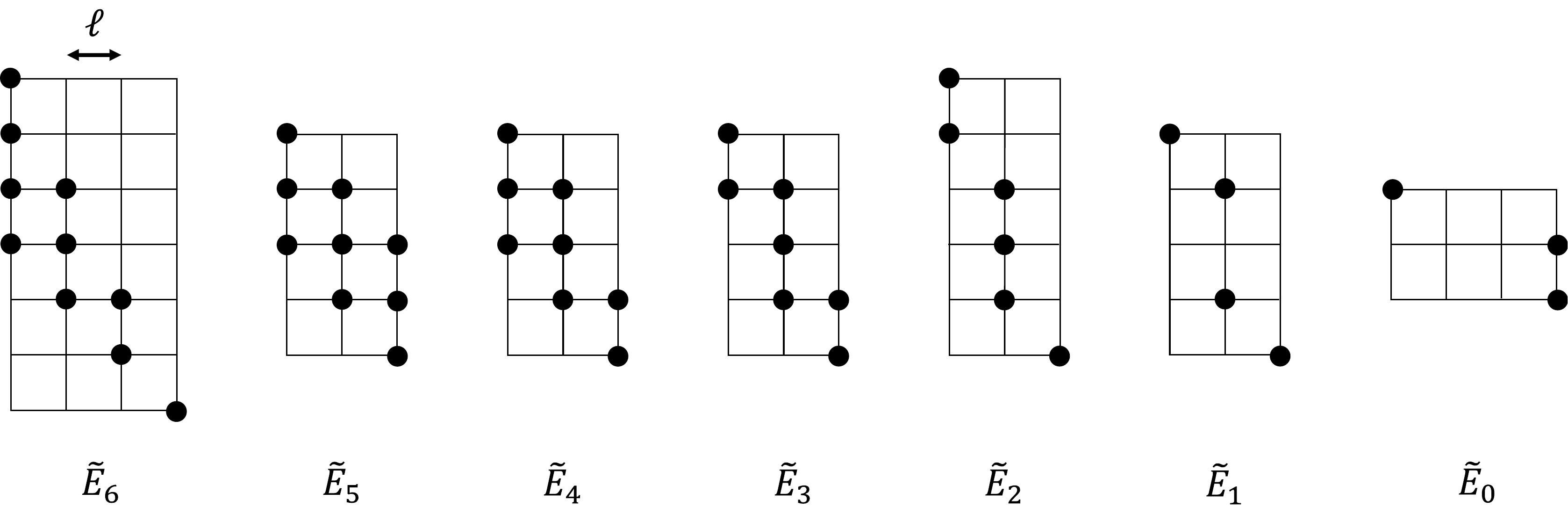}
\par\end{centering}
\caption{Genus one curves in terms of the Newton polygons (or equivalently the toric diagrams). The unit length of the horizontal direction is $\ell$.\label{fig:G1Polygon2}}
\end{figure}
Figure \ref{fig:G1Polygon2} shows the resulting toric diagrams, which we label by $\tilde{E}_{n}$. Notice that even after this procedure, additional vertices sometimes appear. To avoid this, we need to restrict $\ell$ where the condition depends on $\tilde{E}_{n}$.

We start with the $\tilde{E}_{6}$ curve. The $\tilde{E}_{6}$ toric diagram depicted in figure \ref{fig:G1Polygon2} can be regarded as a sum of three $\lp_{0,0,1}^{\left(\ell,1\right)}$ with $\mathrm{gcd}\left(\ell,2\right)=1$. Namely, the $\tilde{E}_{6}$ toric diagram is realized by the following brane configuration with three $\left(p,q\right)$ webs
\begin{equation}
\left[\left(\ell,1\right)+\dfl\right]-\left[\left(\ell,1\right)+\dfl\right]-\left[\left(\ell,1\right)+\dfl\right]-_{\mathrm{p}}.
\end{equation}
The matrix model of this brane configuration is
\begin{align}
 & Z^{\left(\tilde{E}_{6}\right)}\left(\eta^{\left(r\right)},\tilde{z}^{\left(r\right)}\right)\nonumber \\
 & =\frac{1}{\left(N!\right)^{3}}\int\prod_{a}^{N}\frac{d\mu_{a}}{2\pi}\frac{d\nu_{a}}{2\pi}\frac{d\rho_{a}}{2\pi}{\cal Z}_{0,0,1}^{\left(\ell,1\right)}\left(\eta^{\left(1\right)},\tilde{z}^{\left(1\right)};\mu,\nu\right){\cal Z}_{0,0,1}^{\left(\ell,1\right)}\left(\eta^{\left(2\right)},\tilde{z}^{\left(2\right)};\nu,\rho\right){\cal Z}_{0,0,1}^{\left(\ell,1\right)}\left(\eta^{\left(3\right)},\tilde{z}^{\left(3\right)};\rho,\mu\right).
\end{align}
The associated quantum curve is
\begin{align}
\hat{\mathcal{O}}^{\left(\tilde{E}_{6}\right)}\left(\eta^{\left(r\right)},\tilde{z}^{\left(r\right)};\hat{x},\hat{y}\right) & =\hat{\mathcal{O}}_{0,0,1}^{\left(\ell,1\right)}\left(\eta^{\left(3\right)},\tilde{z}^{\left(3\right)};\hat{x},\hat{y}\right)\hat{\mathcal{O}}_{0,0,1}^{\left(\ell,1\right)}\left(\eta^{\left(2\right)},\tilde{z}^{\left(2\right)};\hat{x},\hat{y}\right)\hat{\mathcal{O}}_{0,0,1}^{\left(\ell,1\right)}\left(\eta^{\left(1\right)},\tilde{z}^{\left(1\right)};\hat{x},\hat{y}\right)\nonumber \\
 & =\hat{\mathcal{O}}^{\left(E_{6}\right)}\left(\eta^{\left(r\right)},\tilde{z}^{\left(r\right)};\hat{X},\hat{Y}\right),\label{eq:E6QC2}
\end{align}
where $\hat{\mathcal{O}}^{\left(E_{6}\right)}$ is \eqref{eq:E6QC} and
\begin{equation}
\hat{X}=\hat{x},\quad\hat{Y}=-\hat{x}+\ell\hat{y}.
\end{equation}
The quantum curve $\hat{\mathcal{O}}_{F,F_{\wdl},F_{\wdr}}^{\left(p,q\right)}$ can be read off from \eqref{eq:pqD5-QC}.

The $\tilde{E}_{5}$ toric diagram can be regarded as a sum of two $\lp_{1,0,0}^{\left(\ell,1\right)}$ with arbitrary positive integer $\ell$. Namely, the $\tilde{E}_{5}$ toric diagram is realized by the following brane configuration with two $\left(p,q\right)$ webs
\begin{equation}
\left[\left(\ell,1\right)+\df\right]-\left[\left(\ell,1\right)+\df\right]-_{\mathrm{p}}.
\end{equation}
The matrix model of this brane configuration is
\begin{align}
Z^{\left(\tilde{E}_{5}\right)}\left(\eta^{\left(r\right)},\mathbf{z}^{\left(r\right)}\right) & =\frac{1}{\left(N!\right)^{2}}\int\prod_{a}^{N}\frac{d\mu_{a}}{2\pi}\frac{d\nu_{a}}{2\pi}{\cal Z}_{1,0,0}^{\left(\ell,1\right)}\left(\eta^{\left(1\right)},z^{\left(1\right)},\tilde{z}^{\left(1\right)};\mu,\nu\right){\cal Z}_{1,0,0}^{\left(\ell,1\right)}\left(\eta^{\left(2\right)},z^{\left(2\right)},\tilde{z}^{\left(2\right)};\nu,\mu\right).
\end{align}
The associated quantum curve is
\begin{align}
\hat{\mathcal{O}}^{\left(\tilde{E}_{5}\right)}\left(\eta^{\left(r\right)},\mathbf{z}^{\left(r\right)};\hat{x},\hat{y}\right) & =\hat{\mathcal{O}}_{1,0,0}^{\left(\ell,1\right)}\left(\eta^{\left(2\right)},z^{\left(2\right)},\tilde{z}^{\left(2\right)};\hat{x},\hat{y}\right)\hat{\mathcal{O}}_{1,0,0}^{\left(\ell,1\right)}\left(\eta^{\left(1\right)},z^{\left(1\right)},\tilde{z}^{\left(1\right)};\hat{x},\hat{y}\right)\nonumber \\
 & =\hat{\mathcal{O}}^{\left(E_{5}\right)}\left(\eta^{\left(r\right)},\mathbf{z}^{\left(r\right)};\hat{X},\hat{Y}\right),\label{eq:E5QC2}
\end{align}
where $\hat{\mathcal{O}}^{\left(E_{5}\right)}$ is \eqref{eq:E5QC} and
\begin{equation}
\hat{X}=\hat{x},\quad\hat{Y}=-\hat{x}+\ell\hat{y}.
\end{equation}

The $\tilde{E}_{4}$ toric diagram can be regarded as a sum of $\lp_{1,0,0}^{\left(\ell,1\right)}$ and $\lp_{0,0,1}^{\left(\ell,1\right)}$ with $\mathrm{gcd}\left(\ell,2\right)=1$. Namely, the $\tilde{E}_{4}$ toric diagram is realized by the following brane configuration with two $\left(p,q\right)$ webs
\begin{equation}
\left[\left(\ell,1\right)+\df\right]-\left[\left(\ell,1\right)+\dfl\right]-_{\mathrm{p}}.
\end{equation}
The matrix model of this brane configuration is
\begin{align}
Z^{\left(\tilde{E}_{4}\right)}\left(\eta^{\left(r\right)},\mathbf{z}^{\left(r\right)}\right) & =\frac{1}{\left(N!\right)^{2}}\int\prod_{a}^{N}\frac{d\mu_{a}}{2\pi}\frac{d\nu_{a}}{2\pi}{\cal Z}_{1,0,0}^{\left(\ell,1\right)}\left(\eta^{\left(1\right)},z^{\left(1\right)},\tilde{z}^{\left(1\right)};\mu,\nu\right){\cal Z}_{0,0,1}^{\left(\ell,1\right)}\left(\eta^{\left(2\right)},\tilde{z}^{\left(2\right)};\nu,\mu\right).
\end{align}
After an appropriate similarity transformation, we obtain the quantum curve
\begin{align}
\hat{\mathcal{O}}^{\left(\tilde{E}_{4}\right)}\left(\eta^{\left(r\right)},\mathbf{z}^{\left(r\right)};\hat{x},\hat{y}\right) & =\hat{\mathcal{O}}_{0,0,1}^{\left(\ell,1\right)}\left(\eta^{\left(2\right)},\tilde{z}^{\left(2\right)};\hat{x},\hat{y}\right)\hat{\mathcal{O}}_{1,0,0}^{\left(\ell,1\right)}\left(\eta^{\left(1\right)},z^{\left(1\right)},\tilde{z}^{\left(1\right)};\hat{x},\hat{y}\right)\nonumber \\
 & =\hat{\mathcal{O}}^{\left(E_{4}\right)}\left(\eta^{\left(r\right)},\mathbf{z}^{\left(r\right)};\hat{X},\hat{Y}\right),\label{eq:E4QC2}
\end{align}
where $\hat{\mathcal{O}}^{\left(E_{4}\right)}$ is \eqref{eq:E4QC} and
\begin{equation}
\hat{X}=\hat{x},\quad\hat{Y}=-\hat{x}+\ell\hat{y}.
\end{equation}

The $\tilde{E}_{3}$ toric diagram can be regarded as a sum of $\lp_{1,0,0}^{\left(\ell,1\right)}$ and $\lp_{0,0,0}^{\left(\ell,2\right)}$ with $\mathrm{gcd}\left(\ell,2\right)=1$. Namely, the $\tilde{E}_{3}$ toric diagram is realized by the following brane configuration with two $\left(p,q\right)$ webs
\begin{equation}
\left[\left(\ell,1\right)+\df\right]-\left[\left(\ell,2\right)\right]-_{\mathrm{p}}.
\end{equation}
The matrix model of this brane configuration is
\begin{align}
Z^{\left(\tilde{E}_{3}\right)}\left(\eta^{\left(r\right)},\mathbf{z}^{\left(1\right)}\right) & =\frac{1}{\left(N!\right)^{2}}\int\prod_{a}^{N}\frac{d\mu_{a}}{2\pi}\frac{d\nu_{a}}{2\pi}{\cal Z}_{1,0,0}^{\left(\ell,1\right)}\left(\eta^{\left(1\right)},z^{\left(1\right)},\tilde{z}^{\left(1\right)};\mu,\nu\right){\cal Z}_{0,0,0}^{\left(\ell,2\right)}\left(\eta^{\left(2\right)};\nu,\mu\right).
\end{align}
The associated quantum curve is
\begin{equation}
\hat{\mathcal{O}}^{\left(\tilde{E}_{3}\right)}\left(\eta^{\left(r\right)},\mathbf{z}^{\left(1\right)};\hat{x},\hat{y}\right)=\hat{\mathcal{O}}_{0,0,0}^{\left(\ell,2\right)}\left(\eta^{\left(2\right)};\hat{x},\hat{y}\right)\hat{\mathcal{O}}_{1,0,0}^{\left(\ell,1\right)}\left(\eta^{\left(1\right)},z^{\left(1\right)},\tilde{z}^{\left(1\right)};\hat{x},\hat{y}\right)=\hat{\mathcal{O}}^{\left(E_{3}\right)}\left(\eta^{\left(r\right)},\mathbf{z}^{\left(1\right)};\hat{X},\hat{Y}\right),\label{eq:E3QC2}
\end{equation}
where $\hat{\mathcal{O}}^{\left(E_{3}\right)}$ is \eqref{eq:E3QC} and
\begin{equation}
\hat{X}=\hat{x},\quad\hat{Y}=-\hat{x}+\ell\hat{y}.
\end{equation}

The $\tilde{E}_{2}$ toric diagram can be regarded as a sum of $\lp_{0,1,0}^{\left(\ell,1\right)}$ and $\lp_{0,0,0}^{\left(\ell,3\right)}$ with $\mathrm{gcd}\left(\ell,6\right)=1$. Namely, the $\tilde{E}_{2}$ toric diagram is realized by the following brane configuration with two $\left(p,q\right)$ webs
\begin{equation}
\left[\left(\ell,1\right)+\dfl\right]-\left[\left(\ell,3\right)\right]-_{\mathrm{p}}.
\end{equation}
The matrix model of this brane configuration is
\begin{align}
Z^{\left(\tilde{E}_{2}\right)}\left(\eta^{\left(r\right)},z^{\left(1\right)}\right) & =\frac{1}{\left(N!\right)^{2}}\int\prod_{a}^{N}\frac{d\mu_{a}}{2\pi}\frac{d\nu_{a}}{2\pi}{\cal Z}_{0,0,1}^{\left(\ell,1\right)}\left(\eta^{\left(1\right)},\tilde{z}^{\left(1\right)};\mu,\nu\right){\cal Z}_{0,0,0}^{\left(\ell,3\right)}\left(\eta^{\left(2\right)};\nu,\mu\right).
\end{align}
The associated quantum curve is
\begin{equation}
\hat{\mathcal{O}}^{\left(\tilde{E}_{2}\right)}\left(\eta^{\left(r\right)},z^{\left(1\right)};\hat{x},\hat{y}\right)=\hat{\mathcal{O}}_{0,0,0}^{\left(\ell,3\right)}\left(\eta^{\left(2\right)};\hat{x},\hat{y}\right)\hat{\mathcal{O}}_{0,0,1}^{\left(\ell,1\right)}\left(\eta^{\left(1\right)},\tilde{z}^{\left(1\right)};\hat{x},\hat{y}\right)=\hat{\mathcal{O}}^{\left(E_{2}\right)}\left(\eta^{\left(r\right)},\tilde{z}^{\left(1\right)};\hat{X},\hat{Y}\right),\label{eq:E2QC2}
\end{equation}
where $\hat{\mathcal{O}}^{\left(E_{2}\right)}$ is \eqref{eq:E2QC} and
\begin{equation}
\hat{X}=\hat{x},\quad\hat{Y}=-2\hat{x}+\ell\hat{y}.
\end{equation}

The $\tilde{E}_{1}$ toric diagram can be regarded as a sum of $\lp_{0,0,0}^{\left(\ell,1\right)}$ and $\lp_{0,0,0}^{\left(\ell,3\right)}$ with $\mathrm{gcd}\left(\ell,3\right)=1$. Namely, the $\tilde{E}_{1}$ toric diagram is realized by the following brane configuration with two $\left(p,q\right)$ webs
\begin{equation}
\left[\left(\ell,1\right)\right]-\left[\left(\ell,3\right)\right]-_{\mathrm{p}}.
\end{equation}
The matrix model of this brane configuration is
\begin{align}
Z^{\left(\tilde{E}_{1}\right)}\left(\eta^{\left(r\right)}\right) & =\frac{1}{\left(N!\right)^{2}}\int\prod_{a}^{N}\frac{d\mu_{a}}{2\pi}\frac{d\nu_{a}}{2\pi}{\cal Z}_{0,0,0}^{\left(\ell,1\right)}\left(\eta^{\left(1\right)};\mu,\nu\right){\cal Z}_{0,0,0}^{\left(\ell,3\right)}\left(\eta^{\left(2\right)};\nu,\mu\right).
\end{align}
The associated quantum curve is
\begin{equation}
\hat{\mathcal{O}}^{\left(\tilde{E}_{1}\right)}\left(\eta^{\left(r\right)};\hat{x},\hat{y}\right)=\hat{\mathcal{O}}_{0,0,0}^{\left(\ell,3\right)}\left(\eta^{\left(2\right)};\hat{x},\hat{y}\right)\hat{\mathcal{O}}_{0,0,0}^{\left(\ell,1\right)}\left(\eta^{\left(1\right)};\hat{x},\hat{y}\right)=\hat{\mathcal{O}}^{\left(E_{1}\right)}\left(\eta^{\left(r\right)};\hat{X},\hat{Y}\right),\label{eq:E1QC2}
\end{equation}
where $\hat{\mathcal{O}}^{\left(E_{1}\right)}$ is \eqref{eq:E1QC} and
\begin{equation}
\hat{X}=\hat{x},\quad\hat{Y}=-2\hat{x}+\ell\hat{y}.
\end{equation}

Notice that in this procedure we can also deal with the $E_{0}$ curve. The $\tilde{E}_{0}$ toric diagram can be regarded as $\lp_{0,0,1}^{\left(3\ell,1\right)}$ with $\mathrm{gcd}\left(\ell,2\right)=1$. Namely, the $\tilde{E}_{0}$ toric diagram is realized by the following brane configuration with one $\left(p,q\right)$ web
\begin{equation}
\left[\left(3\ell,1\right)+\dfr\right]-_{\mathrm{p}}.
\end{equation}
The matrix model of this brane configuration is
\begin{align}
Z^{\left(\tilde{E}_{0}\right)}\left(\eta,z\right) & =\frac{1}{N!}\int\prod_{a}^{N}\frac{d\mu_{a}}{2\pi}{\cal Z}_{0,1,0}^{\left(3\ell,1\right)}\left(\eta,z;\mu,\mu\right).
\end{align}
The associated quantum curve is
\begin{align}
\hat{\mathcal{O}}^{\left(\tilde{E}_{0}\right)}\left(\eta,z;\hat{x},\hat{y}\right) & =\hat{\mathcal{O}}_{0,1,0}^{\left(3\ell,1\right)}\left(\eta,z;\hat{x},\hat{y}\right)\nonumber \\
 & =\left(\begin{array}{ccc}
+e^{3\pi\ell\eta}e^{\frac{3}{4}\hat{X}-\frac{3}{4}\hat{Y}} & +0 & +0\\
+0 & +0 & +e^{-3\pi\ell-\pi z}e^{-\frac{1}{4}\hat{X}+\frac{5}{4}\hat{Y}}\\
+0 & +e^{-3\pi\ell+\pi z}e^{-\frac{5}{4}\hat{X}+\frac{1}{4}\hat{Y}} & +0
\end{array}\right),\label{eq:E1QC2-1}
\end{align}
where
\begin{equation}
\hat{X}=\hat{x}-\ell\hat{y},\quad\hat{Y}=\ell\hat{y}.
\end{equation}
The last line is indeed the $E_{0}$ curve in \eqref{eq:E1E0}.

Before closing this section, we remark about the value of $\hbar$, which we still have to restrict to $\hbar=2\pi\ell$ with a positive integer $\ell$. If we start with quantum curves and obtain the matrix models, we do not have to restrict $\hbar$, see e.g. \cite{Marino:2015ixa,Kashaev:2015wia}. However, in this paper the direction is opposite. Namely, we start with matrix models and obtain the quantum curves. The point is that in this paper we focus on the matrix models which come from the gauge theories or the brane setups. In terms of the gauge theory, the restriction $\hbar=2\pi\ell$ comes from the fact that the Chern-Simons levels must be integers, and in terms of the brane setup, the restriction comes from the fact that $p$ and $q$ must be integers for the $\left(p,q\right)$5-brane.

\subsection{Relation with the ABJM theory\label{subsec:p1D-ABJM}}

The Fermi gas formalism was first applied to the ABJM theory, and the quantum curve of the ABJM theory was found to be the $E_{1}$ type \cite{Marino:2011eh,Kashaev:2015wia}. On the other hand, the $E_{1}$ curve can also be obtained in our setup. In this section we study the relation between the quantum curves in our study and those in the ABJM theory.

The ABJM theory a the ${\cal N}=6$ superconformal CS theory with $\mathrm{U}\left(N\right)_{k}\times\mathrm{U}\left(N+L\right)_{-k}$ gauge group \cite{Aharony:2008ug,Hosomichi:2008jb,Aharony:2008gk}. Here $\pm k$ denotes the CS level and $L\geq0$. The $S^{3}$ partition function reduces to the matrix model via the localization, and the Fermi gas formalism of the matrix model is known to be \cite{Awata:2012jb,Honda:2013pea,Honda:2014npa,Kashaev:2015wia}
\begin{equation}
Z_{k}^{\mathrm{ABJM}}\left(N,N+L\right)=Z_{k}^{\mathrm{CS}}\left(L\right)\int\prod_{a}^{N}d\mu_{a}\det\left(\left[\braket{\mu_{a}|\hat{{\cal O}}_{k,L}^{\mathrm{ABJM}}\left(\hat{u},\hat{v}\right)^{-1}|\mu_{b}}\right]_{a,b}^{N\times N}\right).
\end{equation}
$\hat{{\cal O}}_{k,L}^{\mathrm{ABJM}}$ is the quantum curve of the ABJM theory
\begin{equation}
\hat{\mathcal{O}}_{k,L}^{\mathrm{ABJM}}\left(\hat{u},\hat{v}\right)=\left(\begin{array}{cc}
+e^{\frac{1}{2}i\pi\left(L-\frac{1}{2}k\right)}e^{\frac{\hat{u}}{2}-\frac{\hat{v}}{2}} & +e^{-\frac{1}{2}i\pi\left(L-\frac{1}{2}k\right)}e^{\frac{\hat{u}}{2}+\frac{\hat{v}}{2}}\\
+e^{-\frac{1}{2}i\pi\left(L-\frac{1}{2}k\right)}e^{-\frac{\hat{u}}{2}-\frac{\hat{v}}{2}} & +e^{\frac{1}{2}i\pi\left(L-\frac{1}{2}k\right)}e^{-\frac{\hat{u}}{2}+\frac{\hat{v}}{2}}
\end{array}\right),\label{eq:QC-ABJM}
\end{equation}
where $\left[\hat{u},\hat{v}\right]=2\pi ik$. $Z_{k}^{\mathrm{CS}}\left(L\right)$ is the $S^{3}$ partition function of the $\mathrm{U}\left(L\right)$ pure CS theory
\begin{equation}
Z_{k}^{\mathrm{CS}}\left(L\right)=e^{-\frac{i\pi}{6k}\left(L^{3}-L\right)}\frac{1}{k^{\frac{L}{2}}}\prod_{n=1}^{L-1}\prod_{n'=n+1}^{L}2\sin\frac{\pi}{k}\left(n'-n\right).\label{eq:PF-CS}
\end{equation}

Although we obtained the $E_{1}$ curve in \eqref{eq:E1QC2}, the $E_{1}$ curve (with general $\hbar$) can be realized in a different way. We start with the following brane configuration
\begin{equation}
\left[\left(\ell,1\right)+\df\right]-_{\mathrm{p}}.\label{eq:BC-E1}
\end{equation}
The quantum curve corresponding to this brane configuration can be read off from \eqref{eq:pqD5-QC} as
\begin{equation}
\hat{\mathcal{O}}_{1,0,0}^{\left(\ell,1\right)}\left(\eta,z,\tilde{z};\hat{x},\hat{y}\right)=e^{\pi\ell\eta-\pi\tilde{z}}e^{\hat{x}-\frac{\ell}{2}\hat{y}}+e^{\pi\ell\eta+\pi\tilde{z}}e^{-\frac{\ell}{2}\hat{y}}+e^{-\pi\ell\eta-\pi z}e^{\frac{\ell}{2}\hat{y}}+e^{-\pi\ell\eta+\pi z}e^{-\hat{x}+\frac{\ell}{2}\hat{y}}.
\end{equation}
By changing the variables as
\begin{equation}
\hat{X}=\hat{x}-\pi\left(z+\tilde{z}\right),\quad\hat{Y}=-\hat{x}+\ell\hat{y}-2\pi\ell\eta,
\end{equation}
one finds
\begin{equation}
\hat{\mathcal{O}}_{1,0,0}^{\left(\ell,1\right)}\left(\eta,z,\tilde{z};\hat{X},\hat{Y}\right)=\left(\begin{array}{cc}
+e^{\frac{\pi}{2}\left(z-\tilde{z}\right)}e^{\frac{1}{2}\hat{X}-\frac{1}{2}\hat{Y}} & +e^{-\frac{\pi}{2}\left(z-\tilde{z}\right)}e^{\frac{1}{2}\hat{X}+\frac{1}{2}\hat{Y}}\\
+e^{-\frac{\pi}{2}\left(z-\tilde{z}\right)}e^{-\frac{1}{2}\hat{X}-\frac{1}{2}\hat{Y}} & +e^{\frac{\pi}{2}\left(z-\tilde{z}\right)}e^{-\frac{1}{2}\hat{X}+\frac{1}{2}\hat{Y}}
\end{array}\right),\label{eq:QC-E1}
\end{equation}
where $\left[\hat{X},\hat{Y}\right]=2\pi\ell i$.

These two quantum curves \eqref{eq:QC-ABJM} and \eqref{eq:QC-E1} match when
\begin{equation}
k=\ell,\quad\frac{1}{2}i\pi\left(L-\frac{1}{2}k\right)=\frac{\pi}{2}\left(z-\tilde{z}\right).\label{eq:E1-ABJM-Cond}
\end{equation}
We would be able to understand the first condition by comparing the two brane configurations with the $\mathrm{SL}\left(2,\mathbb{Z}\right)$ transformation. The ABJM theory is realized by the following brane configuration \cite{Aharony:2008ug}
\begin{equation}
\left[\left(1,0\right)\right]-\left[\left(1,k\right)'\right]-_{\mathrm{p}}.\label{eq:BC-ABJM}
\end{equation}
Here $\left(1,k\right)'$ denotes the $\left(1,k\right)$5-brane extending along 012$\left[3,7\right]_{\theta}$$\left[4,8\right]_{\theta}$$\left[5,9\right]_{\theta}$ with $\tan\theta=k$ for preserving the ${\cal N}=3$ supersymmetry. Ignoring this point, one can see that the content of the 5-branes of the $E_{1}$ brane configuration \eqref{eq:BC-E1} and the ABJM brane configuration \eqref{eq:BC-ABJM} is related by the $S$ transformation when $k=\ell$.

Even after the $S$ transformation, on the one hand the D5-brane forms $\left(p,q\right)$ web with the NS5-brane in \eqref{eq:BC-E1}, and on the other hand the two 5-branes are separated along the direction 6 in \eqref{eq:BC-ABJM}. This difference can be explained more explicitly when we focus on the $k=\ell=1$ and $L=0$ case, where the second condition of \eqref{eq:E1-ABJM-Cond} turns out to play an important role. In this case, with the definition of $z,\tilde{z}$ in \eqref{eq:z-Def}, the second condition of \eqref{eq:E1-ABJM-Cond} means
\begin{equation}
m=\tilde{m},\quad D-\tilde{D}=\frac{1}{2}.\label{eq:E1-ABJM-Cond2}
\end{equation}
As we see below, this means the removing of the D5-brane of $\left[\left(1,1\right)+\df\right]$ in the $-x^{6}$ direction in the $E_{1}$ brane configuration \eqref{eq:BC-E1}, so that it becomes $\left[\left(0,1\right)\right]-\left[\left(1,1\right)\right]-_{\mathrm{p}}$. Note that $Z_{k}^{\mathrm{CS}}\left(0\right)=1$.

Let us consider slightly more general $\left(p,q\right)$ web, $\underset{1}{-}\left[\left(1,q\right)+\df\right]\underset{2}{-}.$ We move the D5-brane in the first interval. Because the D5-brane cannot break in the direction 5 once it leaves the NS5-brane, we need to adjust the masses to be equal 
\begin{equation}
m=\tilde{m}.\label{eq:Remove-mass}
\end{equation}
The move of the D5-brane to the first interval with the length $x^{6}$ gives a superpotential to the (anti-)fundamental chirals on the second interval $W=x^{6}Q^{\left(2\right)}\tilde{Q}^{\left(2\right)}$ \cite{Brunner:1998jr}. This superpotential gives an additional constrain to the R-charges on the second interval
\begin{equation}
\Delta^{\left(2\right)}+\tilde{\Delta}^{\left(2\right)}=2.\label{eq:Remove-R2}
\end{equation}
With the condition \eqref{eq:R-cond}, one also finds a relation among R-charges on the first interval
\begin{equation}
\Delta^{\left(1\right)}+\tilde{\Delta}^{\left(1\right)}=1.\label{eq:Remove-R1}
\end{equation}
This is the same with the relation among the R-charges of the (anti-)fundamental matters coming from the D5-brane on an interval \eqref{eq:R-cond2}. Motivated by this, we define
\begin{equation}
\frac{1}{2}-\Delta^{\left(1\right)}=-\left(\frac{1}{2}-\tilde{\Delta}^{\left(1\right)}\right)=\bar{D}.
\end{equation}
Under these conditions, factors in the D5-brane factor \eqref{eq:D5wd-MF} becomes
\begin{equation}
\frac{s_{1}\left(\frac{\mu}{2\pi}-\tilde{m}-i\tilde{D}+\frac{i}{4}\right)}{s_{1}\left(\frac{\mu}{2\pi}-m-iD-\frac{i}{4}\right)}=\frac{1}{2\cosh\frac{\mu-2\pi\left(m+i\bar{D}\right)}{2}},\quad\frac{s_{1}\left(\frac{\nu}{2\pi}-m-iD+\frac{i}{4}\right)}{s_{1}\left(\frac{\nu}{2\pi}-\tilde{m}-i\tilde{D}-\frac{i}{4}\right)}=1.
\end{equation}
The second equality is consistent with the fact that the superpotential $Q^{\left(2\right)}\tilde{Q}^{\left(2\right)}$ causes the chirals $Q^{\left(2\right)}$ and $\tilde{Q}^{\left(2\right)}$ to gain a mass, and hence they do not contribute to the low energy theory. Thus the matrix factor for $\left[\left(1,q\right)+\df\right]$ becomes
\begin{equation}
\mathcal{Z}_{\mathrm{D5}}\left(m+iD,\tilde{m}+i\tilde{D};\mu,\nu\right){\cal Z}_{0,0,0}^{\left(1,q\right)}\left(\eta;\mu,\nu\right)=\frac{1}{N!}\int\prod_{a}^{N}d\rho_{a}\mathcal{Z}^{\left(0,1\right)}\left(m+i\bar{D};\mu,\rho\right){\cal Z}_{0,0,0}^{\left(1,q\right)}\left(\eta;\rho,\nu\right).
\end{equation}
The right hand side is indeed the matrix factor for $\left[\left(0,1\right)\right]-\left[\left(1,q\right)\right]$. Now, we find that the first condition of \eqref{eq:E1-ABJM-Cond2} corresponds to \eqref{eq:Remove-mass} and the second condition of \eqref{eq:E1-ABJM-Cond2} with \eqref{eq:R-Para} corresponds to \eqref{eq:Remove-R2} and \eqref{eq:Remove-R1}. Note that we can remove a D5-brane from more general $\left(p,q\right)$ web $\left[\left(1,q\right)+F\df+F_{\wdr}\dfr+F_{\wdl}\dfl\right]$. Indeed, the above identity holds also for ${\cal Z}_{F,F_{\wdr},F_{\wdl}}^{\left(1,q\right)}$.

\section{Conclusion\label{sec:Conclution}}

In this paper we proposed the relation between the $\mathcal{N}=2$ brane configurations in type IIB string theory and the quantum curves via the matrix models computing the $S^{3}$ partition functions. The $\mathcal{N}=2$ brane configuration is a consequence of $\left(p,q\right)$ webs, and we conjectured that the partition function can be written as the product of the matrix factors described by the quantum curves whose Newton polygons are the dual toric diagrams of the $\left(p,q\right)$ webs. We proved the conjecture for the Lagrangian theories by using the localization and the Fermi gas formalism. We also performed the consistency checks with the web deformations and the $\mathrm{SL}\left(2,\mathbb{Z}\right)$ transformations. As an application, we proposed the new matrix models for $\left(p,q\right)$ webs including a $\left(p,q\right)$5-brane with $p\geq2$. Then, by using this new matrix models, we provide the brane configurations and the matrix model representations for the genus one curves, where $\hbar$ is not only $2\pi$ but also $2\pi\ell$ with $\ell\in\mathbb{N}$.

There are various interesting directions for further study.

Because we realized the genus one curves as the 3d theories with the brane configurations and gave the matrix model representations, it would be nice if one could reveal relations between these theories and the topological strings or the $q$-Painlev\'e systems. The correspondence between the topological strings and the spectral theories (TS/ST correspondence) for genus one curves has already been studied in \cite{Grassi:2014zfa,Moriyama:2020lyk}, and this correspondence would be useful to reveal the connection with the topological strings. The conjecture also includes the relation between brane configurations and higher genus quantum curves. The higher genus version of the TS/ST correspondence has also been studied in \cite{Codesido:2015dia,Codesido:2016ixn}, and this would be useful for expanding the story for the genus one curves to higher genus cases. The integrable equations can be used for studying the $S^{3}$ partition functions \cite{Nosaka:2024gle}, and thus we expect that the explicit relation between the genus one 3d theories and the $q$-Painlev\'e systems helps in studying the $S^{3}$ partition functions of them.

It would be interesting to reveal the relation between the quantum curves we obtained and the 5d Seiberg-Witten curves \cite{Witten:1997sc}. The 5d theories are realized by $\left(p,q\right)$ webs, which can be uplifted to M5-branes, and the geometry of the M5-brane is encoded in the 5d Seiberg-Witten curve. On the other hand, we saw that the quantum curve also has the information of the $\left(p,q\right)$ web. This would be more concrete evidence that our quantum curves are related to the 5d Seiberg-Witten curves.

Although we called the dual grid diagram of a $\left(p,q\right)$ web the toric diagram, we do not understand a toric Calabi-Yau manifold behind that. On the one hand, the TS/ST correspondence claims that the quantum curves can be regarded as quantized mirror curves of toric Calabi-Yau threefolds. On the other hand, 3d theories describing M2-branes are holographic dual to M-theory on $\mathrm{AdS}_{4}\times Y_{7}$, where the cone of $Y_{7}$ is a toric Calabi-Yau fourfold, and generalities of this is studied in a context of the brane tilings \cite{Hanany:2008cd,Ueda:2008hx,Imamura:2008qs}. It would be important to clarify this point. Note that in \cite{Jensen:2009xh} the same brane setup was investigated, and the relation between the $\left(p,q\right)$ webs and Calabi-Yau manifolds was discussed.

Our work would give new results to the M-theory via holography. For example, the Fermi gas formalism and especially the quantum curve representation are useful for studying the large $N$ behavior in a small $k$ region, which is the M-theory region \cite{Marino:2011eh}. Because our result gives a quantum curve expression for a wide class of 3d $\mathcal{N}=2$ gauge theories, our result can be used for checking or predicting the behavior of the holographic dual free energies on various backgrounds, which include not only the perturbative effects but also non-perturbative effects.

It would be interesting to generalize our result. For example, although we considered only the uniform ranks case, we can consider more general rank theories. This deformation would be important for providing additional parameters to the quantum curves. We can also consider gauge groups rather than the circular quiver diagrams with unitary groups. For example, we can consider orientifold theories or $D$-type quiver theories. Because we can realize them by inserting orientifold branes, and hence they have a brane interpretation, it would be nice if we could give a direct relation between these brane configurations and the Newton polygon of the quantum curves.

In section \ref{subsec:SL2Ztrans}, we saw that the similarity transformations of the quantum curves become the dualities between the 3d theories. Similarity transformations have been used for introducing symmetries of quantum curves, and new dualities have been discovered via the symmetries \cite{Kubo:2018cqw,Kubo:2019ejc,Kubo:2021enh}. We expect that the symmetries also provide new dualities between 3d theories appearing in our setup.

\section*{Acknowledgements}

We are grateful to Sanefumi Moriyama, Tomoki Nosaka and Yi Pang for valuable discussions. NK is partially supported by National key research and development program under grant No. 2022YFE0134300.

\appendix

\section{The double sine function\label{sec:DoubleSine}}

In this section we summarize the definition and properties of the double sine function (see e.g. \cite{Bytsko:2006ut}). The double sine function is defined by
\begin{equation}
s_{\mathsf{b}}\left(z\right)=\prod_{m,n=0}^{\infty}\frac{m\mathsf{b}+n\mathsf{b}^{-1}+\frac{1}{2}\left(\mathsf{b}+\mathsf{b}^{-1}\right)-iz}{m\mathsf{b}+n\mathsf{b}^{-1}+\frac{1}{2}\left(\mathsf{b}+\mathsf{b}^{-1}\right)+iz}.
\end{equation}
This function satisfies the following relations
\begin{equation}
s_{\mathsf{b}}\left(0\right)=1,\quad s_{\mathsf{b}}\left(z\right)=s_{\mathsf{b}^{-1}}\left(z\right),\quad s_{\mathsf{b}}\left(z\right)s_{\mathsf{b}}\left(-z\right)=1,\quad\overline{s_{\mathsf{b}}\left(z\right)},=s_{\mathsf{b}}\left(-\bar{z}\right).\label{eq:DS-id}
\end{equation}
Especially, the following relation is important
\begin{align}
\frac{s_{\mathsf{b}}\left(z+\frac{i}{2}\mathsf{b}^{\pm1}\right)}{s_{\mathsf{b}}\left(z-\frac{i}{2}\mathsf{b}^{\pm1}\right)} & =\frac{1}{2\cosh\left(\pi\mathsf{b}^{\pm1}z\right)}.\label{eq:DS-Trig}
\end{align}
The asymptotic behavior of the double sine function also plays an important role when we consider the web deformation
\begin{equation}
s_{\mathsf{b}}\left(z\right)\sim\begin{cases}
\exp\left(\frac{i\pi z^{2}}{2}+\frac{\pi i}{24}\left(\mathsf{b}^{2}+\mathsf{b}^{-2}\right)\right) & \left({\rm Re}\left[z\right]\rightarrow\infty\right)\\
\exp\left(-\frac{i\pi z^{2}}{2}-\frac{\pi i}{24}\left(\mathsf{b}^{2}+\mathsf{b}^{-2}\right)\right) & \left({\rm Re}\left[z\right]\rightarrow-\infty\right)
\end{cases}.\label{eq:DSasym}
\end{equation}

\section{Quantum mechanics\label{sec:QMnotation}}

In this appendix, we provide the notation and formulas of the quantum mechanics. Let $\hat{x}$ and $\hat{y}$ be the position and momentum operators, respectively. The commutation relation of these operators is $\left[\hat{x},\hat{y}\right]=i\hbar$. (In the main text $\hbar$ is always $2\pi$ except for section \ref{sec:QC-genh}). The normalization of a position eigenvector $\ket x$ and a momentum eigenvector $\kket y$ are
\begin{align}
 & \braket{x_{1}|x_{2}}=\delta\left(x_{1}-x_{2}\right),\quad\bbrakket{y_{1}|y_{2}}=\delta\left(y_{1}-y_{2}\right),\nonumber \\
 & \brakket{x|y}=\frac{1}{\sqrt{2\pi\hbar}}e^{\frac{i}{\hbar}xy},\quad\bbraket{y|x}=\frac{1}{\sqrt{2\pi\hbar}}e^{-\frac{i}{\hbar}xy}.\label{eq:Normalization}
\end{align}
The eigenvectors satisfy the following relations
\begin{equation}
\kket y=\frac{1}{\sqrt{i}}e^{\frac{i}{2\hbar}\hat{x}^{2}}e^{\frac{i}{2\hbar}\hat{y}^{2}}e^{\frac{i}{2\hbar}\hat{x}^{2}}\ket y,\quad\bbra y=\sqrt{i}\bra ye^{-\frac{i}{2\hbar}\hat{x}^{2}}e^{-\frac{i}{2\hbar}\hat{y}^{2}}e^{-\frac{i}{2\hbar}\hat{x}^{2}}.\label{eq:VecSim}
\end{equation}
The operators satisfy the following relations
\begin{align}
 & e^{-\frac{ic}{\hbar}\hat{x}}f\left(\hat{y}\right)e^{\frac{ic}{\hbar}\hat{x}}=f\left(\hat{y}+c\right),\quad e^{-\frac{ic}{\hbar}\hat{y}}f\left(\hat{x}\right)e^{\frac{ic}{\hbar}\hat{y}}=f\left(\hat{x}-c\right),\nonumber \\
 & e^{-\frac{ic}{2\hbar}\hat{x}^{2}}f\left(\hat{y}\right)e^{\frac{ic}{2\hbar}\hat{x}^{2}}=f\left(\hat{y}+c\hat{x}\right),\quad e^{-\frac{ic}{2\hbar}\hat{y}^{2}}f\left(\hat{x}\right)e^{\frac{ic}{2\hbar}\hat{y}^{2}}=f\left(\hat{x}-c\hat{y}\right).\label{eq:OpSim}
\end{align}
The Campbell-Baker-Hausdorff formula for the position and the momentum operators is
\begin{equation}
e^{c_{1}\hat{x}}e^{c_{2}\hat{y}}=e^{\frac{i}{2}c_{1}c_{2}\hbar}e^{c_{1}\hat{x}+c_{2}\hat{y}}=e^{ic_{1}c_{2}\hbar}e^{c_{2}\hat{y}}e^{c_{1}\hat{x}}.\label{eq:CBHform}
\end{equation}
Thanks to the Fourier transformation formula, $\cosh^{-1}$ function can be written in terms of the quantum mechanics
\begin{equation}
\frac{1}{2\cosh\frac{\mu-\nu}{2c}}=\frac{1}{2\pi}\int_{-\infty}^{\infty}dy\frac{e^{\frac{i}{2\pi c}y\left(\mu-\nu\right)}}{2\cosh\frac{y}{2}}=2\pi c\braket{\mu|\frac{1}{2\cosh\left(\frac{\pi c}{\hbar}\hat{y}\right)}|\nu}.\label{eq:Cosh-op}
\end{equation}

\printbibliography

\end{document}